\documentclass[11pt,letterpaper]{article}
\pdfoutput=1
\usepackage{soul}
\usepackage{jheppub}
\usepackage{amsfonts}
\usepackage{amsmath}
\allowdisplaybreaks[4]         
\usepackage{amssymb}   
\usepackage{euscript}       
\usepackage{color} 
\usepackage{mathrsfs}  
\usepackage{booktabs}
\usepackage{siunitx}
\usepackage{amsmath,latexsym,xcolor,mathtools}
\usepackage{graphicx,verbatim, subcaption, caption,inputenc}

\newcommand{\bea}{\begin{eqnarray}}
\newcommand{\eea}{\end{eqnarray}}
\newcommand{\ba}{\begin{eqnarray}}
\usepackage{braket}
\newcommand{\ea}{\end{eqnarray}}

\newcommand{\beq}{\begin{equation}}
\newcommand{\eeq}{\end{equation} }
\newcommand{\beqa}{\begin{eqnarray}}

\newcommand{\eeqa}{\end{eqnarray}}
\newcommand{\beqar}{\begin{eqnarray*}}
\newcommand{\eeqar}{\end{eqnarray*}}
\newcommand{\e}{\epsilon}

\newcommand{\be}{\begin{equation}}
\newcommand{\ee}{\end{equation}}

\newcommand{\rdiff}{\Delta r}

\newcommand{\X}{\mathcal{X}}
\newcommand{\W}{\mathcal{W}}
\newcommand{\R}{\mathcal{R}}
\newcommand{\Z}{\mathcal{Z}}
\newcommand{\A}{\mathcal{A}}

\newcommand{\C}{\mathcal{C}}
\newcommand{\D}{\mathcal{D}}

\newcommand{\G}{\mathcal{G}}

\renewcommand{\c}{$c$}

\newcommand{\rD}{{r_\Delta}}

\newcommand{\tR}{t_{\rm R}}
\newcommand{\tL}{t_{\rm L}}

\def\ket#1{\mathinner{|{#1}\rangle}}

  \newcommand{\rah}[1]{{\color{green}\bf [Robie: #1]}}

\DeclareMathOperator*{\argmin}{arg\,min}

\title{Holographic complexity of rotating black holes}
\author[a]{Abdulrahim Al Balushi,}
\author[b]{Robie A. Hennigar,} 
\author[b]{Hari K. Kunduri}
\author[a]{and Robert B. Mann}

\emailAdd{a2albalu@uwaterloo.ca}
\emailAdd{rhennigar@mun.ca}
\emailAdd{hkkunduri@mun.ca}
\emailAdd{rbmann@uwaterloo.ca}
\affiliation[a]{Department of Physics and Astronomy, University of Waterloo, Waterloo, Ontario, N2L 3G1, Canada}
\affiliation[b]{Department of Mathematics and Statistics, Memorial University of Newfoundland, St. John's, Newfoundland and Labrador, A1C 5S7, Canada}

\date{\today}

\abstract{Within the framework of the ``complexity equals action" and ``complexity equals volume" conjectures, we study the properties of holographic complexity for rotating black holes. We focus on a class of odd-dimensional equal-spinning black holes for which considerable simplification occurs. We study the complexity of formation, uncovering a direct connection between complexity of formation and thermodynamic volume for large black holes. We consider also the growth-rate of complexity, finding that at late-times the rate of growth approaches a constant, but that Lloyd's bound is generically violated. 
}

\begin{document} 
\maketitle
\flushbottom


\section{Introduction}

Holographic duality within the framework of the anti de Sitter/Conformal Field Theory
 correspondence  (AdS/CFT) \cite{Maldacena1998} continues to be the basis of many interesting connections between quantum information and gravity. Geometric quantities in bulk AdS spacetime can be precisely related to entanglement properties of the boundary CFT, most notably through the Ryu-Takayanagi construction \cite{Ryu2006,Casini2011}. Studies of the growth of the Einstein-Rosen (ER) bridge in AdS black holes have led to speculations of its duality to the growth of complexity of the dual boundary state \cite{Susskind2016}. This was refined to new conjectured entries in the AdS/CFT dictionary: the complexity-volume (CV) conjecture \cite{Susskind:2014rva,Stanford2014} and the complexity-action (CA) conjecture \cite{Brown2016a,Brown2016}.

Complexity of quantum states is a measure of how hard it is to prepare a particular target state $\ket{\psi_T}$ from a given reference state $\ket{\psi_R}$ and an initial set of elementary gates $\mathcal{G}$
\begin{align}
V_n&\equiv g_n\dots g_1g_0
\end{align}
where $g_0,\dots g_n\in\mathcal{G}$. The complexity of a state $\ket{\psi_T}$ is then defined as the minimum number $n$ of elementary gates that can approximate it according to some norm, starting from a fixed reference state $\ket{\psi_R}$
\begin{equation}
\mathcal{C}(\ket{\psi_T}) = \argmin\limits_{n}{||\ket{\psi_T}-V_n\ket{\psi_R}||}^2
\end{equation}
In addition to discrete circuit models, complexity can also be defined for systems with continuous Hamiltonian evolution generated by
\begin{equation}
U(t)=\overleftarrow{\mathcal{T}}\exp\left[-i\int_0^t H(t') dt'\right],\qquad H(t)= \sum_kY^k(t)M_k
\end{equation}
with boundary conditions $U(0)=I$ and $U(1)=V_n$, where $M_k$ are the basis Hermitian generators of the Hamiltonian, and $Y^k(t)$ are the time-dependent control functions specifying the tangent vector $\vec{Y}(t)$ of a trajectory in the space of unitaries \cite{Nielsen2005}. The time-ordering operator $\overleftarrow{\mathcal{T}}$ ensures that earlier terms in the expansion of the evolution operator $U(t)$ act on the state before later terms --- i.e. going from right to left. Thus, continuous Hamiltonian evolution defines a path in the space of unitaries of the circuit whose length is \cite{Nielsen2006,Dowling2006}
\begin{equation}
\mathcal{D}(U)=\int_0^1F(U(t),\dot{U}(t)) dt'
\end{equation}
where the cost function $F(U(t),\dot{U}(t))$ is a local functional of positions along $U(t)$ in the space of unitaries, with the overdot denoting a $t$ derivative.\footnote{The properties of the cost function and its possible forms are discussed in \cite{Nielsen2005}.} Thus, 
\begin{equation}
\mathcal{C}(\ket{\psi_T})=\min\limits_{U(t)}\mathcal{D}(U)
\end{equation}
An ongoing topic of active research is the extension of the concept of complexity to quantum field theories using the above geometric formulation of complexity (for example, see \cite{Jefferson2017,Chapman2017,Sinamuli:2018jhm,Khan2018,Yang2018,Sinamuli:2019utz}). The above definition of complexity clearly has many ambiguities  \cite{Chapman2017,Carmi2017} associated with the choice of reference states, basis operators, and cost function, which is expected to be related to the ambiguities associated with calculating the action in CA proposal \cite{Lehner2016}. 


Complexity was originally discussed in the context of holography as the dual to the volume of the ER bridge in eternal black holes \cite{Susskind2016}. The eternal Schwarzschild-AdS black hole is dual to two copies of the CFT prepared in the thermofield double state~\cite{Maldacena2003}. The volume of the ER bridge continues to grow in time even after the system thermalizes, suggesting at any putative CFT dual to this quantity must be something that continues to evolve after equilibrium is reached~\cite{Hartman:2013qma, Susskind:2014moa}. It was proposed that this growth captures some notion of complexity for the CFT state.


The idea that the growth of the black hole interior is connected to computational complexity has come to be refined into a number of concrete proposals, the most studied of which are the CV and CA conjectures. The CV conjecture proposed that the complexity of the TFD state at boundary section $\Upsilon$ is equal to the volume of the extremal/maximal spacelike 
slice $\mathcal{B}$ anchored at $t_L$ and $t_R$ at the boundaries \cite{Stanford2014}
\begin{equation}
\mathcal{C}_\mathcal{V}(\Upsilon)=\max\limits_{\Upsilon=\partial\mathcal{B}}\left[\frac{\mathcal{V}(\mathcal{B})}{G_NR}\right]
\end{equation}
where $R$ is a length scale associated with the bulk geometry (usually taken to be the AdS length $\ell$) chosen to make the complexity dimensionless. This was generalized to the CA conjecture\footnote{For a discussion of the original motivation of the CA conjecture see \cite{Brown2016}}, where complexity depends on the whole domain of dependence of $\mathcal{B}$ --- a region called the Wheeler-DeWitt (WDW) patch~\cite{Brown2016a}. Explicitly, the CA conjecture asserts that the complexity  of the CFT state is given by the numerical value of the gravitational action evaluated on the WDW patch:
\begin{equation}
\mathcal{C}_\mathcal{A}(\Upsilon)=\frac{I_{\text{WDW}}}{\pi\hbar} \, .
\end{equation}

Both the CV and CA conjectures have received considerable attention and basic properties of each are now well-established. Initially, attention was given to the idea that, within the CA proposal, the late-time growth of complexity for the Schwarzschild-AdS black hole is $\pi \dot{\mathcal{C}}_\mathcal{A} = 2M$~\cite{Brown2016a, Brown2016}. This was a suggestive connection with Lloyd's bound and was argued to support the idea that black holes are the fastest computers in nature~\cite{Lloyd2000}. However, subsequent careful analysis revealed that this late-time value is actually approached from above rather than from below, as Lloyd's bound would require~\cite{Carmi2017}. It is now believed that the assumptions required for Lloyd's bound may be incompatible with holography~\cite{Cottrell:2017ayj, Jordan:2017vqh}. Nonetheless, there have been several rather interesting connections uncovered between complexity and black hole thermodynamics in both proposals, but the situation is especially clear in the CA proposal. For example, in the CA proposal the late-time growth rate of complexity for two-horizon geometries reduces to the difference in internal energies (or enthalpies) between the inner and outer horizons:
\be 
\pi \dot{\mathcal{C}}_\mathcal{A} = U_+ - U_- \equiv (F_+ + T_+ S_+) - (F_- + T_- S_-) \, ,
\ee
where $F$ is the free energy, $S$ the entropy, and $T$ the Hawking temperature, while the $+/-$ corresponds to the outer/inner horizon, respectively. This relationship was first observed in Einstein gravity in~\cite{Brown2016}, and then argued to hold for general theories of gravity in~\cite{Huang:2016fks}, and established rigorously for the full Lovelock family of gravitational theories in~\cite{Cano2018} (see also~\cite{Jiang:2018pfk}). Many other properties have been explored, e.g., the effects of topology~\cite{Reynolds:2017jfs, Fu:2018kcp, Sinamuli:2018jhm, Andrews:2019hvq}. If there are topological identifications in the spacetime then  the complexity is rescaled by a factor dependent on the identifications \cite{Sinamuli:2018jhm}.

In many instances, the properties of complexity are qualitatively similar in both the CV and CA proposals. For example, both proposals account for the expected linear time dependence at late times~\cite{Stanford2014, Brown2016} and both exhibit the switchback effect, which is the expected response of complexity to perturbations of the state at early times~\cite{Stanford2014, Chapman:2018dem, Chapman:2018lsv}. However, there are some situations in which the two proposals differ in their behaviour~\cite{Carmi:2016wjl, Chapman:2018lsv, Fan:2018xwf, Chapman:2018bqj, Andrews:2019hvq,  Bernamonti:2019zyy, Bernamonti:2020bcf}. Understanding universal and divergent aspects of the two proposals is useful as there does not yet exist a first-principles derivation for complexity in the holographic dictionary.

Besides the time-dependent complexity rate of growth, another quantity of interest is the complexity of formation \cite{Chapman2017Form} of a black hole 
\begin{equation}
\Delta\mathcal{C}_\mathcal{A}(\Upsilon)=\frac{1}{\pi\hbar}\left[I_{\text{WDW}}(\text{BH})-2I_{\text{WDW}}(\text{AdS})\right]
\label{formationC}
\end{equation}
which measures the additional complexity present in preparing the thermofield double state in two copies of the CFT compared to two copies of the vacuum alone.  The complexity of formation was first defined and discussed in \cite{Chapman2017Form} for Schwarzschild-AdS black holes in various dimensions, where it was found that it grows linearly with entropy in the high-temperature (equivalently, large black hole) limit --- that is, 
$\Delta\mathcal{C}_A \sim k_dS$, for a constant $k_d$ that depends on the (boundary) dimension $d>3$. These considerations were extended to charged black holes in~\cite{Carmi2017} where it was found that the functional dependence of the complexity of formation is more complicated, but its dependence on the size of the black hole was still found to be controlled by the entropy in the limit of large black holes.

Our purpose here is to  study various aspects of the holographic complexity conjectures for rotating black holes. The study of rotating black holes in the context of AdS/CFT was initiated in \cite{Hawking-rotation,Mann:1999bt,Hawking:1999dp,Berman:1999mh,Das:2000cu}, where the thermodynamic properties of the black holes were compared with those of the boundary CFT. This holographic picture was further developed for astrophysical black holes with the ``Kerr/CFT correspondence" \cite{Guica:2008mu}, which conjectures that quantum gravity near the horizon of an extremal Kerr black hole is dual to a two-dimensional CFT (for reviews see \cite{Bredberg:2011hp,Compere:2012jk}). Rotating black holes are dual to thermofield double states with an additional chemical potential 
\be\label{rTFD}
\ket{\text{rTFD}}=\frac{1}{\sqrt{Z(\beta,\{\mu_i\})}}\sum_n e^{-\beta E_n/2}e^{-\beta\mu J_n/2}\ket{E_n,J_n}_L\otimes\ket{E_n,J_n}_R
\ee
associated with the rotation,
where $\mu\equiv\mu_1+\dots+\mu_{(D-1)/2}$,  and $\mu_i$ is the chemical potential associated with the angular momentum $J_i$ along the $\phi_i$ circle, with $Z(\beta,\{\mu_i\})$  the grand canonical partition function. The time evolution of the state is modified by the chemical potentials
\be\label{fullTFD}
\ket{\text{rTFD}(t_L,t_R)}=e^{-i\left(H_L+\mu J_L\right)t_L-i\left(H_R+\mu J_R\right)t_R}\ket{\text{rTFD}}
\ee
where $(H_L,J_L)$ and $(H_R,J_R)$ are the Hamiltonians and angular momentum operators for the left and right boundaries, respectively.

To date, there have been only a few studies focussing on the effects of rotation in the context of complexity, and these  studies are further limited to a derivation of the late-time rate of growth. The late-time complexity growth of Kerr-AdS black holes in CA conjecture was calculated in \cite{Cai2016}. The effect of a probe string attached to a rotating black hole on its complexity was studied in \cite{Nagasaki:2018csh}. One reason that a more detailed analysis is not straightforward
is the more complicated causal structure of rotating black holes. In the case of rotating spacetimes, carrying out a computation of the action for a WDW patch (or of the volume of a spacelike slice) is a technically formidable task.  The description of null hypersurfaces is somewhat complicated even for 4 spacetime dimensions \cite{AlBalushi2019}, and no generalization to higher-dimensional cases presently exists. Fortunately there is a special case that renders the computations tractable: Myers-Perry-AdS spacetimes in odd dimensions with equal angular momenta in each orthogonal rotation plane.   Compared to the most general Myers-Perry-AdS black holes, these solutions enjoy enhanced symmetry that considerably simplifies the analysis of the causal structure. This particular configuration has some similarities with the charged case \cite{Sinamuli:2019utz,Chapman:2019clq}, however, we shall see that there are interesting differences.



One of our main motivations for  considering  rotating black holes is to help develop an understanding of how the CV and CA proposals behave for less symmetric spacetimes. In the context of the AdS/CFT correspondence, understanding how a quantity responds to deformations of the state or the theory itself has been a fruitful approach in understanding which relationships may be universal and which may be specific to the state or theory. For example, this approach has been used with some success in the context of higher-curvature theories of gravity. Those theories introduce additional parameters into the action, which can then be used to discern between the various possible CFT charges. This method has also been used to understand the limitations of the Kovtun-Son-Starinets bound~\cite{Brigante:2007nu}, argue for the existence of $c$-theorems in arbitrary dimensions~\cite{Myers:2010jv, Myers:2010tj}, and generate conjectures for the universal behaviour of terms in entanglement entropy or partition function~\cite{Mezei:2014zla, Bueno:2015rda, Bueno:2018yzo}. Similarly, our hope here is that  the more complicated metric structure of rotating black holes will help to discern both universal  features of and particular distinctions between  the CV and CA proposals.

Along these lines, one of the main results of this paper concerns a connection between the thermodynamic volume of the black hole and the complexity of formation in both the CV and CA proposals. The thermodynamic volume is a quantity that arises naturally when one extends the definition of Komar mass from the asymptotically flat to asymptotically AdS setting~\cite{Kastor:2009wy, Cvetic:2010jb}. It also appears in the first law of black hole mechanics, governing the response of the mass to variations in the cosmological constant which, in this case, is interpreted as a pressure. In general, the thermodynamic volume is an independent thermodynamic potential. However in certain cases (such as those involving spherical symmetry) the thermodynamic volume and entropy are simply related via $S \propto V^{(D-2)/(D-1)}$. In some instances, the thermodynamic volume can be related to the spacetime volume inside the black hole~\cite{Cvetic:2010jb, Bordo:2020ryp}. This fact has motivated some authors to consider its relevance in the context of holographic complexity. However, the results so obtained have either involved new proposals for complexity~\cite{Couch:2016exn, Fan:2018wnv}, or have used thermodynamic identities to understand results in terms of the thermodynamic volume for interpretational reasons~\cite{Huang:2016fks, Liu:2019mxz, Sun:2019yps}. Our result is, to the best of our knowledge, the first to draw a clear connection between thermodynamic volume and the original CV and CA conjectures. We have reported on this result elsewhere~\cite{Balushi:2020wkt}, and here provide additional details and context. While the meaning of thermodynamic volume in the holographic context is understood (it controls the response of the dual field theory to changes in the number of colours and changes in the volume of the space on which the theory is defined~\cite{Karch:2015rpa}), its utility in holography remains rather undeveloped (though see~\cite{Johnson:2014yja, Kastor:2014dra, Caceres:2016xjz, Sinamuli:2017rhp, Johnson:2018amj, Johnson:2019wcq, Rosso:2020zkk} for progress in this direction). Our result may be viewed as an initial step toward developing the utility of thermodynamic volume in holography.


The paper is organized as follows. In section \ref{sec2}, the geometry and causal structure of the Myers-Perry-AdS spacetimes is given. Section \ref{sec3} describes the terms of the action calculations that needs to be evaluated to calculate the complexity according to the CA conjecture as well as the framework to calculate the extremal volume in CV conjecture. In section \ref{sec4}, we calculate the complexity of formation of the state \eqref{rTFD} in reference to the vacuum AdS state, according to both the CA and CV conjectures. In section \ref{sec5}, we present the full time evolution of complexity rate of growth in both the CA and CV conjectures. We discuss the implications of our results and point toward possible future directions in section \ref{sec6}. A number of technical details and supporting calculations are left to the appendices.

Unless explicitly stated otherwise, we will use natural units $\hbar=c=k_B=1$ below.

\section{Myers-Perry-AdS Spacetimes with Equal Angular Momenta}\label{sec2}

\subsection{Solution and global properties}

The Myers-Perry-AdS solution in odd dimension $D=2N+3$ is a cohomogeneity-$(N+1)$ metric with isometry group $\mathbb{R} \times U(1)^{N+1}$, described by its mass $M$ and $N+1$ independent angular momenta $J_i$~\cite{Gibbons:2004uw}.  In the special case in which all angular momenta $J_i$, $i = 1 \ldots N+1$ are equal,  there are considerable simplifications and  the metric depends only on a single radial coordinate and on the parameters $(m,a)$ \cite{Kunduri:2006qa}:
\begin{equation}\label{metric}
ds^2 = -f(r)^2 dt^2 + g(r)^2 dr^2 + h(r)^2 \left[ d\psi +  A - \Omega(r) dt\right]^2  + r^2 \hat{g}_{ab} dx^a dx^b
\end{equation} where
\begin{equation}
g(r) ^2 = \left( 1+ \frac{r^2}{\ell^2} - \frac{2m \Xi}{r^{2N}} + \frac{2m a^2}{r^{2N + 2}}\right)^{-1} , \quad h(r)^2 = r^2 \left( 1 + \frac{2m a^2}{r^{2N+2}}\right), \quad \Omega(r) = \frac{2ma}{r^{2N} h^2}, 
\end{equation} and
\begin{equation}
f(r) = \frac{r}{g(r) h(r)}, \qquad \Xi = 1 - \frac{a^2}{\ell^2}. 
\end{equation} We take $m  > 0$ and by sending $t \to -t$, we can without loss of generality always choose $a \geq 0$.  The metric $\hat{g}$ is the Fubini-Study metric on $\mathbb{CP}^N$ with curvature normalized so that $\text{Ric}(\hat{g}) = 2 (N+1) \hat{g}$ and $A$ is a 1-form on $\mathbb{CP}^N$ that satisfies $dA =2J$ where $J$ is the K\"ahler form on $\mathbb{CP}^N$.  The isometry of the spacetime is enhanced to $\mathbb{R} \times U(1) \times SU(N+1)$. The metric $g$ satisfies the Einstein equations $G_{ab} + \Lambda g_{ab} =0$ with a negative cosmological constant, normalized such that $\Lambda = -(D-1)(D-2)/2\ell^2$ where $\ell$ is the AdS length scale.  The field equations can then be simply expressed as
\begin{equation}
R_{ab}  = -\frac{(D-1)}{\ell^2} g_{ab} \,.
\end{equation} The solution above describes the exterior region of a stationary, multiply rotating asymptotically AdS  black hole. The basic example is in $D=5$, in which case $N=1$ and we have $\mathbb{CP}^1 \cong S^2$ with the metric
\begin{equation}
\hat{g} = \frac{1}{4} \left( d\theta^2 + \sin^2 \theta d\phi^2 \right), \qquad A = \frac{1}{2} \cos \theta d\phi \Rightarrow J = -\frac{1}{4}  \sin\theta d\theta \wedge d\phi.
\end{equation} 
The asymptotic region is obtained in the limit $r \to \infty$, where we recover the  usual AdS$_{2N+3}$ metric provided we periodically identify $\psi \sim \psi + 2\pi$. The line element above is valid in the exterior region of the spacetime; that is we also take $t \in \mathbb{R}$ and $r_+ < r < \infty$ where $r_+$ is the largest positive root of $g(r)^{-2}$.  We will discuss below how the metric can be extended beyond $r_+$ to all $r > 0$. 
As we will review below, the hypersurface $r = r_+$ is in fact a smooth Killing horizon with null generator 
\begin{equation}
\xi = \frac{\partial}{\partial t} + \Omega_H \frac{\partial}{\partial \psi} , \qquad \Omega_H = \frac{2 m a}{r_+^{2N + 2} + 2m a^2} \, .
\end{equation}  
Horizons are located at the positive roots of $g(r)^{-2}$. They can be more easily studied via the polynomial $P(r^2)$ where 
\begin{equation}
P(x) = \frac{x^{N+2}}{\ell^2}  + x^{N+1}  - 2M \Xi x + 2 m a^2.
\end{equation} 
Since   there are only two sign changes between adjacent coefficients we
 can apply Descartes' rule of signs to argue there can be at most two real positive roots $x_+ > x_- > 0$ assuming $m > 0$.  Thus we expect the causal structure to be qualitatively similar to that of a charged black hole, consisting of an outer (event) horizon and an inner Cauchy horizon. We will show this explicitly below.  We can eliminate $(m, a)$ in terms of $(r_+, a)$
\begin{equation}\label{m}
m = \frac{r_+^{2N + 2} (1 + r_+^2 \ell^{-2} )}{2 (\Xi r_+^2 - a^2 )}\, .
\end{equation}   A similar formula holds for $m$ with $r_-$ replacing $r_+$.   Note that regularity of the event horizon requires that
\begin{equation}
\Omega_H \leq \frac{1}{\ell} \sqrt{1 + \frac{N \ell^2}{(N+1) r_+^2}}
\end{equation} with the bound saturated when the black hole is extremal. 

When $a =0$ the solution is just Schwarzschild-AdS.  Then there is one horizon and beyond this the function $g_{rr} < 0$ and $g_{tt} > 0$. The set $r =0$, which is a spacelike hypersurface,  is then a curvature singularity.  We will focus on the case $ a> 0$, for which the set $r =0$ is still a curvature singularity but now is timelike (i.e. $|dr|^2 \to + \infty$). As $r \to 0$, the geometry of the base $\mathbb{CP}^N$ collapses. However, $h(r)^2  \sim r^{-2N}$ as $r \to 0$ so the $S^1$ grows to an infinite size.  Meanwhile $g_{tt} \sim 2m r^{-2N} $ is also diverging (and $\partial_t$ is spacelike).  The metric still has to be Lorentzian however, since $\det g = -r^{4N + 2} < 0$.  Thus instead of the singularity being a timelike worldline, it is a timelike cylinder (i.e. at constant $t$ it has $S^1$ topology).  

The conserved charges corresponding to mass and angular momentum are \cite{Gibbons:2004uw,Gibbons:2004ai} 
\begin{equation}\label{thermoMass}
M = \frac{\Omega_{2N+1} m }{4\pi G_N} \left( N + \frac{1}{2} + \frac{a^2}{2\ell^2}\right) , \qquad J = \frac{\Omega_{2N+1}}{4\pi G_N} (N+1) m a \, ,
\end{equation} where 
\begin{equation}
\Omega_{2N+1} = \frac{2 \pi^{N+1}}{\Gamma(N+1)} 
\end{equation}  
 is the area of a unit $2N+1$ sphere.
Note that $M > 0$ imposes the constraint $\Xi r_+^2 - a^2 > 0$ from \eqref{m}.  We emphasize that the single angular momentum $J$ corresponds to equal angular momenta $J_i = J/(N+1)$ in each of the $N+1$-orthogonal planes of rotation. 
Next, since the volume associated with $(\mathbb{CP}^N, \hat{g})$ is
\begin{equation}
\text{Vol} (\hat{g}) = \frac{\pi^N}{\Gamma(N+1)}
\end{equation} we can read off the area of a spatial cross section of the event horizon at $r = r_+$ 
\begin{equation}
A_H = \frac{2h(r_+) r_+^{2N} \pi^{N+1}}{\Gamma(N+1)} = \Omega_{2N+1} h(r_+) r_+^{2N}  \, .
\end{equation}  It is easy to check that $h(r_+) = r_+^2 / \sqrt{\Xi r_+^2 - a^2}$. Furthermore, the event horizon has surface gravity
\begin{equation}
 \kappa_+ = \frac{h(r_+)}{2 r_+}\partial_r f(r)^2 \Bigg\vert_{r=r_+} =  \frac{1}{h(r_+)} \left((N+1) \left(1 + \frac{r_+^2}{\ell^2}\right) - \frac{1}{1 - a^2 (\frac{1}{r_+^2} + \frac{1}{\ell^2})}\right) \, .
 \end{equation}  Finally, since
\begin{equation}
g_{tt} = \frac{1}{h^2} \left[ \frac{4m^2a^2}{r^{4N}} - \left(r^2 + \frac{r^4}{\ell^2} - \frac{2m \Xi}{r^{2N-2}} + \frac{2m a^2}{r^{2N}}\right)\right]
\end{equation}  one finds that there is an ergoregion since $g_{tt} > 0$ in a region exterior to the horizon, although for sufficiently large $r$,  $g_{tt} < 0$.  Note that the ergosurface is never tangent to the event horizon.

\subsection{Extended thermodynamics} In addition to the mass $M$, angular momentum $J$, and angular velocity $\Omega_H$ given above, the black hole's entropy and temperature are given by
\be\label{thermos}
S =\frac{\Omega_{2N+1}h(r_+)r_+^{2N}}{4G_N} \qquad T =\frac{1}{2\pi h(r_+)} \left((N+1) \left(1 + \frac{r_+^2}{\ell^2}\right) - \frac{1}{1 - a^2 (\frac{1}{r_+^2} + \frac{1}{\ell^2})}\right) \, .
\ee Within the framework of extended thermodynamics (see, e.g. the review \cite{Kubiznak:2016qmn}) one associates a thermodynamic pressure with the cosmological constant via
\begin{equation}\label{press}
P = -\frac{\Lambda}{8\pi G_N} = \frac{(N+1)(2N + 1)}{8\pi \ell^2G_N} 
\end{equation} 
along with
\begin{equation}
V = \frac{\sqrt{ \Xi r_+^2 - a^2} A_H}{2(N+1) }  + \frac{ 4\pi a J}{(2N+1)(N+1)} = \frac{r_+^{2(N+1)} \Omega_{2N+1}}{2(N+1)} + \frac{4 \pi a J}{(2N+1)(N+1)}
\end{equation} 
which is its   conjugate thermodynamic volume.
One can then check that the following first law of extended thermodynamics holds for the Myers-Perry-AdS family \cite{Cvetic:2010jb} 
\begin{equation}
dM = T dS + \Omega_H dJ  + V dP  
\end{equation}   
along with the
 Smarr relation 
\begin{equation}\label{smarrEq}
2N M = (2N+1) (TS + \Omega J) - 2 {VP} \;. 
\end{equation}

\begin{figure}
\centering
\includegraphics[width=0.45\textwidth]{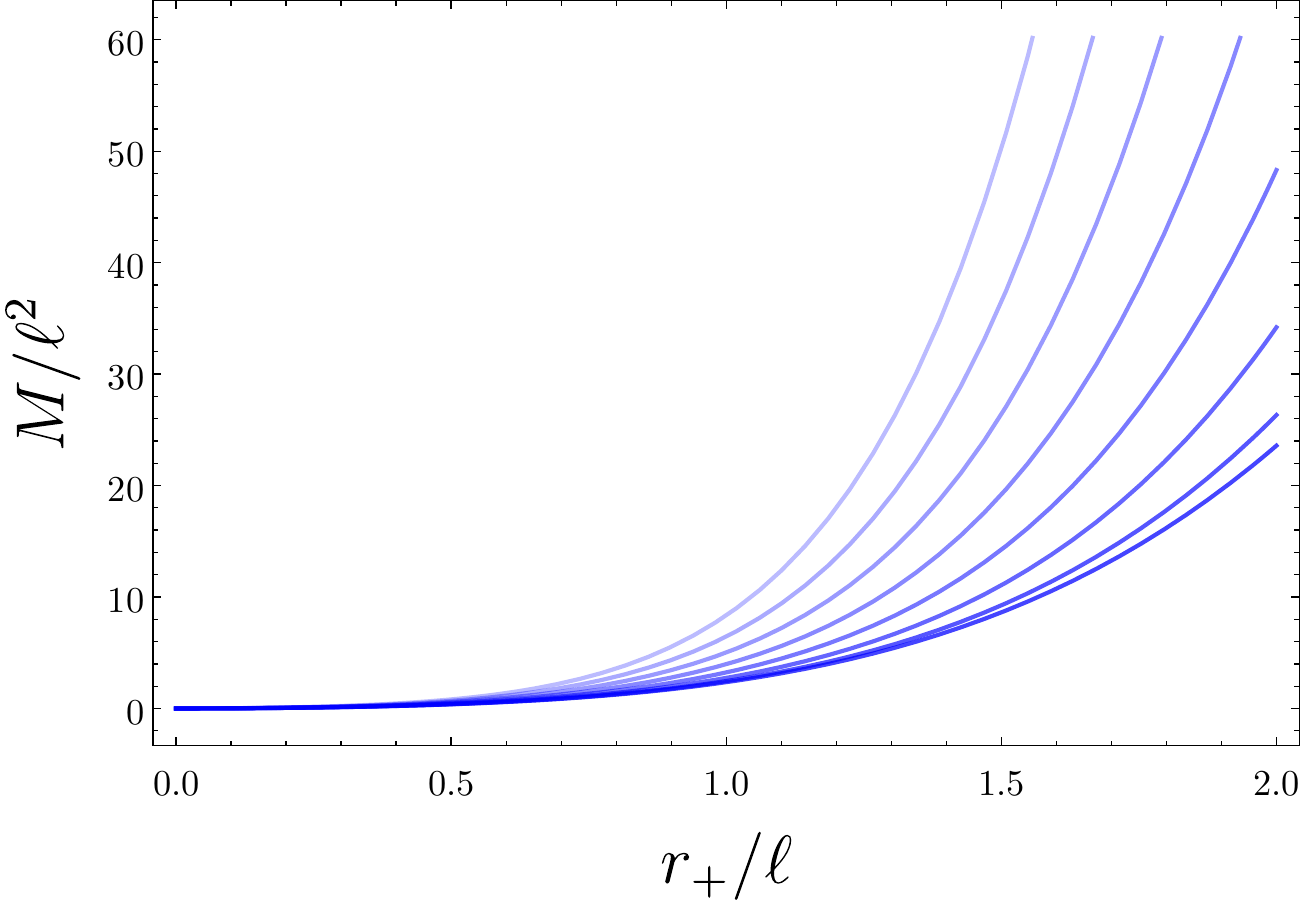}
\quad
\includegraphics[width=0.45\textwidth]{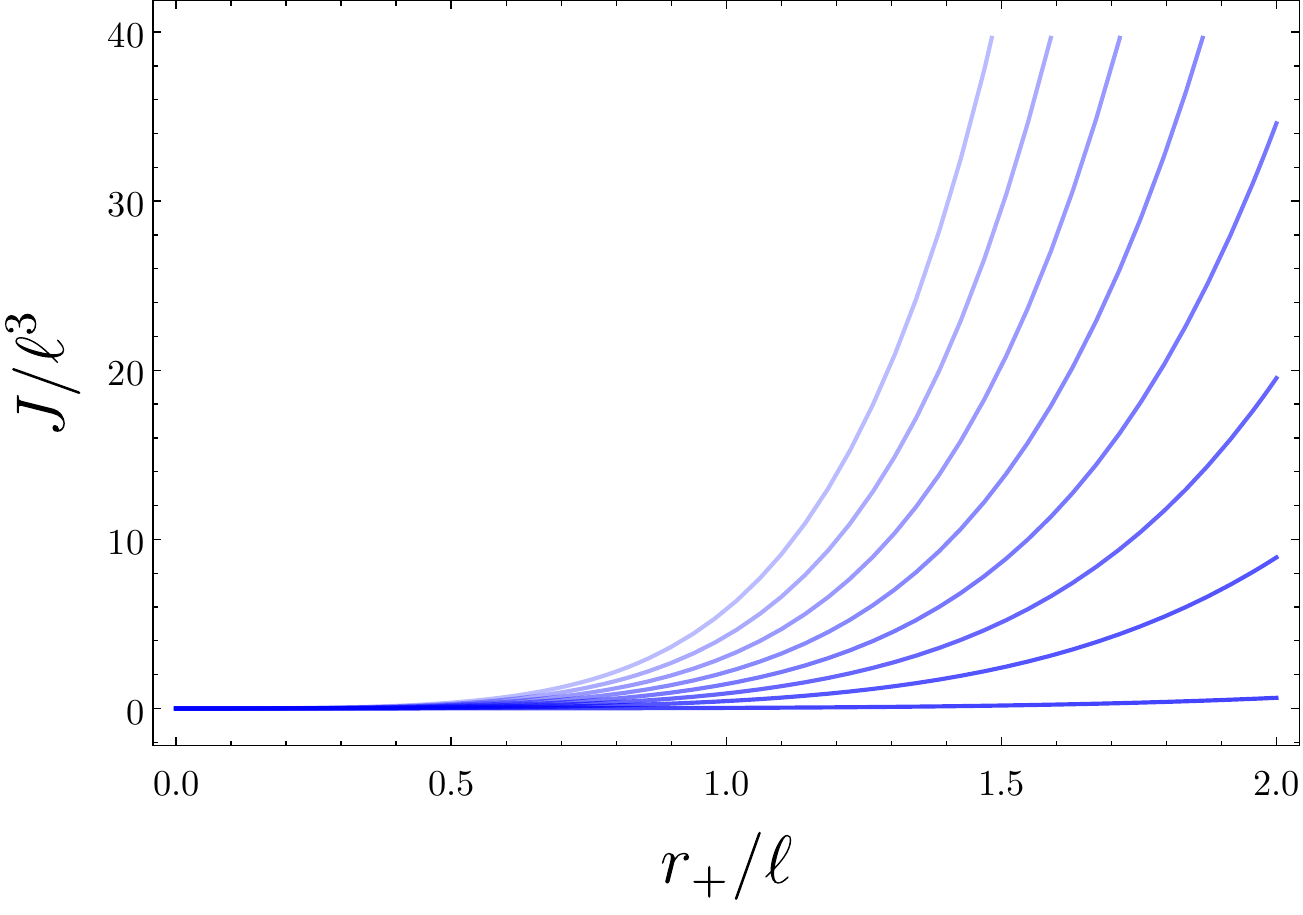}
\caption{Left: A plot of the mass as a function of horizon radius for several values of $r_-/r_+$. Right: A plot of the angular momentum as a function of the horizon radius for several values of $r_-/r_+$. In each case, the lower dark blue curve corresponds to $r_-/r_+ = 1/100$, and this value increases in increments of $1/8$ as one moves vertically in the plot (lines of decreasing opacity). }
\label{MJplots}
\end{figure}

In what follows, it will often be convenient to work in terms of the parameters $(r_+, r_-)$ rather than $(m, a)$. To make the connection between these quantities and the physical parameters of the black hole more explicit, in figure~\ref{MJplots} we plot the mass and angular momentum as functions of $r_+/\ell$ for different values of the ratio $r_-/r_+$. The basic conclusion is that, for large black holes, both the mass and angular momentum grow with increasing $r_+/\ell$. However, for black holes closer to extremality, the growth is stronger. Although we show this pictorially only for five dimensions, the plots are qualitatively similar in higher dimensions.

\begin{figure}
\centering
\includegraphics[width=0.45\textwidth]{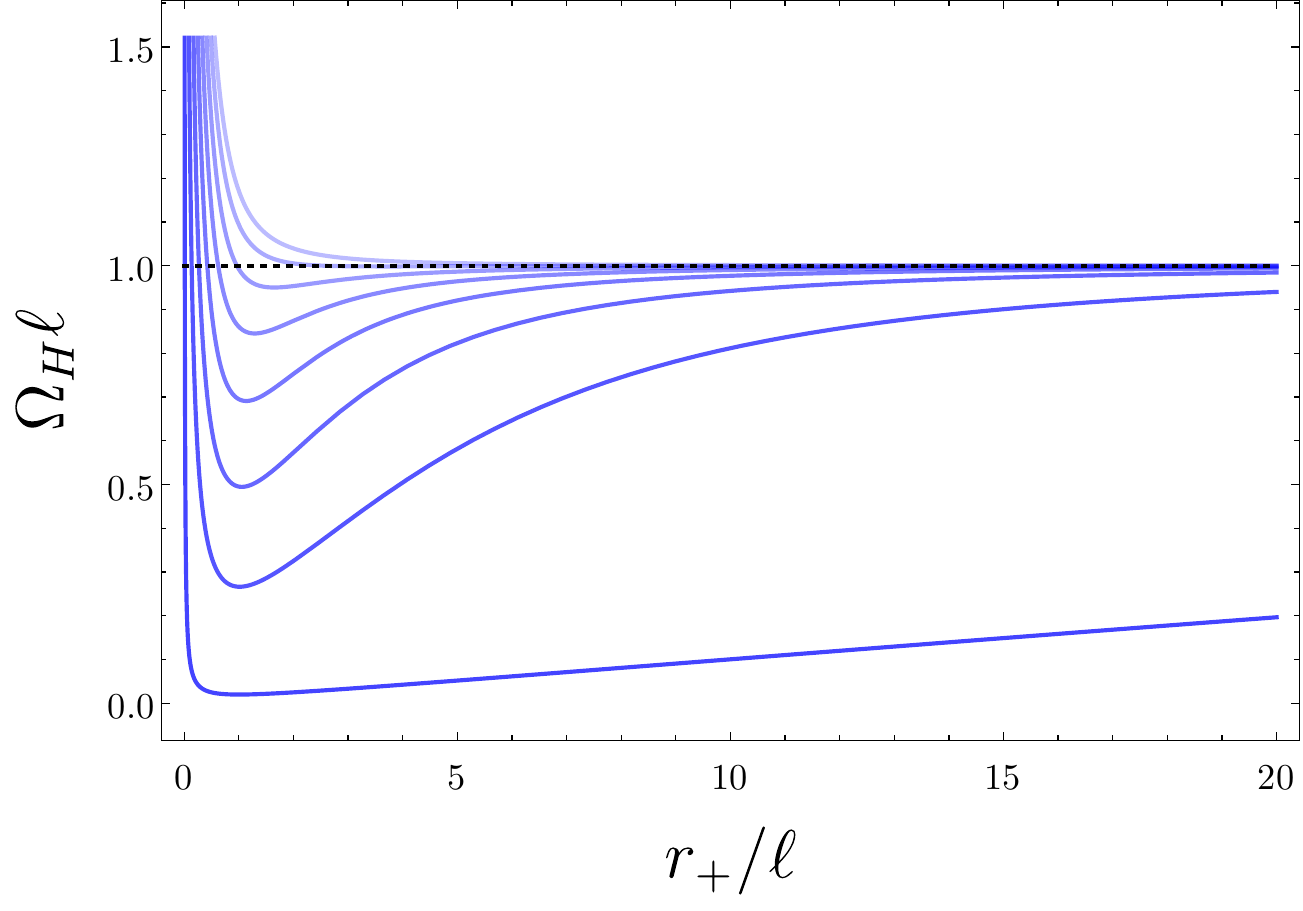}
\quad
\includegraphics[width=0.45\textwidth]{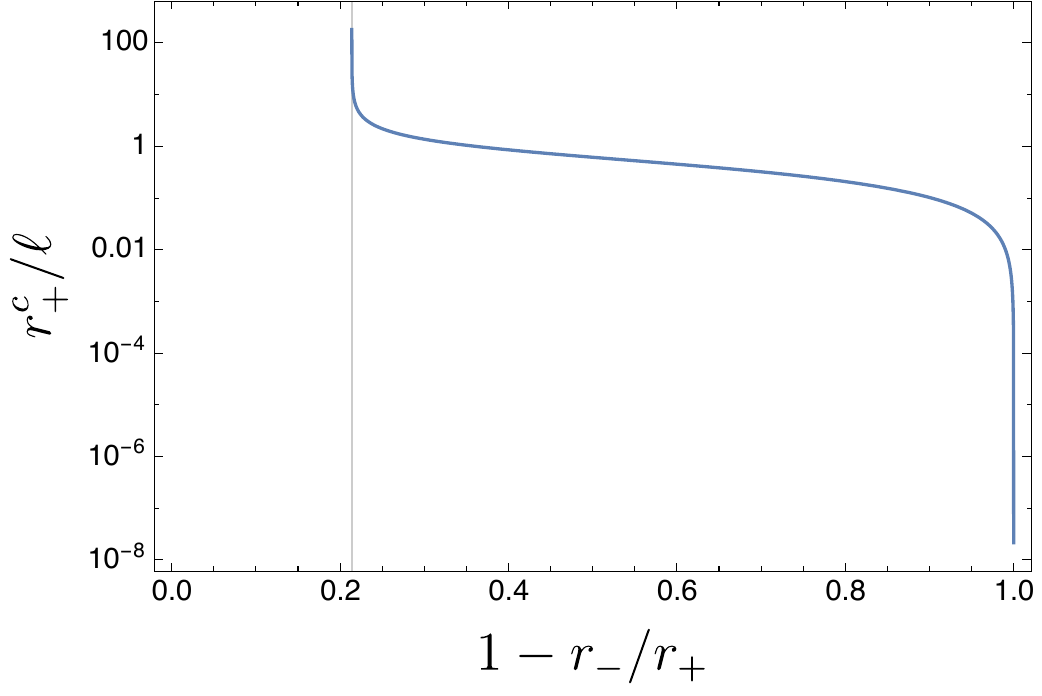}
\caption{Left: Here we show the angular velocity of the horizon as a function of $r_+/\ell$ for several values of $r_-/r_+$. The curves correspond to $r_-/r_+ = 1/100$ (dark blue, bottom) and increase in increments of $1/8$ from bottom to top. The dashed black curve corresponds to the critical angular velocity $\Omega_H^c = 1/\ell$. Right: A plot of the value of $r_+/\ell$ at which the black hole rotates superluminally. The vertical line corresponds to $r_-/r_+ = \sqrt{\sqrt{5}-1}/\sqrt{2}$. }
\label{OmPlots}
\end{figure}

We show also in figure~\ref{OmPlots} the angular velocity of the horizon as a function of $r_+/\ell$, again for different values of the ratio $r_-/r_+$. In the left plot, the dashed black line corresponds to the case where the black hole rotates at the speed of light with respect to an observer situated at infinity. For a ratio $r_-/r_+$ sufficiently below unity, the  angular velocity exhibits a minimum for some intermediate value of $r_+/\ell$ and then increases.  When this minimum coincides with the critical angular velocity $\Omega_H^c = 1/\ell$, the minimum disappears and the angular velocity is a monotonically decreasing function of the horizon radius, asymptoting to $\Omega_H^c = 1/\ell$ from above. The minimum of the angular velocity coincides with the critical value when
\be \label{critRot} 
1 - 2\left(\frac{r_-}{r_+}\right)^2  + \left(\frac{r_-}{r_+}\right)^{2N+4} = 0 \, .
\ee 
Although it is not possible to obtain a simple-closed form, for five-dimensions it occurs when $r_-/r_+ = \sqrt{\sqrt{5}-1}/\sqrt{2}$ and decreases with increasing spacetime dimension, asymptoting to $r_-/r_+ = 1/\sqrt{2}$ in the limit $N \to \infty$. All black holes with $r_-/r_+$ above this threshold rotate faster than light. Provided that $r_-/r_+$ is less than the value corresponding to the solution of \eqref{critRot}, the location of the minimum of the angular velocity occurs at
\be 
\frac{r_+}{\ell} = \sqrt{ \frac{y^2 - y^{2N+2} + (y^2-1) \sqrt{1-y^{2N}} }{1-2y^2 + y^{2N+4} } } \quad \text{where}\quad y \equiv \frac{r_-}{r_+} \, .
\ee

The equally-rotating Myers-Perry-AdS black holes considered here are unstable to linearized gravitational perturbations when they rotate faster than light~\cite{Kunduri:2006qa}. The instability is `superradiant'  in the sense that certain perturbations are trapped by the AdS potential barrier and are reflected back to the black hole, creating an amplification process \cite{Hawking:1999dp}. Note that extreme black holes in this class always rotate faster than the speed of light and are hence unstable.  The endpoint of these instabilities are expected to be stationary, nonaxisymmetric black hole. Although it will not be particularly important for the considerations we are interested in here, it would be interesting to investigate the relation of our findings to known results on the dynamical stability of rotating, asymptotically AdS black holes.

\subsection{Causal structure}

Next let us discuss the global structure of the spacetime.  In general the causal structure of spacetimes with nontrivial rotation is far more complicated than that of their static counterparts. The reason for this, at least partly, is because  in general rotating spacetimes the null hypersurfaces are no longer effectively two dimensional as they are in the static case. However, for the special case of odd-dimensional rotating black holes with equal angular momenta some of these difficulties can be circumvented, as first emphasized in~\cite{Andrews:2019hvq}. Let us illustrate this, following the methods of~\cite{Pretorius1998, AlBalushi2019,Imseis:2020vsw}.  For convenience we will focus on the non-extreme case $r_+ \neq r_-$. 
 
Our task is to construct a suitable family of null hypersurfaces. We start with an ansatz
\be 
v = t + r^*(r, \psi_i)
\ee
where $\psi_i$ stands for the various angular coordinates and $r^*$ denotes a suitable `tortoise' coordinate.  We then demand that $d v$ --- the one-form normal to surfaces of constant $v$ --- is null, i.e. $g^{-1}(dv, dv) = 0$. A direct computation reveals that this condition admits an additively separable solution:
\be 
r^* = R(r) + \sum_i \Psi_i(\psi_i) \, .
\ee
Using an appropriate choice of integration constants the dependence on the angular coordinates can be eliminated, leaving 
\be 
\frac{dr^*}{dr} = \frac{g(r)}{f(r)} = \frac{g(r)^2 h(r)}{r}   
\ee  
or in other words,  $r^*$ is a function only of the radial variable, somewhat akin to the static case. These rotating black holes possess the ``simplest'' causal structure, and are therefore natural candidates for a first foray into the properties of complexity in rotating backgrounds. 
 
Unfortunately, the tortoise coordinate cannot be obtained in a useful closed form and numerical techniques are required for its evaluation. However, for later convenience, here we note both the asymptotic form of the tortoise coordinate, and   that the integral can be massaged into a form much more amenable to numerical evaluation. 
 
Working to the leading order at which differences between the tortoise coordinate for the black holes differs from that for global AdS we find
\be 
r^*(r) =  \sum_{k=0}^{N+1} \frac{(-1)^{k+ 1}\ell^{2k + 2} }{(2k+1) r^{2k+ 1}} + \frac{\ell^2(a^2 - 2 \ell^2) m}{(2N+3) r^{2N+3} } + \mathcal{O}(r^{-2N - 5}) \, .
\ee
Of course, the tortoise coordinate will exhibit logarithmic singularities at the event and inner horizons. To better understand the behaviour of the tortoise coordinate it is useful to define
\be 
g(r)^2 = \frac{G(r)}{(r^2-r_+^2)(r^2-r_-^2)} \, ,
\ee
where $G(r) > 0$ will be completely regular at both horizons. We can series expand the integrand in the vicinity of the horizon to obtain the behaviour near the poles. We find
\be 
\frac{dr^*}{dr} = \frac{G(r_+)h(r_+)}{2 r_+^2 (r_+^2-r_-^2)(r-r_+)} + \mathcal{O}(1) \quad \text{as} \quad r \to r_+ \, ,
\ee
and 
\be 
\frac{dr^*}{dr} = -\frac{G(r_-)h(r_-)}{2 r_-^2 (r_+^2-r_-^2)(r-r_-)} + \mathcal{O}(1) \quad \text{as} \quad r \to r_- \, .
\ee
Noting this behaviour, we can then perform a splitting of the integral, subtracting the pole contributions from the integrand to leave a completely convergent integral, and then handle the poles separately. We choose
\begin{align} 
\frac{dr^*}{dr} =& \left(\frac{G(r)h(r)}{r(r^2-r_+^2)(r^2-r_-^2)} + \frac{G(r_-)h(r_-)}{r_-(r_+^2 - r_-^2)(r^2-r_-^2)} -  \frac{G(r_+)h(r_+)}{r_+(r_+^2 - r_-^2)(r^2-r_+^2)} \right) 
\nonumber\\
&- \frac{G(r_-)h(r_-)}{r_-(r_+^2 - r_-^2)(r^2-r_-^2)} +  \frac{G(r_+)h(r_+)}{r_+(r_+^2 - r_-^2)(r^2-r_+^2)} \, .
\end{align}
where we have kept $(r^2 - r_\pm^2)$ in the denominator to ensure that, when integrated, these  terms converge also as $r \to \infty$. Note that the term in parentheses is now completely regular at $r = r_\pm$.  The integrals involving the divergent parts can then be evaluated directly and we obtain
\begin{align}
\label{goodTort} 
r^*(r) &= \frac{G(r_+)h(r_+)}{2 r_+^2 (r_+^2 - r_-^2)} \log \frac{|r-r_+|}{r+r_+}  - \frac{G(r_-)h(r_-)}{2 r_-^2 (r_+^2 - r_-^2)} \log \frac{|r-r_-|}{r+r_-} 
\nonumber\\
&+ \int^{\infty}_r \frac{r_- r_+ (r_-^2 - r_+^2) G(r)h(r) - r r_+ G(r_-)h(r_-)  (r^2- r_+^2) + r r_- G(r_+)h(r_+) (r^2-r_-^2)}{r  (r^2 - r_-^2)(r^2-r_+^2)(r_+^2 - r_-^2) r_- r_+} dr \, .
\end{align}
Here we emphasize that the integrand in the last term is completely regular at both horizons. Furthermore, in so doing we have extended the integration at infinity, choosing $r^* \to  0$ as $r \to \infty$. This form of the tortoise coordinate is much more amenable to numerical evaluation. 

By expressing the surface gravities at the inner and outer horizons in terms of $G(r)$ we find
\begin{equation}
\kappa_\pm = \pm \frac{r_\pm^2}{h(r_\pm)} \frac{(r_+^2 - r_-^2)}{G(r_\pm)}
\end{equation} which allows us to write the tortoise function in the simple form
\begin{equation}\label{tort}
r^*(r) = \mathcal{R}(r)   + \frac{1}{2\kappa_+} \log \left(\frac{|r - r_+|}{r + r_+}\right) + \frac{1}{2\kappa_-} \log \left(\frac{|r- r_-|}{r + r_-}\right).
\end{equation} where $\mathcal{R}(r)$ is a smooth function defined by the integral term in \eqref{goodTort}. 

So far we have shown that the null sheets $v=$constant in the equal-angular momenta Myers-Perry-AdS solution have a particularly simple form compared to the general situation. We next turn to investigating the causal structure of the solution. To begin, we will construct horizon-penetrating ingoing coordinates adapted to these light sheets. We first pass to corotating coordinates
\begin{equation}
T = t, \qquad \psi^+ = \psi - \Omega_H t \,,
\end{equation} so that the null generator of the event horizon $\xi = \partial_T$.  Next we introduce new coordinates $(v, r, \Psi^+)$ by setting 
\begin{equation}
v = T + r^*, \qquad \Psi^+ = \psi^+ - (\Omega_H - \Omega(r)) r^*, 
\end{equation} so that the metric becomes
\begin{equation}
ds^2 = -\frac{r^2 dv^2}{h(r)^2 g(r)^2} + \frac{2 r}{h(r)} dv dr + h(r)^2 (d\Psi^+ + A + (\Omega_H - \Omega(r)) dv)^2 + r^2 \bar{g} \,.
\end{equation} The metric is clearly smooth and non-degenerate at both horizons (i.e. at poles of $g(r)$). The coordinates cover one exterior region, and can be continued through the event horizon, beyond the inner horizon, and finally to the timelike singularity at $r =0$. However, as in the well known Reissner-Nordstrom case, to determine the maximal analytic extension, the ingoing coordinates are not sufficient.  To construct the required Kruskal-like coordinates, we first define a new chart $(v,u, \Psi^+)$ where $u = v - 2r^*$ to obtain the metric in `double null coordinates' 
\begin{equation}\label{doublenull}
ds^2 = -\frac{r^2}{h(r)^2 g(r)^2} du dv + h(r)^2 \left(d \Psi^+ + A + (\Omega_H - \Omega(r))dv)\right)^2 + r(u,v)^2 \hat{g} , 
\end{equation} where $r^* = (v- u)/2$.  The metric \eqref{doublenull} is clearly degenerate at both the event and inner horizons. As $r \to r_+$ we see from \eqref{tort} that $r^* \to -\infty$ whereas as $r \to r_-$, $r^* \to  \infty$.  Therefore in a neighbourhood of the event horizon as $r \to r_+$,  $v \to -\infty$ or $u \to \infty$ at the rate
\begin{equation}
v - u \to \frac{1}{\kappa_+} \log \left( \frac{|r - r_+|}{2r_+}\right)
\end{equation} which implies that  $| r- r_+ | \to 2 r_+ e^{\kappa_+(v- u)}$ as $r\to r_+$. We next define coordinates
\begin{equation}
U^+: = -e^{-\kappa_+ u} < 0, \qquad V^+ := e^{\kappa_+ v} > 0.
\end{equation}  Therefore as we approach the event horizon, 
\begin{equation}
-\frac{r^2}{h(r)^2 g(r)^2} du dv \to -\frac{4r_+^4 (r_+^2 - r_-^2)}{h(r_+)^2 G(r_+)}e^{\kappa_+ (v - u) }du dv = - \frac{4 r_+^2}{\kappa_+} dU^+ dV^+ \, , 
\end{equation} Furthermore it is easily checked that $(\Omega_H - \Omega(r)) dv$ is smooth as $r \to r_+$. This demonstrates that the metric 
\begin{equation}\label{Kruskal+}
ds^2 = -\frac{r^2 e^{-2\kappa_+ r^*}}{h(r)^2 g(r)^2 \kappa_+^2} dU^+ dV^+ + h(r)^2 \left(d\Psi^+ + A + \frac{1}{\kappa_+V^+}(\Omega_H - \Omega(r))dV^+ \right)^2 + r^2 \hat{g}
\end{equation} is smooth and non-degenerate at the event horizon in the $(U^+, V^+, \Psi^+)$ chart and we can analytically continue the chart through the event horizon ($U^+ =0$ or $V^+ =0$) to a new region $U^+ > 0, V^+ < 0$ so that the metric \eqref{Kruskal+} is regular for $r_- < r < \infty$.  The chart covers four regions (quadrants in the $(U^+, V^+)-$plane) with a bifurcation $S^3$ at $(U^+, V^+) = (0,0)$.   The coordinate system breaks down near the inner horizon as $r \to r_-$ and there are radial null geodesics that reach this null hypersurface in finite affine parameter.   We can extend beyond this coordinate singularity by reversing the above coordinate transformations to return to the ingoing coordinates $(v, r, \Psi^+)$, which are regular at both horizons.  Define
\begin{equation}
\Psi^- = \Psi^+ + (\Omega_H - \Omega(r_-)) v
\end{equation} so that in the $(v, r, \Psi^-)$ chart, the Killing field $\partial_v$ is corotating with the inner horizon $r = r_-$.  Introduce a second double null coordinate system $(\hat{v}, \hat{u})$ with
\begin{equation}
\hat{v} = v, \qquad \hat{u} = v - 2r^* ,,
\end{equation} so that in particular $r^* = (\hat{v} - \hat{u})/2$.  The metric in the $(\hat{v}, \hat{u}, \Psi^-)$ coordinate chart will resemble  \eqref{doublenull} with the obvious replacements and hence will be degenerate at $r = r_-$.  We then introduce a second pair of Kruskal-like coordinates adapted to the inner horizon by setting
\begin{equation}
U^- = -e^{-\kappa_- \hat{u}} < 0, \qquad V^- = -e^{\kappa_- \hat{v}}  < 0\,.
\end{equation} By repeating the above computations we find the metric in the $(V^-, U^-, \Psi^-)$ chart is
\begin{equation}
ds^2 = \frac{r^2 e^{-2\kappa_- r^*}}{h(r)^2 g(r)^2 \kappa_-^2} dU^- dV^- + h(r)^2 \left( d\Psi^- + A + (\Omega(r_-) - \Omega(r))\frac{dV^-}{\kappa_- V^-}\right)^2 + r^2 \hat{g} \, ,
\end{equation} which is indeed smooth and non-degenerate at $r = r_-$ using the fact that $(r- r_-) e^{-2\kappa_- r^*} \to 2 r_-$ as $r \to r_-$ and $(\Omega(r_-) - \Omega(r)) / V^- = \mathcal{O}(1)$.  In this coordinate system, the inner horizon corresponds to either $U^-= 0$ or $V^- =0$ and we may analytically continue the metric in this chart to allow  $U^- \geq 0$ and $V^- \geq 0$, corresponding to $0< r < r_-$.  This region contains a timelike coordinate singularity at $r = 0$ , or $U^-V^- = e^{2\kappa_- \mathcal{R}(0)}$.  Since this region is actually isometric to a region for which the event horizon lies to the future, we can introduce new coordinates $(\hat{U}^+, \hat{V}^+)$  and analytically continue the metric into new exterior regions $r > r_+$ that are isometric to the original asymptotically AdS regions described by the $(U^+, V^+)$ coordinate chart.  We can repeat this procedure indefinitely both to the future and past to produce a maximal analytic extension with infinitely many regions, qualitatively similar to the familiar maximal analytic extension of the non-extreme rotating BTZ black hole \cite{Banados:1992gq}.  Note that in contrast to the Kerr black hole, and generic members of the Myers-Perry(-AdS) black holes, one cannot continue into a region of spacetime for which $r ^2< 0$. 


\section{Framework for Complexity Computations}\label{sec3}

\subsection{Framework for Action calculations}

Given a $D-$dimensional bulk region $\mathcal{M}$, the gravitational action, including all the various terms for boundary surfaces and joints
\cite{Booth:2001gx}, over this region is given by\footnote{Note that we follow the conventions of~\cite{Carmi:2016wjl} with the minor correction pointed out in~\cite{Chapman:2018dem}. } \cite{Lehner2016}
\begin{align}
I_{\text{grav}}&=\frac{1}{16\pi G_N}\int_{\mathcal{M}}\sqrt{-g}\left(R+\frac{(D-1)(D-2)}{\ell^2}\right) d^{D}x +\frac{1}{8\pi G_N}\int_{\mathcal{B}}\sqrt{|h|}K d^{D-1}x \nonumber\\&+\frac{1}{8\pi G_N}\int_{\mathcal{B}'}\sqrt{\gamma}\kappa d\lambda d^{D-2}\theta +\frac{1}{8\pi G_N}\int_{\mathcal{J}}\sqrt{\sigma}\eta d^{D-2}x +\frac{1}{8\pi G_N}\int_{\mathcal{J}'}\sqrt{\sigma}\tilde{a} d^{D-2}x . 
\label{I}
\end{align}
The first term is the Einstein-Hilbert bulk action with cosmological constant, which from \eqref{press} is
\be 
\Lambda \equiv - \frac{(D-1)(D-2)}{2 \ell^2}
\ee 
integrated over $\mathcal{M}$. The second term is the Gibbons-Hawking-York boundary term \cite{York1972,Gibbons1977} that contributes at spacelike/timelike boundaries $\mathcal{B}$. The convention adopted here for the extrinsic curvature is that the normal \textit{one-form} is directed outward from the region of interest.  The third term is the contribution of the null boundary surface $\mathcal{B}'$ of $\mathcal{M}$. For a null boundary segment with normal $k^\alpha$ the parameter $\kappa$ is defined in the usual way: $k^\beta \nabla_\beta k^\alpha = \kappa k^\alpha$, while $\gamma$ is the determinant of the induced metric on the $(D-2)$-dimensional cross-sections of the null boundary and the parameter $\lambda$ is defined according to $k^\alpha = \partial x^\alpha/\partial \lambda$. The fourth term is the Hayward term \cite{Hayward1993,Booth:1999se} for joints $\mathcal{J}$ between non-null boundary surfaces --- these terms will play no role in our construction. The last term is the contribution of joints $\mathcal{J}'$ from the intersection of at least one null boundary surface \cite{Booth:2001gx}. The parameter $\tilde{a}$ is defined according to
\begin{align}
\label{ajntterm1}
\text{timelike/null}: \quad \tilde{a} & \equiv \epsilon \log | \mathbf{t}_1 \cdot \mathbf{k}_2 | \quad &\text{with} \quad &\epsilon = - {\rm sign}(\mathbf{t}_1 \cdot \mathbf{k}_2) {\rm  sign} (\hat{\mathbf{n}}_1 \cdot \mathbf{k}_2 ) \, ,
\\ \label{ajntterm2}
\text{null/spacelike}: \quad \tilde{a} & \equiv \epsilon \log | \mathbf{k}_1 \cdot \mathbf{n}_2 | \quad &\text{with} \quad &\epsilon = - {\rm sign}(\mathbf{k}_1 \cdot \mathbf{n}_2) {\rm  sign} (\mathbf{k}_1 \cdot \hat{\mathbf{t}}_2 ) \, ,
\\ \label{ajntterm3}
\text{null/null}: \quad \tilde{a} & \equiv \epsilon \log | (\mathbf{k}_1 \cdot \mathbf{k}_2) /2 | \quad &\text{with} \quad &\epsilon = - {\rm sign}(\mathbf{k}_1 \cdot \mathbf{k}_2) {\rm  sign} (\hat{\mathbf{k}}_1 \cdot \mathbf{k}_2 ) \, ,
\end{align}
where $\mathbf{k}_i$ is a null normal, $\mathbf{t}_i$ is a timelike unit normal, and $\mathbf{n}_i$ is a spacelike unit normal. Additionally, depending on the intersecting boundary segments, auxillary vectors --- indicated with a hat --- are required. These unit vectors are defined by the conditions of living in the tangent space of the appropriate boundary segment and pointing outward \textit{as a vector} from the joint of interest.

The action as presented above is ambiguous when the spacetime region of interest contains null boundaries. Namely, the action is not invariant under reparameterizations of the normals to the null boundary segments. To ensure this invariance we add to the above the following counterterm~\cite{Lehner2016}:
\be
I_{\text{ct}}=\frac{1}{8\pi G_N}\int_{\mathcal{B}'}  \Theta\log l_{\text{ct}}\Theta  \sqrt{\gamma}  d\lambda d^{D-2}\theta
\ee
where $l_{\text{ct}}$ is an arbitrary length scale and 
\be
\Theta=\partial_\lambda\log\sqrt{\gamma}
\ee
is the expansion scalar of the null boundary generators, which depends only on the intrinsic geometry of the null boundary surfaces. While this term is not required to have a well-defined variational principle, it is known to have important implications for holographic complexity --- for example, it is crucial for reproducing the switchback effect in the complexity equals action conjecture~\cite{Susskind:2014jwa, Chapman:2018dem, Chapman:2018lsv}.

A further difficulty is that the gravitational action is divergent. To control these divergences (and allow for appropriate regularization in the complexity of formation calculations) we introduce a UV cut-off $\delta$ at the boundary CFT and integrate the radial dimension in the bulk up to $r=r_{\text{max}}(\delta)$ \cite{Skenderis2002,DeHaro2001}. When calculating the complexity of formation, the choice of $r_{\text{max}}(\delta)$ for the black hole spacetime should be consistent with that in vacuum AdS. This subtlety can be resolved~\cite{Chapman2017Form}  by expanding the metrics of both geometries in the Fefferman-Graham canonical form \cite{Fefferman2007} and setting in both cases the radial cut-off surface at $z=\delta$. We discuss the Fefferman-Graham form of the rotating metrics in Appendix~\ref{fgForm}.

\begin{figure*}[t]
\centering
\begin{subfigure}{0.4\textwidth}
\includegraphics[width=1.3\linewidth]{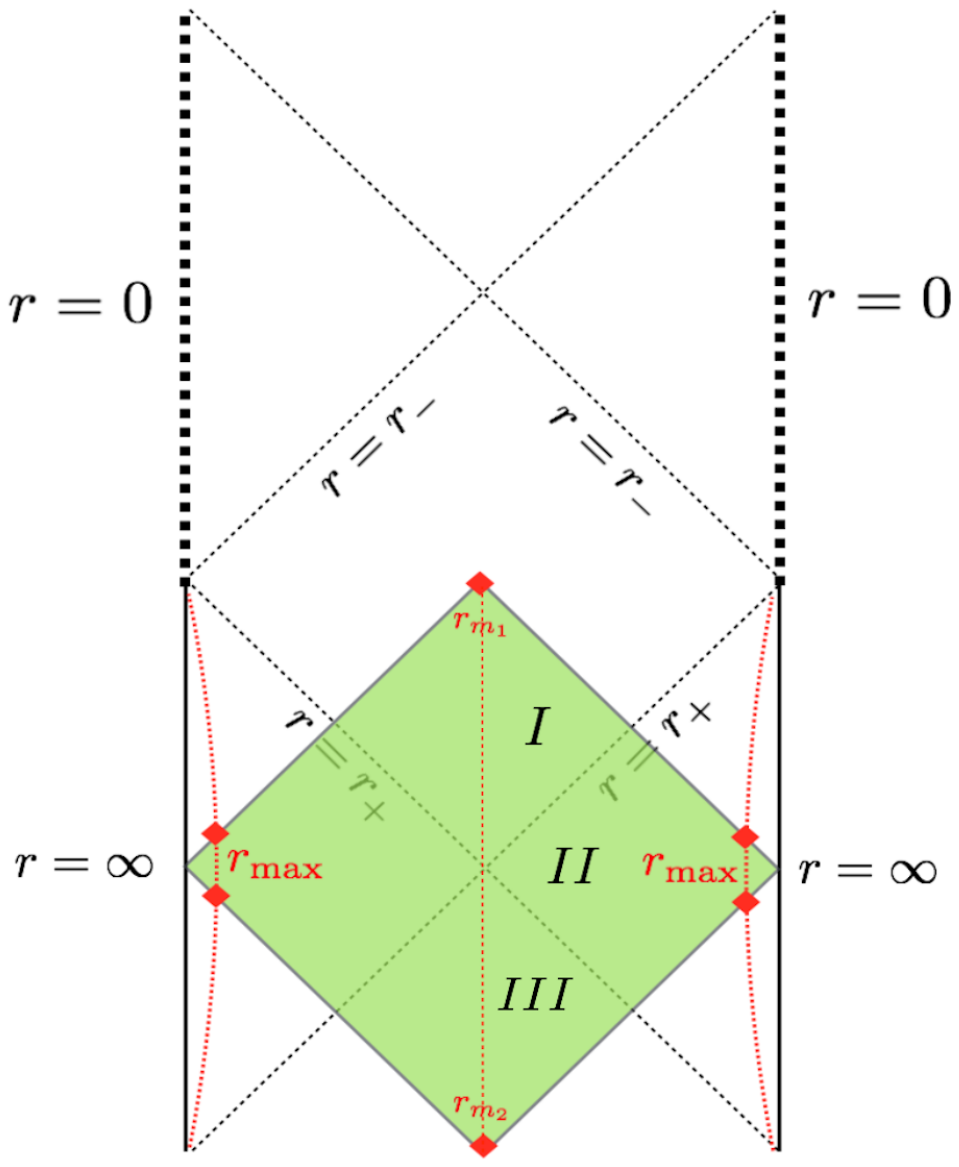}
 \caption{WDW patch at $-t_L=t_R=0$}
\label{fig1:p1}
\end{subfigure}
\quad 
\begin{subfigure}{0.4\textwidth}
 \includegraphics[width=1.3\linewidth]{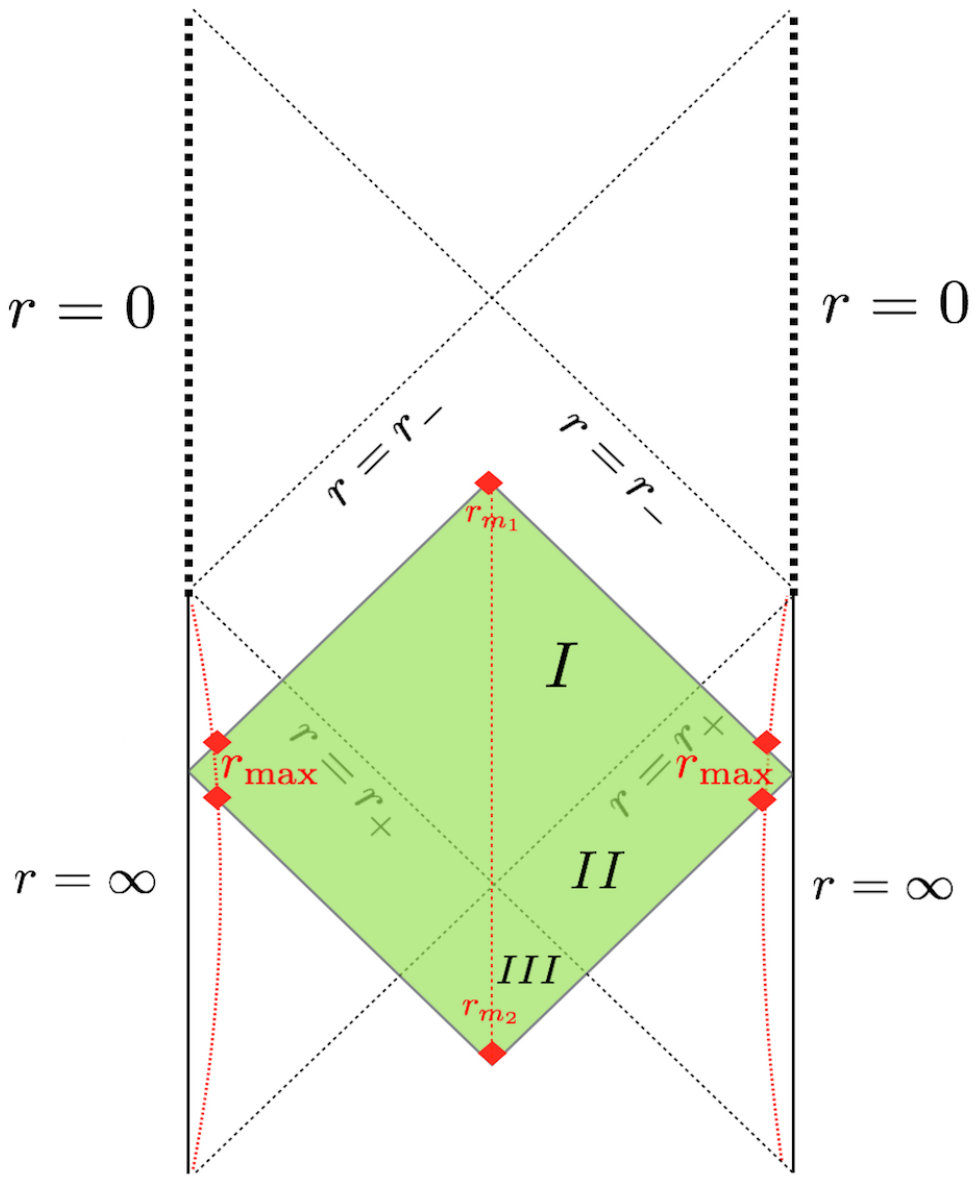}
\caption{WDW patch at $-t_L=t_R=\tau/2>0$}
\label{fig1:p2}
\end{subfigure}
\caption{Penrose diagrams for the rotating black holes in \eqref{metric} with the WDW patches corresponding to the TFD state at $-t_L=t_R=0$ in (a) and $-t_L=t_R=t/2>0$ in (b) as the green shaded region. Joints contributing to the action terms are highlighted with red diamonds. Also shown are three regions $I$, $II$, and $III$ defined for the bulk contribution to $I_{\text{WDW}}$. The vertical dotted red line divides the WDW patch into two symmetric parts and define the regions I, II, and III. The curved dotted red line is the cut-off surface $r=r_\text{max}(\delta)$.}
\label{penrose} 
\end{figure*}

To evaluate the complexity within the CA conjecture, we must evaluate the gravitational action and counterterm on the Wheeler-DeWitt patch of spacetime. Using the boost invariance of the spacetime, it is always possible to shift the WDW patch so that it intersects the left and right boundaries at the same times: $t_L = t_R \equiv \tau/2$. We show the structure of the WDW patch in Figure~\ref{penrose}, which has the same structure for all the rotating black holes considered here. Of particular importance are the joints where the future/past boundaries of the WDW patch meet.

Let us determine the past meeting points of the boundaries of the WDW patch. We denote the future meeting point as $r_{m_1}$ and the past meeting point as $r_{m_2}$. Consider first the past meeting point, and denote its coordinates inside the horizon as $(t_{m_2}, r_{m_2})$.  From the right side of the Penrose diagram, this point lies along a $u = constant$ surface, while from the left it lies along a $v = constant$ surface. These facts translate into two equations:
\be
t_{m_2} + r^*(r_{m_2}) = \tL + r^*_\infty \, , \quad t_{m_2} - r^*(r_{m_2}) = \tR - r^*_\infty  \, ,
\ee
where $\tL$ and $\tR$ denote the timeslices at which the lightsheets intersect the left and right boundaries, respectively. Note that $t_{m_2}$ is the same in both equations as those points lie in a common patch of the diagram. Eliminating $t_{m_2}$ from these equations we obtain
\be 
 \tR - \tL + 2 \left( r^*(r_{m_2}) - r^*_\infty \right) = 0 \, .
\ee
Upon noting that $\tL = - \tR$ (which implies $t_{m_2}=0$)
 and setting $\tR = \tau/2$ we obtain
\be\label{rmeet-1}
\frac{\tau}{2} + r^*(r_{m_2}) - r^*_\infty = 0 \, .
\ee
An analogous derivation holds for $r_{m_1}$, the only difference being a sign in the last two terms:
\be\label{rmeet-2}
\frac{\tau}{2} - r^*(r_{m_1}) + r^*_\infty = 0 \, .
\ee
Note that here we have chosen to use the time $\tau$ instead of $t$ to avoid possible confusion of this quantity with the $t$ appearing in the metric (which, when considering the patches outside of the horizon, would be either $t_L$ or $t_R$). 

In general the values of $r_{m_{1,2}}$ must be obtained numerically. However, let us note that it is possible, starting from eq.~\eqref{goodTort}, to obtain an asymptotic form of this quantity valid for early times in the limit $r_-/r_+ \to 0$. This can be obtained by evaluating the integral appearing in~\eqref{goodTort} perturbatively in $r_-/r_+$. Here we note the result only in five dimensions:
\be 
r_m \approx r_- \left\{1 + \exp \left[- \frac{\pi (r_+^2 + \ell^2) \ell +  (r_+/\ell) \sqrt{\ell^2 + r_+^2} (\ell^2 + 2 r_+^2)  \varepsilon \tau }{r_+ (\ell^2 + 2 r_+^2)} \frac{r_+^2}{r_-^2} + \cdots \right] + \cdots \right\} \, ,
\ee
where the dots denote subleading terms and $\varepsilon = +1$ for $r_{m_1}$  and $-1$ for $r_{m_2}$. This expression reveals that, as $r_-/r_+ \to 0$, the value of $r_m$ tends exponentially towards the inner horizon --- consistent with the discussion of charged black holes in~\cite{Carmi2017}, albeit with a slightly different rate of approach.


\subsection{Evaluating the Action}

\subsubsection*{Bulk Action}

The bulk contributions to the action are very simple in this case since the black holes are vacuum solutions. In particular, we have
\be 
R = \frac{2 D \Lambda}{(D-2)} 
\ee
and thus
\be 
R - 2 \Lambda = \frac{4 \Lambda}{(D-2)} \, .
\ee
The bulk action is then simply the spacetime volume of the WDW patch weighted by this dimension-dependent prefactor:
\be 
I_{\rm bulk} = \frac{\Lambda}{4 (D-2) \pi G_N} \int_{\rm WDW} \sqrt{-g}  d^D x \, .
\ee
To evaluate the bulk contribution, we recall that the determinant of the metric is  
\be 
\sqrt{-g} = r^{2N+1} d \Omega_{2N+1} \, .
\ee 
We then split the integration domain into three regions where the $(t,r)$ coordinates are valid, as shown in Figure~\ref{penrose}.
In region I, the integration over $\tR$ is between $0$ (i.e. $t_{m_1}$) and
\be 
\tR = \frac{\tau}{2} + r^*_\infty - r^*(r) \, .
\ee
In region II the integration over $\tR$ is between 
\be 
\tR =  \frac{\tau}{2} - r^*_\infty + r^*(r) \, , \quad \text{and} \quad  \tR = \frac{\tau}{2} + r^*_\infty - r^*(r) \, .
\ee
Finally, the integration in region III occurs between 
\be 
\tR = \frac{\tau}{2} + r^*(r) - r^*_\infty 
\ee
and $0$.  We then have
\begin{align}
I_{\rm bulk}^I &= \frac{\Lambda \Omega_{2N+1}}{4 (D-2) \pi G_N} \int_{r_{m_1}}^{r_+} r^{2N+1} \left(\frac{\tau}{2} + r^*_\infty - r^*(r) \right)\,dr  \, , 
\\
I_{\rm bulk}^{II} &= \frac{2 \Lambda \Omega_{2N+1}}{4 (D-2) \pi G_N} \int_{r_+}^{r_{\rm max}} r^{2N+1} \left(r^*_\infty - r^*(r) \right)\, dr \, , 
\\
I_{\rm bulk}^{III} &= \frac{\Lambda \Omega_{2N+1}}{4 (D-2) \pi G_N} \int_{r_{m_2}}^{r_+}  r^{2N+1} \left(- \frac{\tau}{2} + r^*_\infty - r^*(r) \right)\,dr \, .
\end{align}
The total bulk action is then twice the sum of these three terms.

\subsubsection*{Surface contributions}
There are two cut-off surfaces at $r=r_{\rm max}$, which each contribute a term  
\be
I_{\rm GHY}=\frac{1}{8\pi G_N}\int_{\mathcal{B}}\sqrt{|h|}K d^{D-1}x\; .
\ee
The normal to the timelike surface $r=r_\text{max}$ is 
\begin{equation}
n^{\mu}=\left(0,\sqrt{g^{rr}},0,0\right)
\end{equation}
and the induced metric to the timelike surface of constant $r=r_{\text{max}}$ has the determinant 
\begin{align}
\sqrt{|h|}&=\frac{\sqrt{-g}}{\sqrt{g_{rr}}}\; .
\end{align}
The trace of the extrinsic curvature of the boundary surface is then
\begin{align}
K&=\nabla_{\mu}n^{\mu} =\frac{1}{\sqrt{-g}}\partial_{\mu}\left(\sqrt{-g}n^{\mu}\right) =\frac{1}{\sqrt{-g}}\partial_{r}\left(\sqrt{|h|}\right).
\end{align}
This gives a contribution for the two boundary surfaces at $r=r_{\text{max}}$ of the form
\begin{align}\label{GHYterm}
I_{\text{GHY}}&=\frac{\Omega_{D-2}}{2\pi G_N}r^{D-2}\left[\frac{(D-2)}{rg(r)^2}-\frac{g'(r)}{g(r)^3}\right] \left(r^*_\infty-r_*(r)\right)\bigg|_{r=r_{\text{max}}}.
\end{align}
Note that this term is time-independent, so it does not contribute to the complexity rate of change $d\mathcal{C}_A/d\tau$. Furthermore, it does not contribute to the complexity of formation $\Delta\mathcal{C}_A$ because it is cancelled by the contribution made by the $AdS_D$ vacuum --- as shown explicitly in appendix \ref{GHYappend}.

\subsubsection*{Joint contributions}

There are two different types of joint contributions that arise here. First, there are the intersections of the null boundaries of the WDW patch with the regulator surface at $r = r_{\rm max}$. There are four of these joints in total. Second, there are the intersections of the null sheets of the WDW patch in the future and in the past. Let us begin with the first case.

Considering the future, right boundary of the WDW patch near the regulator surface $r = r_{\rm max}$, the relevant null normal is given by
\be 
k_R = \alpha \left[dt + dr^* \right] \, ,
\ee
while the outward pointing normal to the surface $r = r_{\rm max}$ is
\be 
n = g(r) dr \, .
\ee
We need also a vector $\tilde{t}$ that is a future-pointing unit time-like vector directed outwards from the region. In this case the correct choice is 
\be 
\hat{t} = - f(r) dt \, 
\ee
where we have written it as a form, but the sign is chosen so that the corresponding vector is outward directed. The relevant dot products are easily computed
\be 
k_R \cdot n = \frac{\alpha}{f(r)} \, , \quad k_R \cdot \hat{t} = \frac{\alpha}{f(r)} \, , 
\ee
and since $f(r) > 0$ near the boundary we obtain $\epsilon = - 1$. We have then
\be 
\tilde{a} = - \log \frac{\alpha}{f(r)} 
\ee
from \eqref{ajntterm2}, yielding 
\be 
I_{\rm jnt}^{r_{\rm max}} = \frac{\Omega_{2N + 1} r^{2N} h(r)}{16 \pi G_N} \log \frac{f(r)^2}{\alpha^2} \, ,
\ee
where we have made use of the fact that 
\be 
\sqrt{\gamma} = r^{2N} h(r) d \Omega_{2N + 1} 
\ee
on the joint. Note that   by  $d \Omega_{2N+1}$ we mean the volume form on the usual, round $2N+1$ sphere --- when integrated over the angles this gives
\be 
\Omega_{2N+1} \equiv \int d \Omega_{2N+1} = \frac{2 \pi^{N+1}}{\Gamma\left[N+1\right]} \, .
\ee
An analogous computation for the remaining three joints can be shown to yield the same answer as that presented here. 

\begin{figure*}[t]
\centering
\begin{subfigure}{0.4\textwidth}
\includegraphics[width=0.8\linewidth]{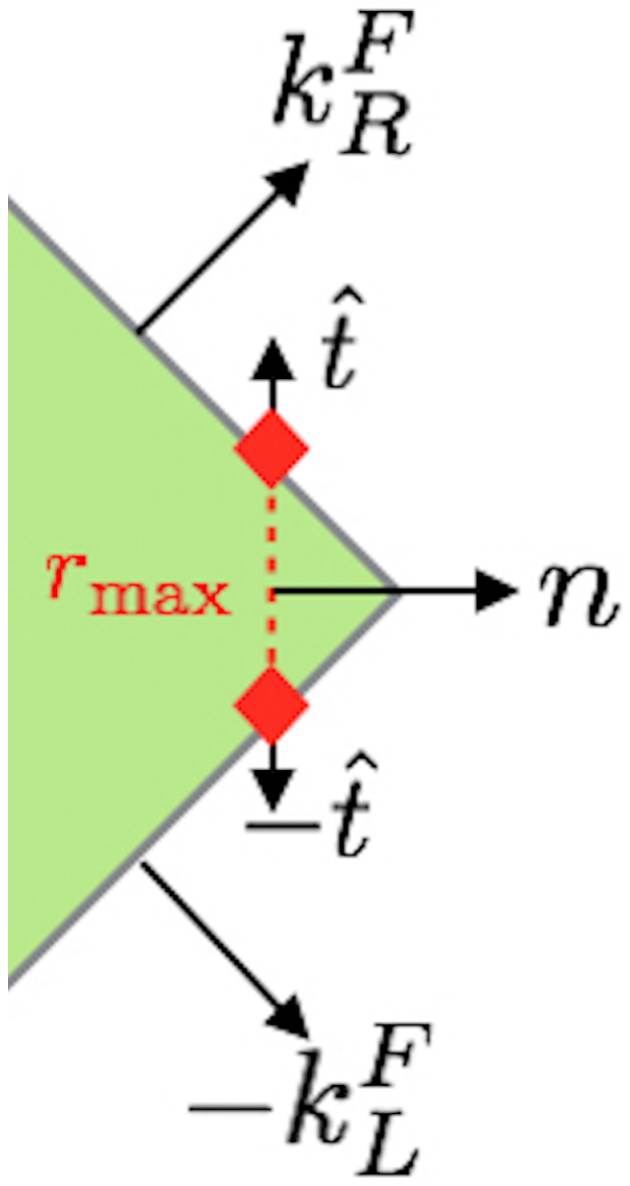}
\label{joints:p1}
\end{subfigure}
\begin{subfigure}{0.3\textwidth}
 \includegraphics[width=1.\linewidth]{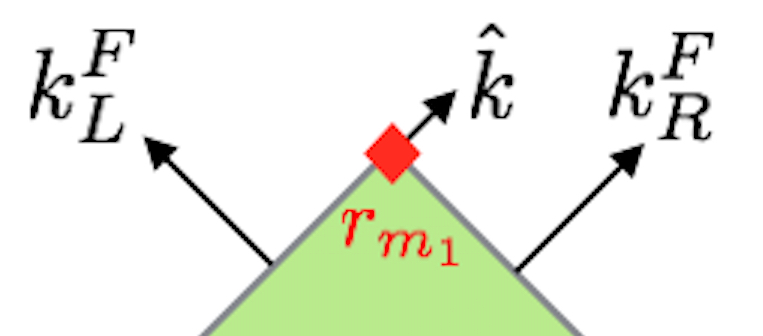}
\label{joints:p2}
\end{subfigure}
\captionsetup{justification=centering}
\caption{The joints in the WDW patch in figure \ref{penrose} and the associated vectors needed to calculate their contribution to $I_{\text{WDW}}$, according to \eqref{ajntterm1}-\eqref{ajntterm3}.}
\label{fig-joints} 
\end{figure*}

Next let us consider the joints at the future and past meeting points of the WDW patch. The determinant of the induced metric at the intersection of the lightsheets is given by 
\be 
\sqrt{\gamma} =h(r_m) r_m^{2N}  d \Omega_{2N+1}  \, ,
\ee
where $r_m$ is the value of $r$ at the point of intersection. At the future meeting point  the relevant null normals are
\be 
k_R^F = \alpha \left[dt + dr^* \right] \, , \quad k_L^F = \alpha \left[-dt + dr^*  \right]  
\ee
where, in determining the relevant signs, it is important to recognize that $t$ increases from left to right inside the future horizon and $dr^*$ points in the negative $dr$ direction inside the horizon. Note also that the $dt$ appearing in these normals is the $t$ that appears in the metric, not the boundary time. We need also $\hat{k}$ --- a null vector, living in the tangent space of the right sheet of  the WDW patch that is orthogonal to $k_{R}^F$ and outward pointing as a vector. In this case, a one-form that points in the \textit{negative} $k_{R}^F$ direction yields a vector with the correct properties. We take
\be 
\hat{k} = - \left[dt + dr^* \right] \, .
\ee
We then find for the dot products
\be 
k_R^F \cdot k_L^F = \frac{2 \alpha^2}{f(r)^2} \, , \quad \hat{k} \cdot k_L^F = - \frac{2 \alpha}{f(r)^2} \, ,
\ee
yielding  $\epsilon = +1$ and from \eqref{ajntterm3}
\be 
\tilde{a} = - \log \frac{|f(r_{m_1})^2|}{\alpha^2} \, .
\ee
Putting this all together we obtain for the joint contribution at the future meeting point
\be 
I_{\rm jnt}^F = - \frac{(r_{m_1})^{2N} h(r_{m_1}) \Omega_{2N+1}}{8 \pi G_N} \log \frac{|f(r_{m_1})^2|}{\alpha^2}  \, .
\ee 
A completely analogous calculation gives an identical form for the joint contribution at the past meeting point, with $r_{m_1}\to r_{m_2}$.

\subsubsection*{Null boundaries}

Since the normals to the lightlike boundaries of the WDW patch are affinely parameterized, the boundary term on these surfaces makes no contribution. Nonetheless, we consider here the contribution from the counterterm for null boundaries that ensures the total action does not depend on the parameterization used for the null generators.

Considering the future segment on the right of the Penrose diagram, we have
\be 
\frac{\partial}{\partial \lambda} = \frac{\alpha}{g(r)f(r)} \frac{\partial}{\partial r} \, ,
\ee
which yields
\be 
\Theta = \frac{1}{\sqrt{\gamma}} \frac{dr}{d \lambda} \frac{d \sqrt{\gamma}}{d r} = \frac{1}{h(r) r^{2N}} \frac{dr}{d \lambda}  \frac{d \left( h(r) r^{2N} \right)}{dr}  = \frac{\alpha }{f(r) g(r)} \left[\frac{2N}{r} + \frac{h'}{h} \right]\, .
\ee
We therefore have for the counterterm
\be 
I_{\rm ct}^{F, R} = \frac{\Omega_{2N+1}}{8 \pi G_N} \int_{r_{m_1}}^{r_{\rm max}}  \log \ell_{ct} \Theta  \frac{d \left( h(r) r^{2N} \right)}{dr}\, dr \, .
\ee
We can use integration by parts to express this object in terms of two contributions at the joints    and an integral independent of $\alpha$ and $\ell_{ct}$:
\be 
I_{\rm ct}^{F, R} = \frac{\Omega_{2N+1}}{8 \pi G_N} \left[r^{2N} h(r) \log \ell_{ct} \Theta \right]_{r_{m_1}}^{r_{\rm max}} - \frac{\Omega_{2N+1}}{8 \pi G_N} \int_{r_{m_1}}^{r_{\rm max}}  r^{2N} h(r) \frac{ \Theta'}{\Theta}\, dr
\ee
where here we have used the shorthand $\Theta' = d \Theta/dr$. It can easily be confirmed that the counterterm evaluates to the same result for the future left segment. Additionally, the result for the past segments is equivalent with the substitution $r_{m_1} \to r_{m_2}$.

\subsection{Framework for Complexity equals Volume calculations}

We will compare our results obtained for the action with the results within the ``Complexity equals Volume'' framework~\cite{Susskind:2014rva,Stanford2014}.\footnote{A related proposal, called the complexity=volume 2.0, was put forward in \cite{Couch:2016exn}, which suggests that the complexity volume is the spacetime volume of the associated WDW patch.} According to the CV proposal, the complexity of a holographic state at the boundary time slice $\Upsilon$ is related to the volume of an extremal codimension-one slice $\mathcal{B}$ by
\begin{equation}
\mathcal{C}_\mathcal{V}(\Upsilon)=\max\limits_{\Upsilon=\partial\mathcal{B}}\left[\frac{\mathcal{V}(\mathcal{B})}{G_NR}\right] . 
\end{equation} 
The fact that the CV conjecture requires an (arbitrary) length scale $R$ was originally used as an argument in favour of CA over CV. However, there is as yet no universally accepted prescription for computing the bulk complexity, and useful information can be gleaned by comparing different proposals.\footnote{Moreover, it was subsequently realized that the CA proposal also possesses an ambiguous length scale, namely the one associated with the counter-term for null boundaries.}

To find the volume of the extremal codimension-one slice $\mathcal{B}$, write the metric \eqref{metric} in ingoing coordinates $x^\mu=\left(v,r,\vec{\Omega}\right)$, and parameterize the surface with coordinates $y^a=\left(\lambda,\vec{\Omega}\right)$, where $\vec{\Omega}$ are the angular coordinates.\footnote{This choice is possible due to the enhanced symmetry of the equal-spinning black holes studied here.} Below, we choose the symmetric case of boundary times $t_R=t_L\equiv\tau/2$. The induced metric on the codimension-one slice is then
\be
\sigma_{ab}=e_a^\mu e_b^\nu g_{\mu\nu},\quad e_a^\mu\equiv\frac{\partial x^\mu}{\partial y^a}
\ee
where $g_{\mu\nu}$ is the MP-AdS metric \eqref{metric}. The volume functional of this slice can be shown to be
\begin{align}\label{MP-V}
\mathcal{V}&=\int \sqrt{|\sigma|} d^{D-1}x =\Omega_{D-2}\int h(r)r^{D-3}\sqrt{-f(r)^2\dot{v}^2+2g(r)f(r)\dot{v}{\dot{r}}}\, d\lambda
\end{align}
where $v=v(\lambda)$ and $r=r(\lambda)$. We assume\footnote{This is possible because the volume functional \eqref{MP-V} is reparametrization-invariant --- that is, it is invariant under $\lambda\rightarrow\tilde{\lambda}(\lambda)$.} a parametrization where
\be\label{choice}
h(r)r^{D-3}\sqrt{-f(r)^2\dot{v}^2+2g(r)f(r)\dot{v}{\dot{r}}}=1.
\ee
This Lagrangian is independent of $v$ and hence there is a conserved quantity (analogous to energy) given by
\be\label{E}
E=-\frac{\partial\mathcal{L}}{\partial \dot{v}} = h(r)^2r^{2(D-3)} \left(f(r)^2\dot{v}-g(r)f(r)\dot{r}\right).
\ee
Furthermore, we have from \eqref{choice} and \eqref{E}
\be
h(r)^{-2}r^{-2(D-3)}E^2+f(r)^2=h(r)^2r^{2(D-3)}g(r)^2f(r)^2\dot{r}^2 . 
\ee
The volume of this extremal surface is obtained by integrating \eqref{MP-V} on-shell:
\begin{align}
\mathcal{V}&=2\Omega_{D-2}\int_{r_{\rm min}}^{r_{\rm max}} \frac{dr}{\dot{r}}=2\Omega_{D-2}\int_{r_{\rm min}}^{r_{\rm max}}\frac{h(r)^2r^{2(D-3)}g(r)f(r)}{\sqrt{h(r)^{2}r^{2(D-3)}f(r)^2+E^2}} \, dr
\end{align}
where we included a factor of $2$ to include the left half of the surface. Here we wish to take $r_{\rm max}$ to be infinity, but this will yield a divergent result in general. A finite result can be obtained by studying the time derivative of the volume (as relevant for the growth rate), or by performing a carefully matched subtraction of the AdS vacuum (as relevant for the complexity of formation). Here $r_{\rm min}$ is the turning point of the surface, determined by the condition $\dot{r}  = 0$:
\be\label{rmin}
h(r_{\rm min})^{-2}r_{\rm min}^{-2(D-3)}E^2+f(r_{\rm min})^2=0 \, .
\ee
A simple calculation shows that $r_{\rm min}$ will be on or inside the (outer) horizon, and so we have that, using \eqref{E}, $f(r_{\rm min})^2<0,\dot{v}(\lambda_{\rm min})>0\Rightarrow E<0$ and we recall that $f(r)^2 < 0$ in the region between the inner and event horizon. 

\section{Complexity of Formation}\label{sec4}
In this section, we study the complexity of formation for rotating black holes in both the CA and CV conjectures. In both cases, we verify  convergence to the static limit and study the dependence of the complexity of formation on thermodynamic parameters near the extremal limit.

\subsection{Complexity Equals Action}\label{sec4.1}

Within the CA conjecture, the complexity of formation is given by the difference between the action of the WDW patch and the action of the global AdS vacuum both evaluated at the $\tau = 0$ timeslice. 

Let us now put together the various pieces accumulated so far. First, consider the sum of the joint and counterterm contributions. As we know from the general arguments in~\cite{Lehner2016}, this result must be independent of the parameterization of the null generators, i.e. independent of $\alpha$. We find that
\begin{align} 
I_{\rm jnt}^{F} + 2 I_{\rm jnt}^{r_{\rm max}} &+  I_{\rm ct}^{F, R} + I_{\rm ct}^{F, L} = \frac{\Omega_{2N + 1}}{8 \pi G_N} \bigg[r_{\rm max}^{2N} h(r_{\rm max}) \log \frac{\ell_{ct}^2 \Theta(r_{\rm max})^2 |f(r_{\rm max})^2|}{\alpha^2}
\nonumber\\
&-(r_{m_1})^{2N} h(r_{m_1}) \log \frac{\ell_{ct}^2 \Theta(r_{m_1})^2 |f(r_{m_1})^2|}{\alpha^2} \bigg] - \frac{\Omega_{2N+1}}{4 \pi G_N} \int_{r_{m_1}}^{r_{\rm max}}  r^{2N} h(r) \frac{ \Theta'}{\Theta}\, dr \, .
\end{align}
Note that this expression is completely independent of $\alpha$ --- $\Theta$ is proportional to $\alpha$ and thus all $\alpha$ dependence precisely cancels out. This is, of course, necessary, but it nonetheless provides a consistency check of our computations. It can further be shown --- assuming the scale $\ell_{ct}$ is the same for both the AdS vacuum and the black hole solutions --- that the first term evaluated at $r_{\rm max}$ cancels precisely with the corresponding ones occurring in the global AdS vacuum. A completely analogous computation holds for the past sheets of the WDW patch yielding the same result as above with the substitution $r_{m_1} \to r_{m_2}$. However, in this case we can further simplify matters by noting that, since $\tau = 0$ for the complexity of formation, $r_{m_1} = r_{m_2} \equiv r_{m_0}$.  Noting that for the AdS vacuum we have
\be 
\Theta_{\rm AdS} = \frac{(2N+1) \alpha}{r} \, ,
\ee
and combining the above with the relevant background subtraction we obtain for the joint and counterterms:\footnote{In obtaining this we have made use of the fact that the caustics at the future meeting point of the WDW patch do not contribute for global AdS.}
\begin{align}
\Delta \left(I_{\rm jnt} + I_{\rm ct} \right) =& -\frac{\Omega_{2N+1} (r_{m_0})^{2N+1}}{2 \pi G_N (2N+1)} -\frac{\Omega_{2N + 1}}{4 \pi G_N} (r_{m_0})^{2N} h(r_{m_0}) \log \frac{\ell_{ct}^2 \Theta(r_{m_0})^2 |f(r_{m_0})^2|}{\alpha^2}  
\nonumber\\
&- \frac{\Omega_{2N+1}}{2 \pi G_N} \int_{r_{m_0}}^{\infty}  r^{2N} \left[  h(r) \frac{\Theta'}{\Theta} + 1 \right]\, dr \, .
\end{align} 
where we have extended the range of integration to   infinity in the last term since the subtraction has made the integral convergent. Note also that $r_{m_0}$ is obtained by solving the equation
\be 
r^*(r_{m_0}) - r^*_\infty = 0 \, .
\ee
\begin{figure}[t]
\centering
\includegraphics[width=0.6\textwidth]{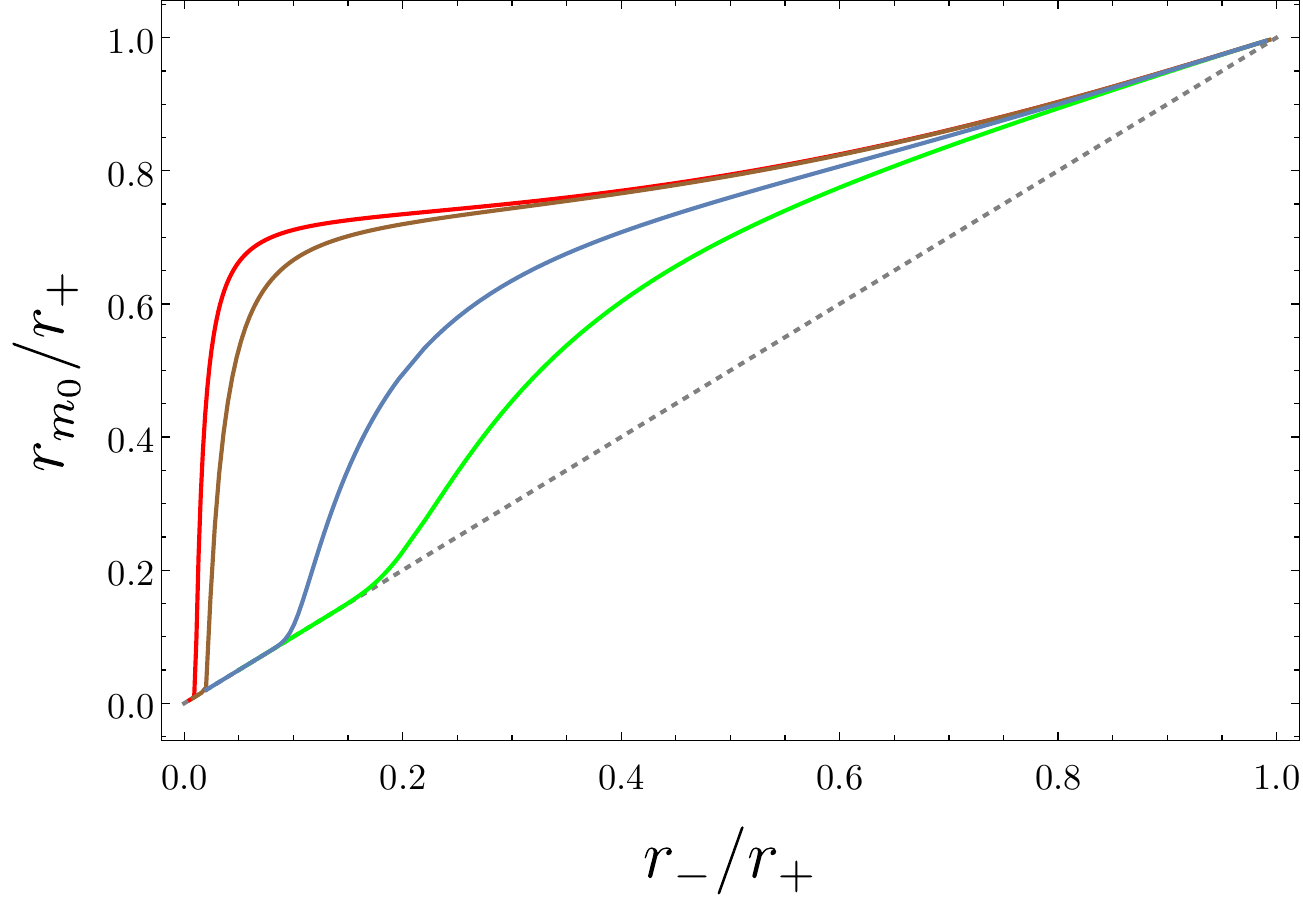}
\caption{A plot showing the value of $r_{m_0}$ vs. $r_-/r_+$ for several different values of $r_+/\ell$. The curves correspond to $r_+/\ell = 5, 10, 50, 100$ in order from bottom to top.}
\label{rMeetPlot}
\end{figure}

For the case of the complexity of formation,  additional simplifications occur for the bulk integral. It becomes (including the necessary factor of two)
\be 
I_{\rm bulk} = \frac{\Lambda \Omega_{2N+1}}{(2N+1) \pi G_N} \int_{r_{m_0}}^{r_{\rm max}} r^{2N+1} \left[r^*_\infty - r^*(r) \right] dr \, .
\ee
Since $r^*$ must be computed numerically, followed by a numerical evaluation of this integral, it is actually more convenient to use integration by parts to eliminate the appearance of $r^*(r)$ inside this expression, leaving only a single integral to evaluate numerically. Doing so, we find that
\be 
I_{\rm bulk} = \frac{\Lambda \Omega_{2N+1}}{2 (N+1) (2N+1) \pi G_N} \left( r^{2(N+1)} \left[r^*_\infty - r^*(r) \right]\bigg|_{r_{m_0}}^{r_{\rm max}} + \int_{r_{m_0}}^{r_{\rm max}} r^{2N+1} g(r)^2 h(r)  dr \right)   \, .
\ee
Note that the evaluation of the first term at $r_{m_0}$ vanishes by virtue of the equation defining $r_{m_0}$. It can further be shown, using the asymptotic form of the tortoise coordinate, that the evaluation at $r_{\rm max}$ cancels with the analogous one coming from the global AdS vacuum. Taking this into account and performing the background subtraction we obtain the result
\begin{align} 
\Delta I_{\rm bulk} = \frac{\Lambda \Omega_{2N+1}}{2 (N+1) (2N+1) \pi G_N} \bigg[&
\int_{r_{m_0}}^{\infty} r^{2N+1} \left( g(r)^2 h(r)  - \frac{r}{1 + r^2/\ell^2} \right)  dr 
\nonumber\\
&- \int_0^{r_{m_0}} \frac{r^{2(N+1)}}{1 + r^2/\ell^2}dr  \bigg] \, ,
\end{align}
where we have extended the range of the first integral to $r = \infty$ since the subtraction has made it convergent.
\begin{figure}[t]
\centering
\includegraphics[width=0.6\textwidth]{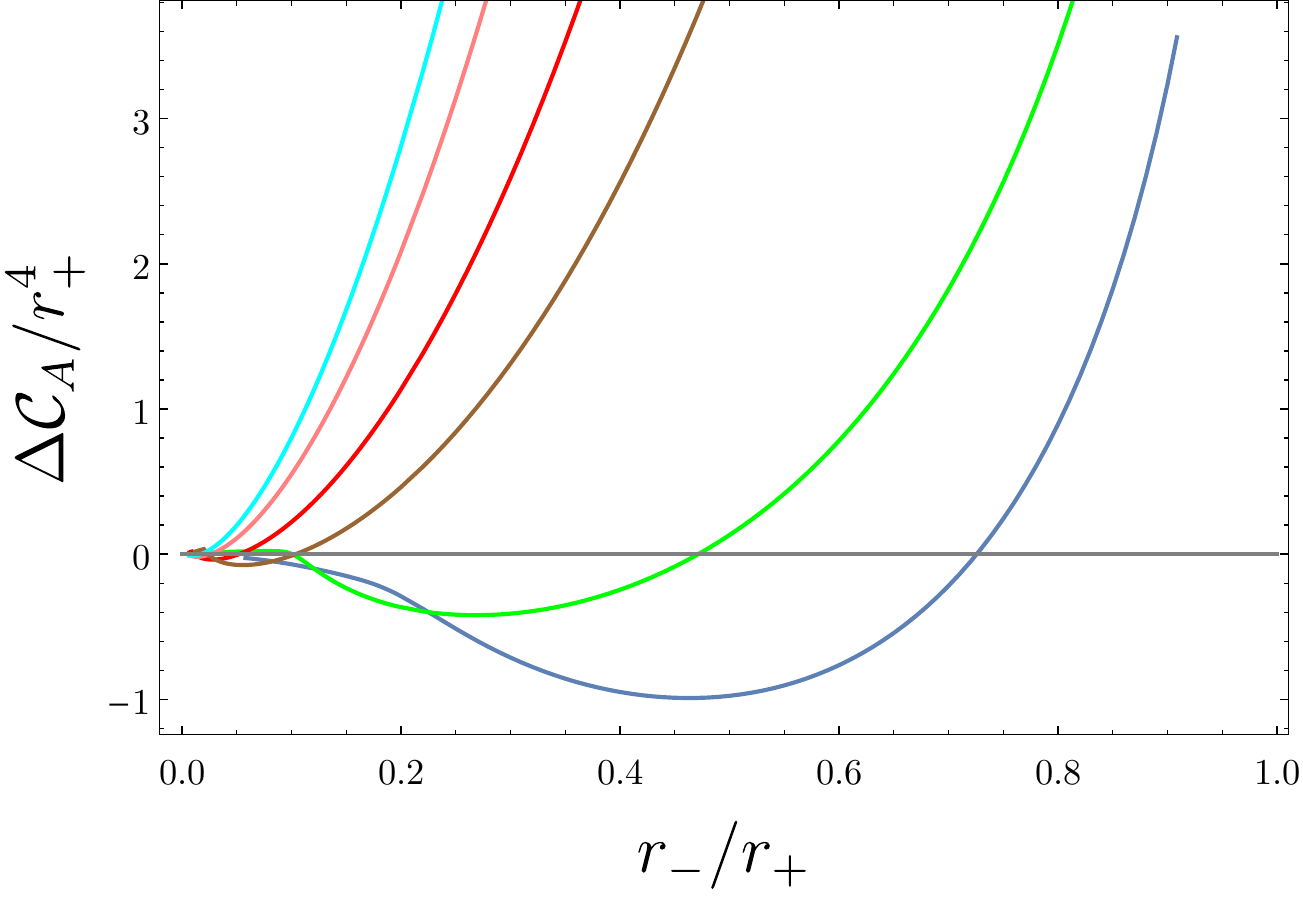}
\caption{A plot showing the complexity of formation in the action formalism for different values of $r_+/\ell$ and for $D=5$. The curves correspond to $r_+/\ell = 5, 10, 50, 100, 200, 300$ in order from bottom to top in a vertical slice on the right side of the plot. Here we have set $\ell_{\rm ct} = \ell$.}
\label{CAformPlot}
\end{figure}

The most complicated aspect of determining the complexity of formation within the action framework is computing the value of $r_{m_0}$ numerically. We show in figure~\ref{rMeetPlot} the resulting curves for several values of $r_+/\ell$. The difficulty arises in determining accurate results in the limit where $r_-/r_+$ becomes small. As discussed previously, in this limit the value of $r_{m_0}$ can be worked out perturbatively and, for five dimensions, reads
\be 
r_{m_0} = r_- \left[1 + \exp \left(- \frac{\pi  \ell (r_+^2 + \ell^2) }{ r_+ (\ell^2 + 2r_+^2) } \frac{r_+^2}{r_-^2} + \cdots \right)  + \cdots \right] \, .
\ee 
Thus, as $r_-/r_+ \to 0$, the difference between $r_{m_0}$ and $r_-$ tends to zero like $\exp(-1/r_-^2)$, and so increasing numerical precision is required in this limit. For sufficiently small $r_-/r_+$ the problem effectively becomes numerically intractable and we are forced to resort to perturbative techniques. 

In figure \ref{CAformPlot}, we show the complexity of formation $\Delta\mathcal{C}_A$ for five-dimensional rotating black holes with different values of $r_+/\ell$. There are a few noteworthy things here. The basic structure of the curves is qualitatively similar for different values of $r_+/\ell$. A somewhat strange feature is that there is a range of parameter values over which the complexity of formation actually becomes negative. While strange, it must be kept in mind that complexity of formation is a relative quantity: it is computed by subtracting one (infinite) result from another. Moreover, in some cases, namely involving gravitational solitons, a negative complexity of formation has been previously observed~\cite{Reynolds:2017jfs, Andrews:2019hvq}, and so this result in and of itself is not new.  While there is an intermediate regime in which the complexity of formation is negative, it is always positive at the two extremes of the plot: in the extremal and nonrotating limits. That the former is true is obvious from the plot, but the static limit is subtle and requires additional scrutiny.

The static limit is examined in detail in appendix~\ref{appStaticLimit}. Here, for conciseness, we will present a discussion relevant to the five-dimensional case. In the static limit $r_-/r_+ \to 0$ all contributions to the corner/joint term vanish except for the term involving the logarithm,
\be 
-\frac{\Omega_{2N + 1}}{4 \pi G_N} (r_{m_0})^{2N} h(r_{m_0}) \log \frac{\ell_{ct}^2 \Theta(r_{m_0})^2 |f(r_{m_0})^2|}{\alpha^2} \, .
\ee
Using the perturbative expansion for $r_{m_0}$ shown above, we can work out that this term yields a finite limit 
\be 
-\frac{\Omega_{2N + 1}}{4 \pi G_N} (r_{m_0})^{2N} h(r_{m_0}) \log \frac{\ell_{ct}^2 \Theta(r_{m_0})^2 |f(r_{m_0})^2|}{\alpha^2} \to \frac{\pi^2 r_+^2 (r_+^2 + \ell^2)^{3/2}}{2G_N(\ell^2 + 2 r_+^2)}
\ee
and we reiterate that here we are considering the case of five dimensions ($N = 1$). This result is \textit{exactly half} the contribution arising from the GHY terms on the future/past singularity in the Schwarzschild-AdS geometry. A similar analysis can be carried out for the bulk term, which in the static limit (see appendix for details) yields
\be 
\lim_{r_-/r_+ \to 0} \Delta I_{\rm bulk} = \Delta I_{\rm bulk}^{\rm Schw} \, .
\ee
That is, the bulk contribution of the rotating black hole limits to exactly the bulk contribution for the non-rotating black hole. As a result, there is an order of limits problem for the action computation: taking the static limit of the action result gives an answer that does not agree with the direct computation done for the Schwarzschild-AdS black hole.

\begin{figure}[t]
\centering
\includegraphics[width=0.45\textwidth]{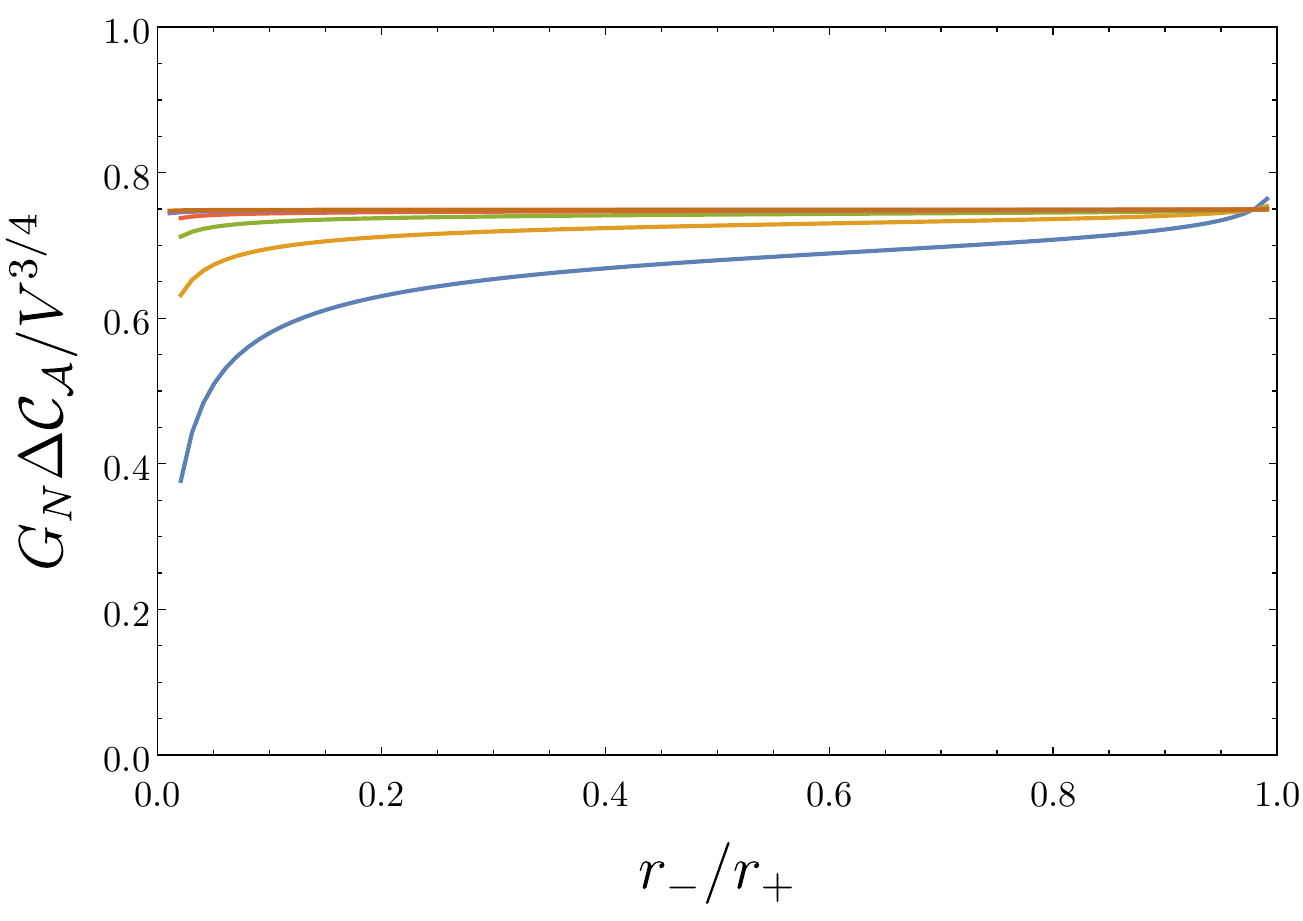}
\quad 
\includegraphics[width=0.45\textwidth]{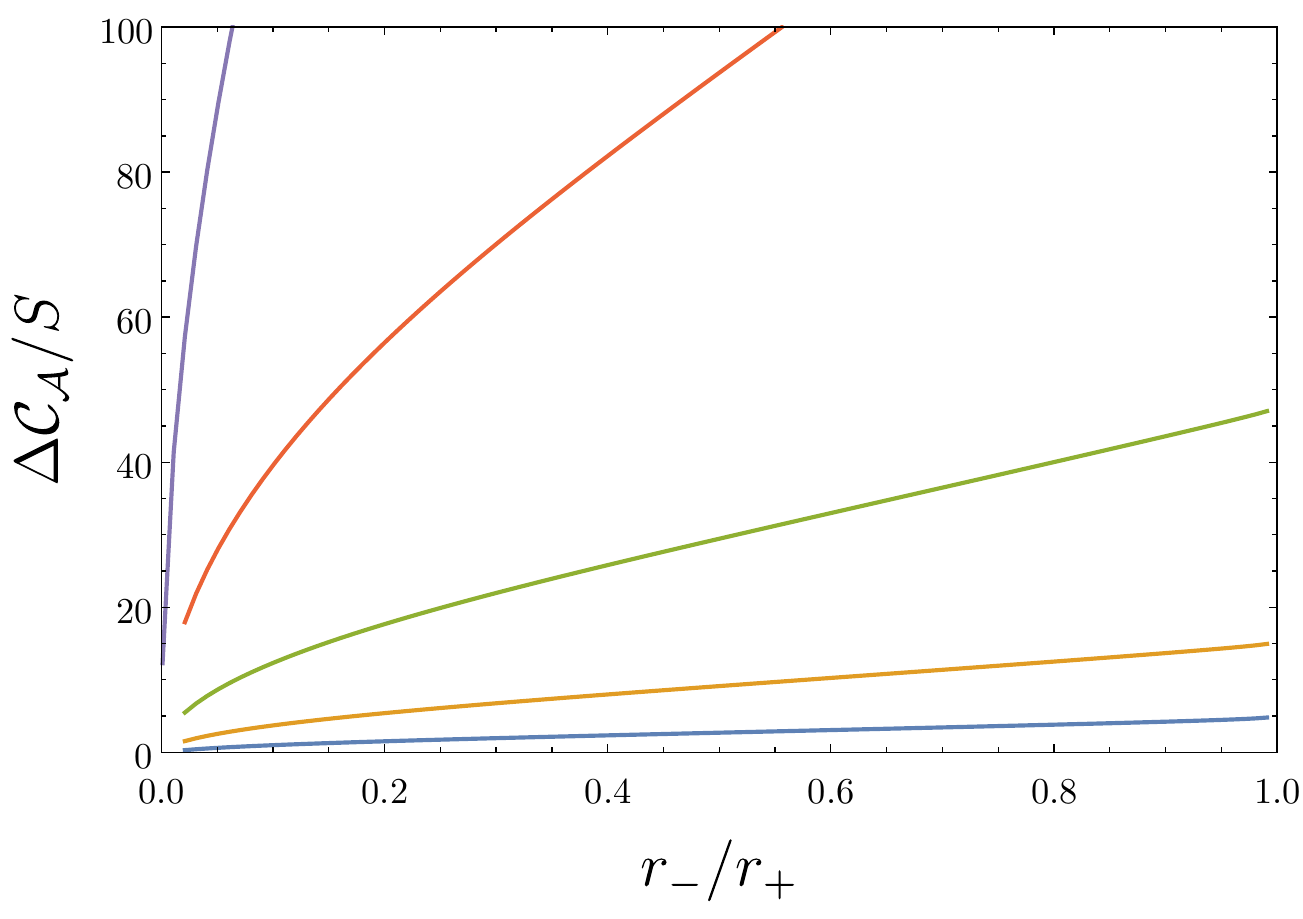}
\caption{The complexity of formation is shown as a function of $r_-/r_+$ normalized by a power of the thermodynamic volume (left) and the entropy (right). The curves, in order from bottom to top, correspond to $r_+/\ell = 10^3, 10^4, 10^5, 10^6, 10^7$ and $10^8$. }
\label{CAformExt}
\end{figure}


It is insightful here to consider how this limit compares with the analogous neutral limit for charged black holes. Again, we consider this in full detail and in all dimensions in appendix~\ref{appStaticLimit}. For the charged black hole, the joint term reduces to a fraction of the Schwarzschild-AdS GHY term in the neutral limit, while the bulk action for charged black holes reproduces the full Schwarzschild-AdS bulk action \textit{along with the remaining fraction of the GHY term}. Thus, for charged black holes, there is not an order of limits problem. However, the manner in which the various terms conspire to give the neutral limit is rather nontrivial. The main difference here in the rotating case is that the limit of bulk term  does not include an additional fraction of the GHY term. This can be traced, mathematically, to the behaviour of the metric function $h(r)$ in this limit.

It should be noted that while when $a = 0$ the metric is simply the usual static AdS black hole, the limit considered here is  different and this is the mathematical reason behind the order of limits issue. Effectively, here we are simultaneously zooming in on the inner horizon while taking the limit $r_- \to 0$. In this limit the metric function $h$ is not simply $r$ (as it would be for the static black hole), but instead it limits to a constant value. As discussed in Appendix~\ref{appStaticLimit}, this behaviour is the source of the order of limits issue, which in general dimensions becomes: 
\be
\lim_{r_-/r_+\to 0} \pi \Delta \mathcal{C}_\mathcal{A} = \frac{I_{\rm GHY}^{\rm Schw}}{N+1} + \Delta I_{\rm Bulk}^{\rm Schw} \neq \pi  \Delta \mathcal{C}_{\rm form}^{\rm Schw}
\ee
where the complexity of formation of the static black hole $\Delta \mathcal{C}_{\rm form}^{\rm Schw}$ is the sum of the bulk $\Delta I_{\rm Bulk}^{\rm Schw}$ and surface $I_{\rm GHY}^{\rm Schw}$ contributions.

There are (at least) two perspectives one could have on this issue. First, it could be viewed as simply a genuine feature of the CA proposal. The CA proposal is highly sensitive to the detailed causal structure of spacetime, and the order of limits issue found here is not the first of its kind. For example the rate of growth of complexity for dilaton black holes was found to be highly sensitive to the details of the causal structure~\cite{Goto:2018iay}.
 Moreover in the usual framework the complexity growth rate for magnetic black holes is precisely zero~\cite{Goto:2018iay, Brown:2018bms}, leading to an obvious order of limits problem (though it is possible to remedy this case through the addition of an electromagnetic counterterm). Furthermore, the growth rate of complexity for charged black holes in higher-curvature theories  exhibits an order of limits problem in the neutral limit~\cite{Cai2016, Cano2018, Fan:2019aoj}. Thus there is precedent for subtle behaviour of the CA conjecture, and it would be interesting to better understand whether this is consistent with CFT expectations.

An alternate perspective is that this order of limits issue is a problem that must be resolved. One means to do so is to consider an alternative regularization scheme for the WDW patch --- which we explain in detail in appendix~\ref{appOtherReg}. The basic idea is to introduce space-like regulator surfaces cutting off the future and past tips of the WDW patch at $r=r_{m_0}+\Delta r$. This could be motivated from the perspective that the inner Cauchy horizon is expected to be unstable to generic perturbations~\cite{Balasubramanian:2019qwk, Hollands:2019whz, Hartnoll:2020rwq}, and therefore this cutoff would encode some level of agnosticism of what happens precisely at the inner horizon. This leads to a well-defined static limit to the complexity
\be
\lim_{\Delta r\to 0}\lim_{r_-/r_+\to 0} \pi \Delta \mathcal{C}_\mathcal{A} = I_{\rm GHY}^{\rm Schw} + \Delta I_{\rm Bulk}^{\rm Schw} = \pi  \Delta \mathcal{C}_{\rm form}^{\rm Schw}
\ee
but it must be noted that the limits do not commute. Moreover, for sufficiently small $\Delta r$ there is no appreciable effect of this term on the results when both $r_-$ and $r_+$ are sufficiently large, but it becomes important in the limit $r_-/r_+ \to 0$.\footnote{It is also worth noting that there appears to be no simple modification of the action proposal itself that would account for the order of limits problem. For example, if one considers only the bulk action as the relevant term then there would be no order of limits issue for rotating black holes, but it would introduce one for charged black holes --- see appendix~\ref{appStaticLimit}.}


Let us now leave aside this issue of limits and consider in more detail some further interesting properties of the complexity of formation. Our focus here is primarily on the scaling behaviour of complexity in the limit of large ($r_+/\ell \gg 1$) black holes. For neutral and charged static black holes this behaviour is governed by the entropy~\cite{Chapman2017Form, Carmi2017}, leading to the idea that the complexity of formation is effectively controlled by the number of degrees of freedom possessed by the system. We can schematically write this relationship for charged black holes as:
\be 
\Delta \mathcal{C}^{\rm charged}_{\mathcal{A}} \underset{\frac{r_+}{\ell} \gg 1}{\sim} S \log \frac{\mu}{T} + f\left(\frac{\mu}{T}\right) S 
\ee 
where $\mu$ is the chemical potential. The function $f(\mu/T)$ has a smooth, non-vanishing limit as $\mu \to 0$. The relationship above is schematic and so neglects possible constant terms in the coefficients and so on. However it conveys the important features: the complexity of formation exhibits a logarthmic singularity near extremality and the general form is controlled by the entropy.

We consider the analogous problem in detail for rotating black holes in appendix~\ref{appExtremal}. Again, there is a logarthmic singularity in the extremal limit that is controlled by the entropy. However, the general behaviour is markedly different. The schematic form for the complexity of formation for large rotating black holes takes the form
\be \label{C-vol}
\Delta \mathcal{C}_\mathcal{A} \underset{\frac{r_+}{\ell} \gg 1}{\sim} S \log \frac{\Omega_H}{T} +  f\left(\frac{\Omega_H}{T}\right) V^\frac{D-2}{D-1} \, ,
\ee
where $\Omega_H$ is the angular velocity of the horizon, $V$ is the thermodynamic volume and again $f$ is some function of the ratio $r_-/r_+$ (which can, of course, be expressed as a function of $\Omega_H/T$). Examining the curves in figure~\ref{CAformExt}, we see that  the second term in \eqref{C-vol} dominates over a larger range of temperature. For  smaller values of $r_+/\ell$, the logarithmic divergence becomes manifest in the limit of extremality. The implication of the above relationship is that at a given fixed temperature, and for sufficiently large black holes, the complexity of formation is always controlled by the thermodynamic volume rather than the entropy. The validity of this conclusion can be seen clearly in the plots shown in figure~\ref{CAformExt} for five dimensions --- see also figure~\ref{CAplot}. We emphasize that this observation is possible due to the independence of the thermodynamic volume and the entropy for rotating black holes. In the case of static (charged or neutral) black holes, these quantities are not independent and one is free to write the final result in terms of either $S$ or $V$ as the two quantities are related by 
\be 
S \underset{\text{static limit}}{\sim} V^{(D-2)/(D-1)}  \, .
\ee
We will return to discuss the implications of this result in the discussion.


\subsection{Comparison with Complexity=Volume Conjecture}

\begin{figure}[t]
\centering
\includegraphics[width=0.44\textwidth]{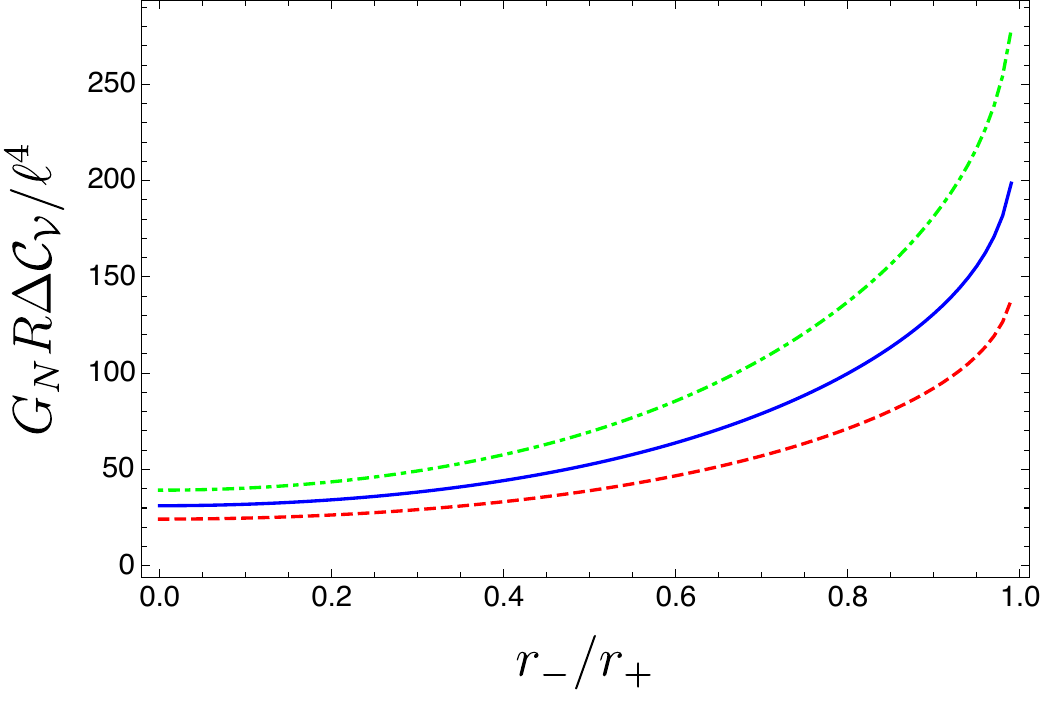}
\quad
\includegraphics[width=0.44\textwidth]{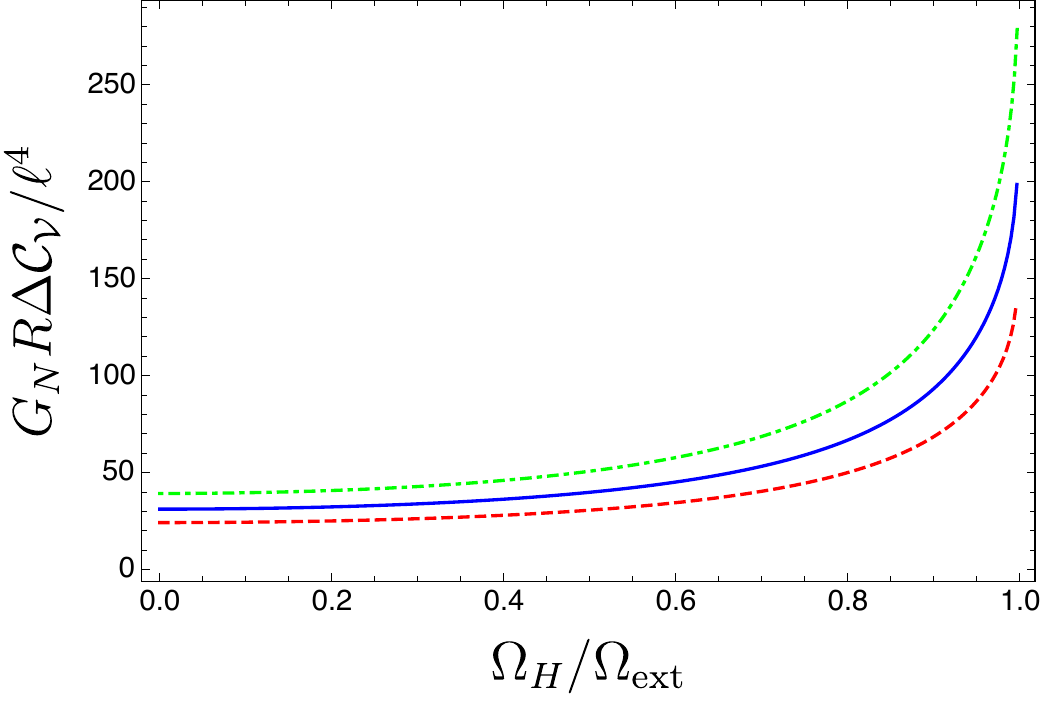}
\caption{Here we show the complexity of formation within the CV proposal. {\it Left}: The curves correspond to $r_+/\ell = 9/10, 1, 11/10$ from bottom to top for $D=5$. The complexity of formation diverges in the extremal limit and tends to a constant  ($r_+/\ell$-dependent) value as $r_-/r_+ \to 0$. {\it  Right}: The same curves now plotted against the rotation of the boundary CFT in the form of the ratio $\Omega_H/\Omega_{\rm ext}$ where $\Omega_H$ is the angular velocity of the boundary CFT and $\Omega_{\rm ext}$ is the zero temperature limit of this angular velocity.}
\label{CVformPlot}
\end{figure}

The complexity of formation in the CV proposal is straightforward to calculate. The volume of the maximal slice in vacuum $AdS_D$ is
\be
\mathcal{V}_0=\Omega_{D-2}\int_{0}^{r_{\rm max}^{\rm AdS}} \frac{r^{D-2}}{\sqrt{f_0(r)}} \,dr ,\quad f_0(r)=1+\frac{r^2}{\ell^2} \, .
\ee
In the black hole geometry, we are interested in the maximal slice at $\tau = 0$. In this case we have  $r_{\rm min}=r_+$ which gives $E=0$ from \eqref{E}. The complexity of formation is then
 \begin{align}\label{CVforma}
 \Delta\mathcal{C}_\mathcal{V}=\frac{\mathcal{V} - 2 \mathcal{V}_0}{G_NR} =\frac{2\Omega_{D-2}}{G_NR}\left[\int_{r_{+}}^{r_{\rm max}} h(r)r^{D-3}g(r)\, dr-\int_{0}^{r_{\rm max}^{\rm AdS}} \frac{r^{D-2}}{\sqrt{f_0(r)}}\, dr\right] \, .
 \end{align}
The integral can be evaluated numerically in a straightforward manner, and we show some representative examples in figure~\ref{CVformPlot}. The qualitative structure of the curves is independent of the value of $r_+/\ell$. Though, since $\Delta \mathcal{C}_\mathcal{V}$ is not a homogeneous function of $r_+/\ell$, there is no simple factor that collapses the different curves to a single line for all values of $r_+/\ell$. When $r_-/r_+ \to 0$, the complexity of formation tends to a constant value, whereas it diverges in the extremal limit. This divergence is consistent with results obtained previously for charged black holes~\cite{Carmi2017}.

In the CA framework we encountered an order of limits issue when taking $r_-/r_+ \to 0$. Here there is no such issue, which is due to the fact that the CV proposal is less sensitive to the detailed properties of the causual structure than the CA proposal. In the static limit, the complexity of formation \eqref{CVforma} reduces directly to that of the static black hole $\Delta\mathcal{C}_\mathcal{V}^{\text{Schw}}$ \cite{Chapman2017Form} since
\be
\lim\limits_{a\rightarrow 0}h(r)g(r)=\frac{r}{\sqrt{f^{\text{Schw}}(r)}}
\ee
where $f^{\text{Schw}}(r)$ is the metric function of the Schwarzschild-AdS spacetime.

\begin{figure}[t]
\centering
\includegraphics[width=0.44\textwidth]{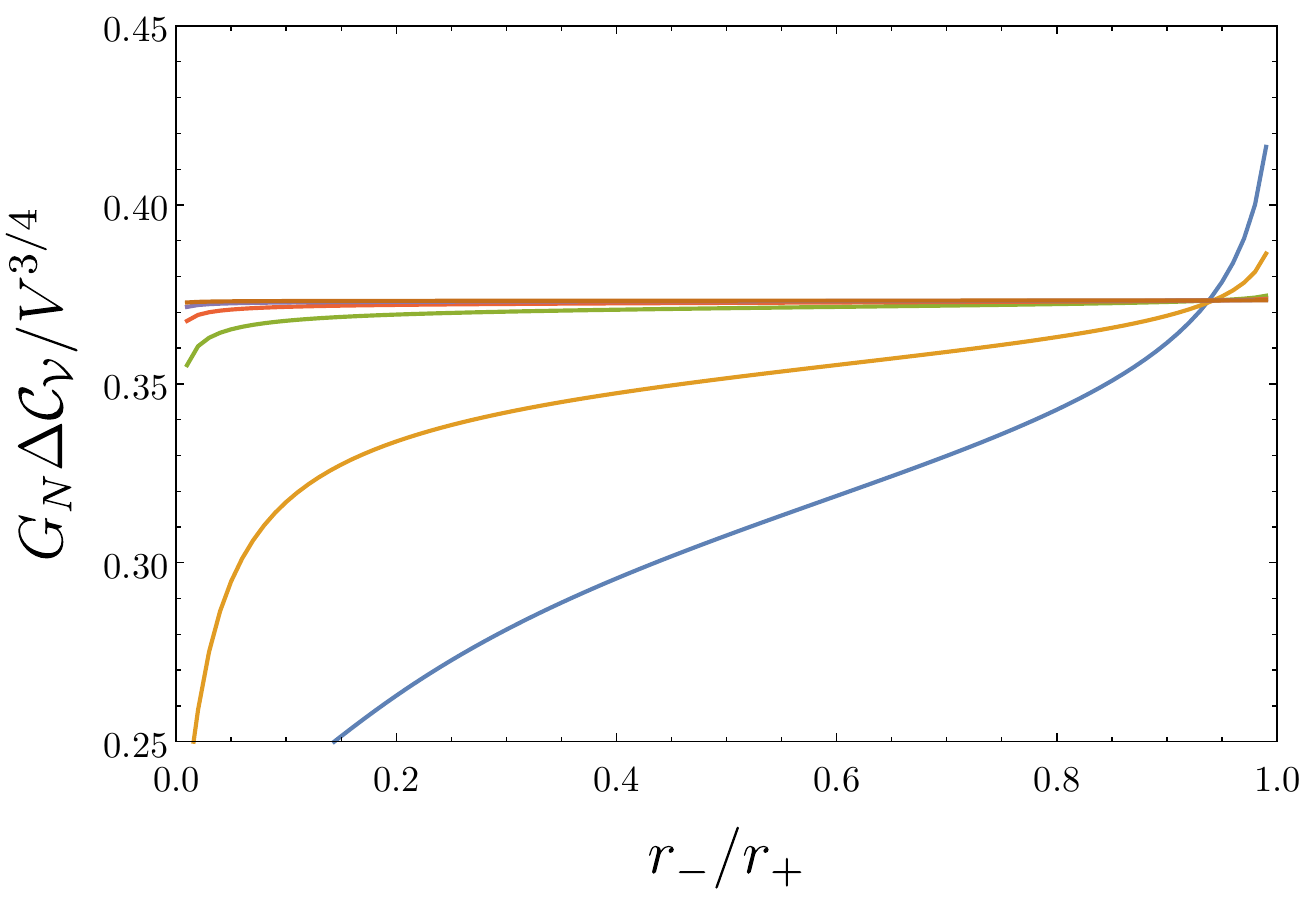}
\quad
\includegraphics[width=0.44\textwidth]{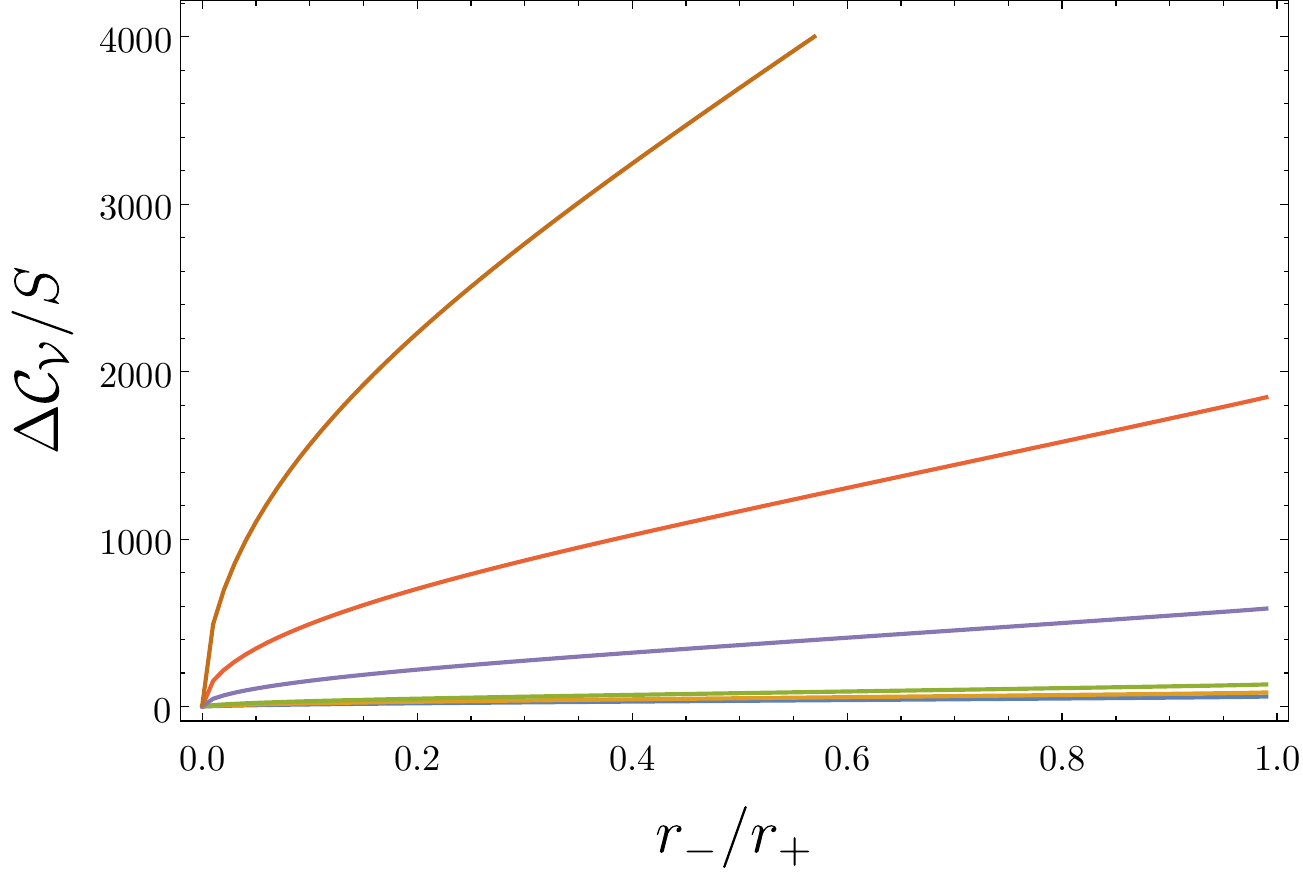}
\caption{Here we show the complexity of formation within the CV proposal normalized by the thermodynamic volume (left) and by the entropy (right).  In each case the curves correspond to $r_+/ \ell = 10, 10^2, 10^3, 10^4, 10^5, 10^6, 10^7$  (in order from bottom to top) for $D=5$. }
\label{CVformThermo}
\end{figure}

It is interesting to further compare the general behaviour of the complexity of formation of large black holes within the CV proposal to the CA proposal. The details of this analysis are presented in appendix~\ref{appExtremal}, but the conclusion is the same. The complexity of formation exhibits a logarithmic singularity near extremality that is controlled by the entropy, while the non-logarthmic terms are controlled by the thermodynamic volume. Thus we once again arrive at the result that for sufficiently large black holes the complexity of formation is controlled by the thermodynamic volume:
\be 
\Delta \mathcal{C}_\mathcal{V} \underset{\frac{r_+}{\ell} \gg 1}{\sim} S \log \frac{\Omega_H}{T} + \tilde{f}\left(\frac{\Omega_H}{T} \right) V^\frac{D-2}{D-1} \, .
\ee
The validity of this can be seen directly in figure~\ref{CVformThermo} --- see also figure~\ref{CV-logPlot}.

\section{Growth Rate of Holographic Complexity}\label{sec5}
In this section, we use the CA and CV proposals to study the full time evolution of holographic complexity of the boundary state \eqref{fullTFD} dual to the MP-AdS black hole geometry. Our interest here will be in understanding the growth rate of complexity, and how this quantity evolves in time.

\subsection{Complexity Equals Action}
As before, we begin our considerations with the action conjecture. The various terms appearing in the computation were assembled in section~\ref{sec3}, and here we proceed and use these directly.  Taking the time derivative of all action terms, we see that only the bulk and joint terms contribute, giving

\begin{align}
\frac{dI_{\text{WDW}}}{d\tau}&=-\frac{(D-1)\times\Omega_{D-2}}{8\pi G_N\ell^2}\int_{r_{m_1}}^{r_{m_2}} r^{2N+1}\, dr\nonumber\\&+\frac{\Omega_{D-2}}{8\pi G_N}\left[\partial_r\left(r^{2N} h(r)\right)\log\left|\frac{\alpha^2}{\ell_{ct}^2{\Theta}^2 f^2}\right|-2 r^{2N} h(r) \frac{(f^2)'}{f^2}\right]_{r= r_{m_2}} \frac{dr_{m_2}}{dt}
\nonumber
\\&+\frac{\Omega_{D-2}}{8\pi G_N}\left[\partial_r\left(r^{2N} h(r)\right)\log\left|\frac{\alpha^2}{\ell_{ct}^2{\Theta}^2 f^2}\right|-2 r^{2N} h(r) \frac{(f^2)'}{f^2}\right]_{r= r_{m_1}} \frac{dr_{m_1}}{dt}.
\label{blah}
\end{align}
The first line in the above is the time derivative of the bulk action, while the second and third lines correspond to the time derivatives of the combined joint and counterterm contributions at the future and past tips of the WDW patch. We recall that since $\Theta \propto \alpha$, this result is actually independent of the parameterization of the null vectors normal to the WDW patch, as it must be. From \eqref{rmeet-1} and \eqref{rmeet-2}, 
\be\label{lll}
\frac{dr_{m_2}}{d\tau}=-\frac{1}{2}\frac{f(r_{m_2})}{g(r_{m_2})},\quad \frac{dr_{m_1}}{d\tau}=\frac{1}{2}\frac{f(r_{m_1})}{g(r_{m_1})} \, ,
\ee
and so once the values of $r_{m_1}$ and $r_{m_2}$ are known, it is possible to evaluate directly the growth rate of complexity. 

\begin{figure}
\centering
\includegraphics[width=0.45\textwidth]{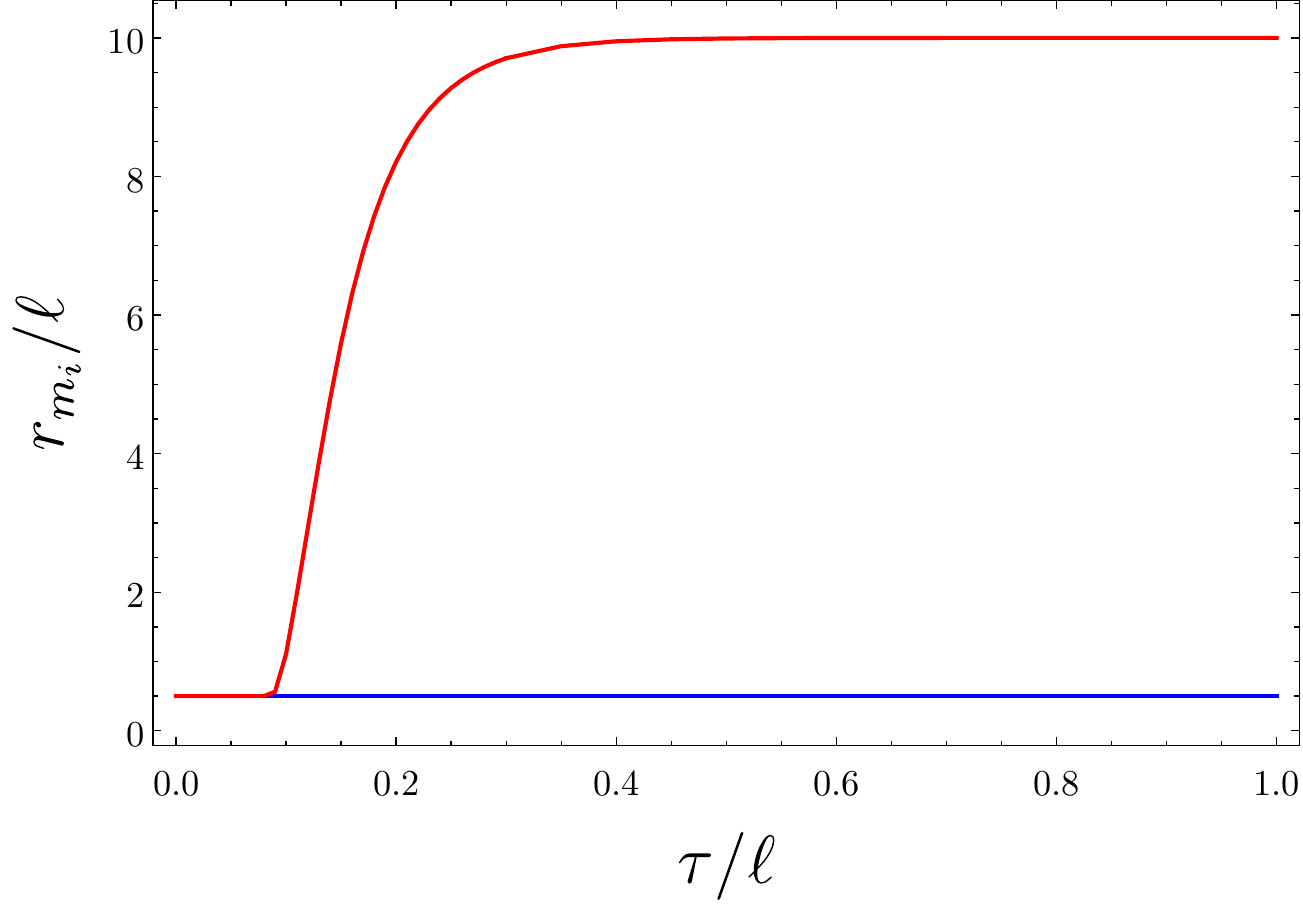}
\quad 
\includegraphics[width=0.45\textwidth]{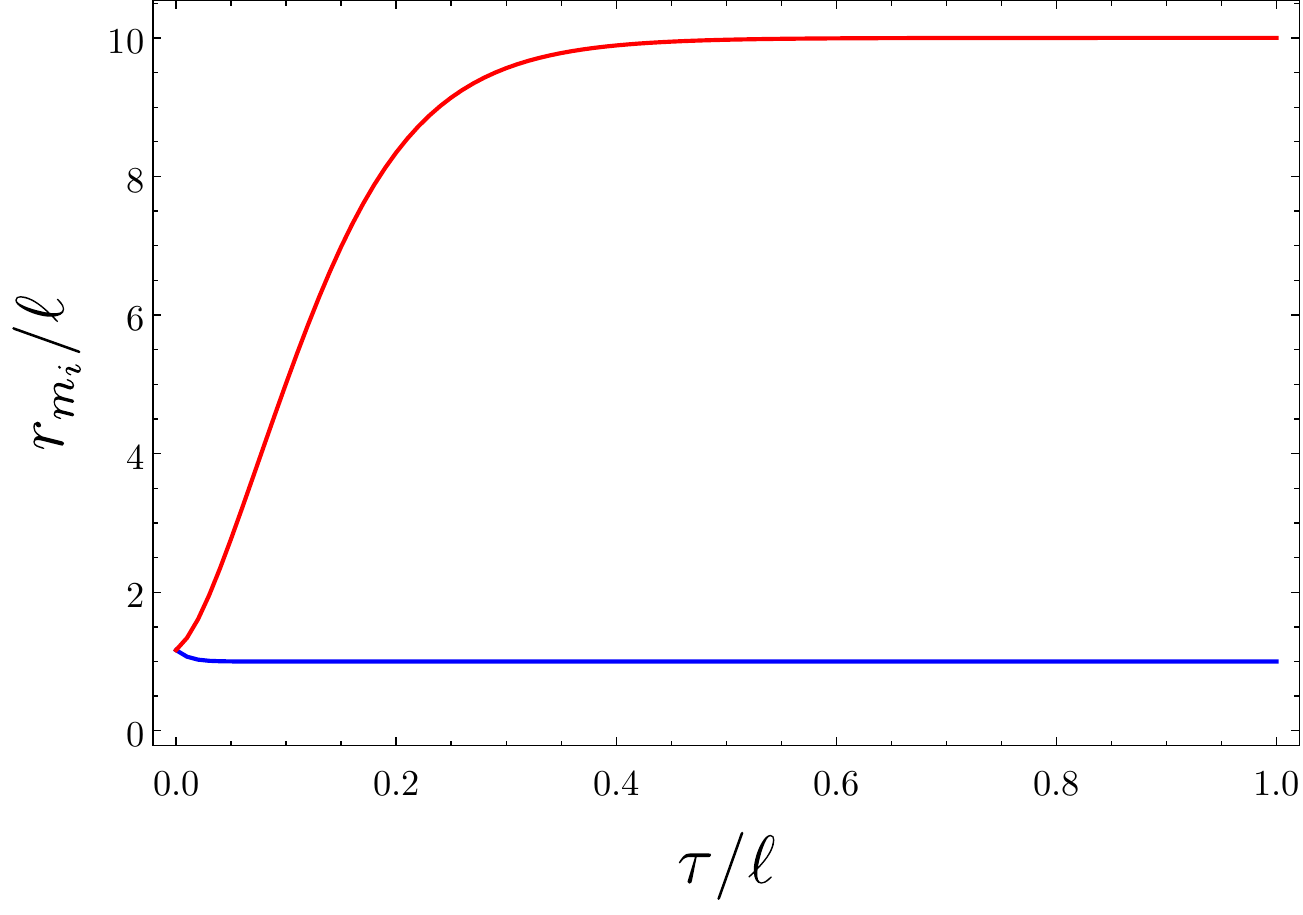}
\includegraphics[width=0.45\textwidth]{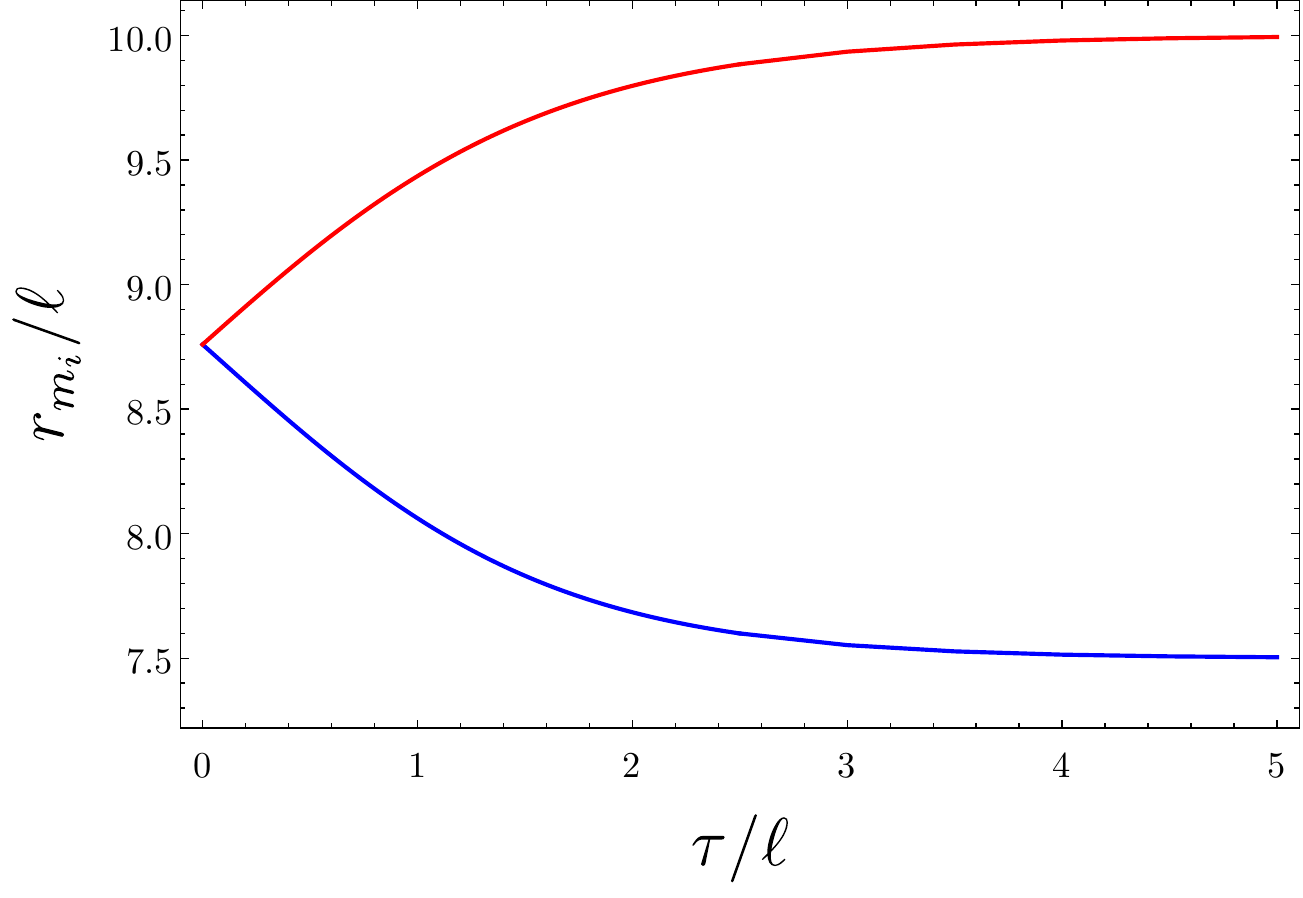}
\caption{Here we show plots of $r_{m_1}$ (blue) and $r_{m_2}$ (red) as a function of time. In each plot we have set $r_+/\ell = 10$, while the different plots correspond to $r_-/r_+ = 1/20, 1/10, 3/4$ (left to right). }
\label{rMeet-time}
\end{figure}

Just as in the case of the complexity of formation, the difficulty here arises in determining the values of $r_{m_i}$, which is a numerically subtle problem. We show some representative results in figure~\ref{rMeet-time}. While we show the results here for a particular value of $r_+/\ell$, this is unimportant for understanding the general behaviour which depends much more strongly on the value of $r_-/r_+$. We see from the top-left figure that, when $r_-/r_+$ is a small value, $r_{m_1}$ and $r_{m_2}$ present a phase where they are effectively constant. The implication of this is a period in the growth rate where the complexity effectively stalls and does not exhibit significant dependence on time. As $r_-/r_+$ increases, $r_{m_i}$ exhibit stronger time dependence, but generally become ``squished'' in a smaller interval (since they must lie between $r_-$ and $r_+$). In all cases, $r_{m_1}$ and $r_{m_2}$ asymptote to the inner and outer horizons, respectively.

\begin{figure}
\centering
\includegraphics[width=0.45\textwidth]{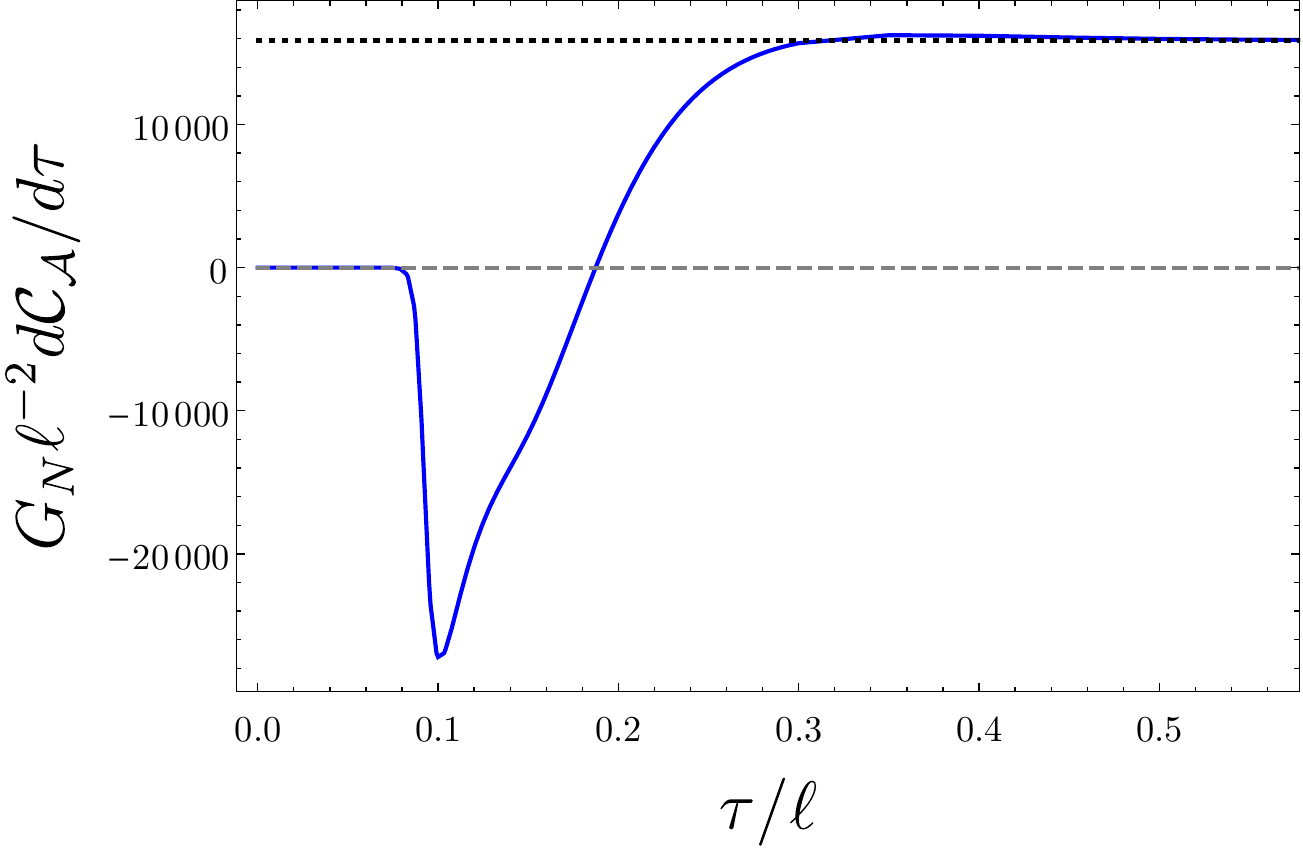}
\quad 
\includegraphics[width=0.45\textwidth]{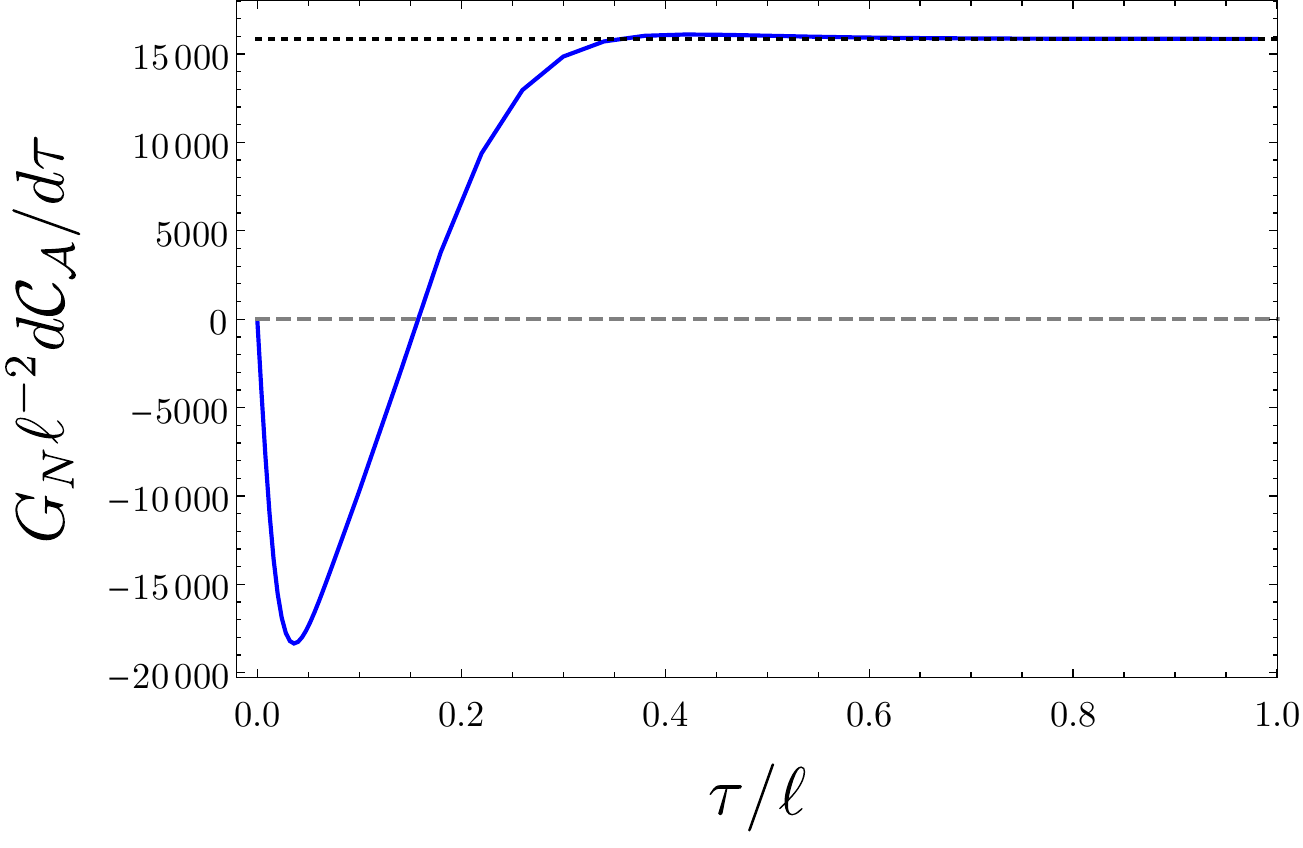}
\includegraphics[width=0.45\textwidth]{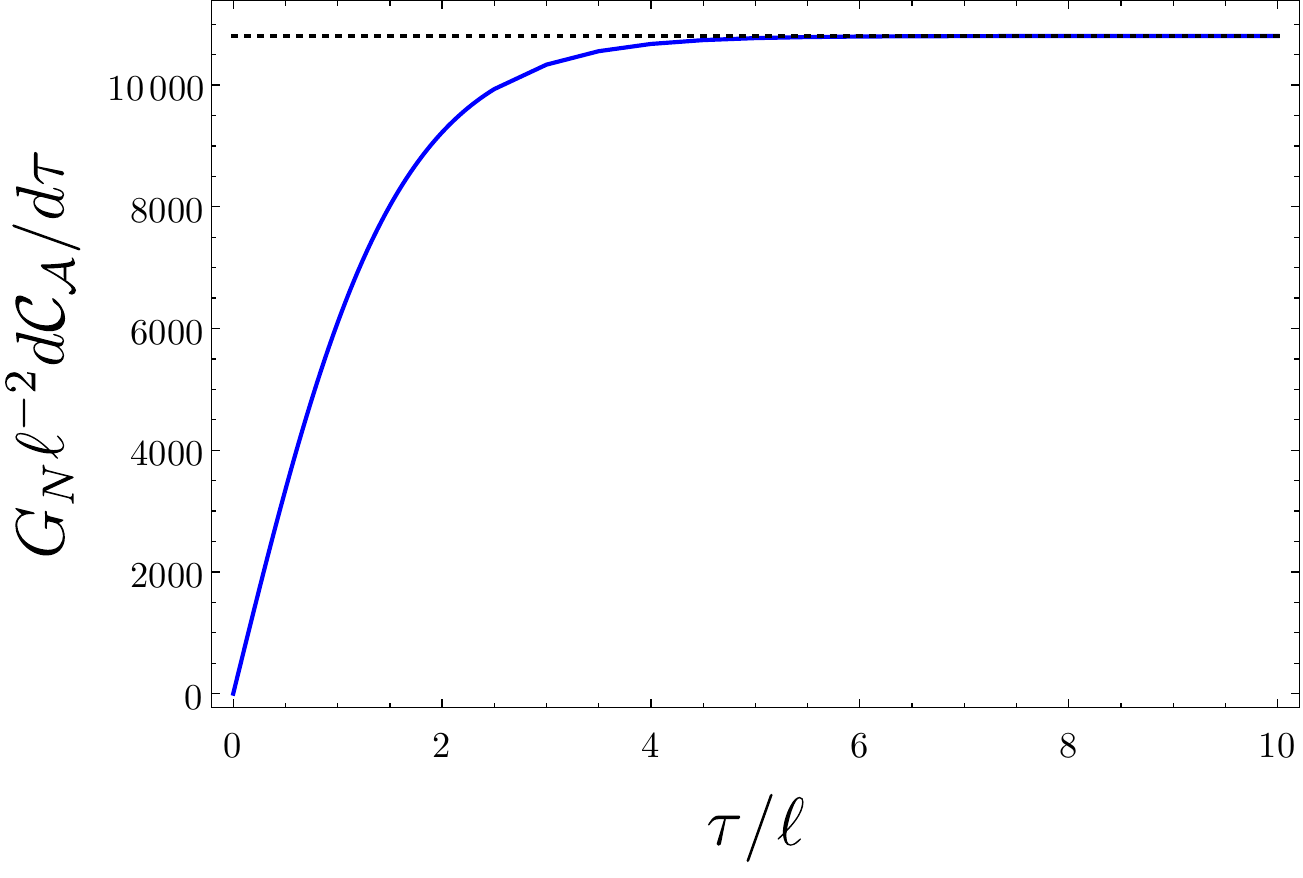}
\caption{Here we show plots of the growth rate of complexity as a function of time. In each plot we have set $r_+/\ell = 10$, while the different plots correspond to $r_-/r_+ = 1/20, 1/10, 3/4$ (left to right). We have set $\ell_{\rm ct} = 1$. The dotted black line shows the growth rate of complexity in the limit $\tau \to \infty$.}
\label{growthRateAction}
\end{figure}
\begin{figure}
\centering
\includegraphics[width=0.65\textwidth]{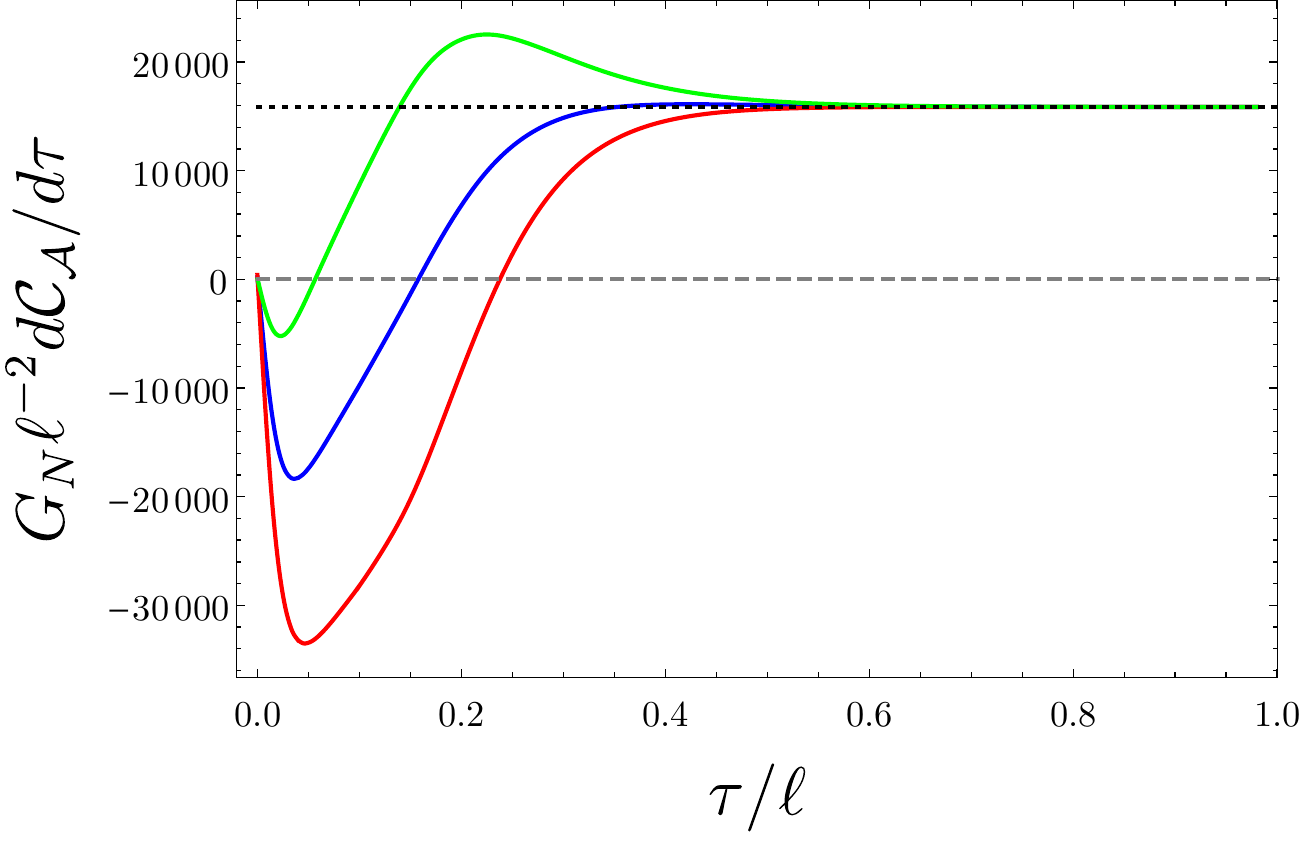}
\caption{Here we show a plot of the growth rate of complexity for the parameter choices $r_+/\ell = 10$, $r_-/r_+ = 1/10$ for the choices $\ell_{\rm ct} = 10, 1, 1/10$ corresponding to red, blue, green curves, respectively --- or bottom to top. The dotted black line shows the growth rate of complexity in the limit $\tau \to \infty$.}
\label{Lct-effect}
\end{figure}

Once the values of $r_{m_i}$ have been determined, it is straightforward to determine the growth rate as a function of time. We show representative results in figure~\ref{growthRateAction} for the same cases for which we displayed $r_{m_i}$ in figure~\ref{rMeet-time}. The results are qualitatively similar to what has been previously observed for charged black holes (c.f. figure 10 of~\cite{Carmi2017}). There are some general features that can be remarked on. First, we note that in the limit of small rotation (equivalently, small $r_-/r_+$) the growth rate develops a minimum. As the rotation is decreased, the minimum becomes sharper and deeper. Moreover, in the same case, the growth rate exhibits a phase where it is close to zero before this oscillatory behaviour manifests. These observations are consistent with the growth rate limiting to that of the static black holes~\cite{Carmi2017}. As the rotation is increased, both the late-time limit of the growth rate decreases and the transient oscillations become less significant. The ultimate limiting case is the extremal limit, where the late-time growth actually goes identically to zero (this will be justified below). While we have shown the growth rate for the particular choice of $\ell_{\rm ct} = 1$, the precise value of this parameter affects significantly only the early-time behaviour --- we show an example of this in figure~\ref{Lct-effect}.

\subsubsection*{Perturbative expansion at late times}

Having presented numerical computations for the full time-dependent growth rate of complexity, let us now turn to discuss some general features at late times. At large $\tau$, using \eqref{goodTort}, we can solve \eqref{rmeet-1} and \eqref{rmeet-2} perturbatively to find that
\begin{align}\label{late-meet}
r_{m_1}(\tau)&=r_-\left(1+c_1\exp\bigg[-{\frac{r_-^2(r_+^2-r_-^2)\tau}{G(r_-)h(r_-)}}\bigg]\right)+\dots,\nonumber\\ 
r_{m_2}(\tau)&=r_+\left(1-c_2\exp\bigg[-{\frac{r_+^2(r_+^2-r_-^2)\tau}{G(r_+)h(r_+)}}\bigg]\right)+\dots
\end{align}
where the dots indicate subleading terms in the large $\tau$ expansion and
\begin{align}
c_1&=2{\left(\frac{r_+-r_-}{r_++r_-}\right)}^{\frac{r_-^2G(r_+)h(r_+)}{r_+^2G(r_-)h(r_-)}}\exp\bigg[\frac{2r_-^2(r_+^2-r_-^2)}{G(r_-)h(r_-)}\int_{r_-}^\infty H(r')\, dr'\bigg]\nonumber,\\
c_2&=2{\left(\frac{r_+-r_-}{r_++r_-}\right)}^{\frac{r_+^2G(r_-)h(r_-)}{r_-^2G(r_+)h(r_+)}}\exp\bigg[-\frac{2r_+^2(r_+^2-r_-^2)}{G(r_+)h(r_+)}\int_{r_+}^\infty H(r')\, dr'\bigg]
\end{align} 
where $H(r)$ is the integrand of $\mathcal{R}(r)$ defined in \eqref{tort}. In the limit $\tau\rightarrow\infty$, it can be shown that
\be\label{ll}
\frac{g_{,r}^{tt}(r)}{g^{tt}(r)}\frac{f(r)}{g(r)}\bigg|_{r\rightarrow r_\pm}=-\frac{r_\pm \mathcal{G}'(r_\pm)}{2h(r_\pm)} = - 2\pi T_\pm
\ee
where we have introduced the notation $\mathcal{G}(r)\equiv {g(r)}^{-2}$, $T_\pm$ is the temperature of the black hole at the horizon $r_\pm$ given in \eqref{thermos}, and
\be
\lim\limits_{r_{m_1}\rightarrow r_-}\frac{f(r_{m_1})}{g(r_{m_1})}\log\left|\frac{\alpha^2 g^{tt}(r_{m_1})}{\ell_{ct}^2{\Theta(r_{m_1})}^2}\right|= \lim\limits_{r_{m_2}\rightarrow r_+}\frac{f(r_{m_2})}{g(r_{m_2})}\log\left|\frac{\alpha^2 g^{tt}(r_{m_2})}{\ell_{ct}^2{\Theta(r_{m_2})}^2}\right|=0 . 
\ee
Expanding \eqref{blah} in this limit using \eqref{late-meet} gives
\begin{align}\label{latetimeIWDW}
\frac{dI_{\text{WDW}}}{d\tau}&=\frac{dI_{\text{WDW}}}{d\tau}\bigg|_{\tau\rightarrow\infty}+\frac{\Omega_{D-2}}{8\pi G_N}{\left(r_+^2-r_-^2\right)}^2\tau \nonumber\\&\times\bigg(\frac{r_+^{D+1}\left((D-3)h(r_+)+r_+h'(r_+)\right)}{G(r_+)^2h(r_+)^2}c_2\exp\bigg[-{\frac{r_+^2(r_+^2-r_-^2)\tau}{G(r_+)h(r_+)}}\bigg]\nonumber\\&-\frac{r_-^{D+1}\left((D-3)h(r_-)+r_+h'(r_-)\right)}{G(r_-)^2h(r_-)^2}c_1\exp\bigg[-{\frac{r_-^2(r_+^2-r_-^2)\tau}{G(r_-)h(r_-)}}\bigg]\bigg)+\dots
\end{align}
where the dots indicate subleading terms in $\tau$, and
\begin{align}\label{IWDWinfty}
\frac{dI_{\text{WDW}}}{d\tau}\bigg|_{\tau\rightarrow\infty}&=\frac{\Omega_{D-2}}{8\pi G_N}\left[-\frac{r_+^{2N+2}-r_-^{2N+2}}{\ell^2}+\frac{r_+^{2N+1}\mathcal{G}'(r_+)-r_-^{2N+1}\mathcal{G}'(r_-)}{2}\right]\nonumber\\
&=\frac{\Omega_{D-2}}{8\pi G_N}\ 2ma^2\left[\frac{1}{r_-^2}-\frac{1}{r_+^2}\right](N+1)  \nonumber \\
&=\frac{\Omega_{D-2}}{8\pi G_N}2ma^2\left[\frac{2m}{r_-^{2N+2}+2ma^2}-\frac{2m}{r_+^{2N+2}+2ma^2}\right](N+1) \nonumber\\
&=  (\Omega_--\Omega_+)J \; .
\end{align}
It is easiest to see the equality of the second and third lines by writing the parameters $(m,a)$ in the bracket in the third term in terms of $(r_+,r_-)$, which yields the second term. Furthermore, this agrees with
\be\label{bulkWDW}
\frac{dI_{\text{bulk}}}{d\tau}\bigg|_{\tau\rightarrow\infty}=M-TS-\Omega J\bigg|_{r_-}^{r_+}
\ee
which is the difference in thermodynamic free energy between the outer and inner horizons, and
\be\label{jntWDW}
\frac{dI_{\text{jnt}}}{d\tau}\bigg|_{\tau\rightarrow\infty}=TS\bigg|_{r_-}^{r_+}
\ee
where we note that $S_\pm$ is given by \eqref{thermos}. Therefore, the late-time complexity rate of growth is simply the difference in internal energy between the outer and inner horizons\footnote{This thermodynamic interpretation of complexity rate of growth was first noted in \cite{Huang:2016fks}. In \cite{Cano2018} it was shown to hold for charged black holes in Lovelock gravity. Thus, we expect that \eqref{bulkWDW} and \eqref{jntWDW} exhibit a universal feature of complexity growth in black holes with two horizons.}
\be\label{CAthermo}
\pi\frac{d\mathcal{C}_\mathcal{A}}{d\tau}\bigg|_{\tau\rightarrow\infty}=\left(F_++T_+S_+\right)-\left(F_-+T_-S_-\right)=U_+-U_-
\ee
where $F_\pm$ and $U_\pm$ are the free and internal energies, respectively, of the outer and inner horizons. The second term in \eqref{latetimeIWDW} was checked for various dimensions and found that it is always positive and less than 1. This strongly suggests that the late-time limit of action rate of growth \eqref{IWDWinfty} is always approached from above.

Using the fact that the Smarr relation \eqref{smarrEq} holds for both outer and inner horizons, we can rewrite \eqref{CAthermo} as
\be
\pi\frac{d\mathcal{C}_\mathcal{A}}{d\tau}\bigg|_{\tau\rightarrow\infty} = T_+S_+ - T_-S_- - \frac{2}{2N+1}P\Delta V
\ee
where $\Delta V=V_+-V_-$ is the difference between the thermodynamic volumes of the outer and inner horizons. Interestingly, in the limit of large black holes, the $TS$ factors and $P\Delta V$ term become proportional to each other and one can show that
\be
\lim\limits_{r_+/\ell\rightarrow\infty}\pi\frac{d\mathcal{C}_\mathcal{A}}{d\tau}\bigg|_{\tau\rightarrow\infty} = \frac{2N+2}{2N+1}P\Delta V.
\ee
As will be shown below, a similar result also holds for the complexity rate of growth in CV conjecture. 

\subsection{Comparison with Complexity=Volume Conjecture}
We will compare the complexity rate of growth according to the CV conjecture $d\mathcal{C}_\mathcal{V}/d\tau$ with the results found according to the CA conjecture. The volume of the extremal codimension-one slice was found in \eqref{MP-V}. To relate to boundary time, note first that
\be
v_{max}=t_R+r_*(\infty),\quad v_{min}=t_{min}+r_*(r_{min})
\ee
where $t_{min}=0$ by left-right symmetry (we have left-right symmetry because the functional $\mathcal{V}$ is invariant under $t\rightarrow -t$ and $a\rightarrow -a$), and $r_{min}$ is defined by \eqref{rmin}. Therefore,
\begin{align}\label{tR}
t_R+r_*(\infty)-r_*(r_{min})&=\int_{v_{min}}^{v_{max}}dv=\int_{r_{min}}^{r_{max}} \frac{g(r)}{f(r)}\left[\frac{E}{\sqrt{h(r)^{2}r^{2(D-3)}f(r)^2+E^2}}+1\right] \, dr
\end{align}
where we used \eqref{E}. Note that the integrand here is convergent at $r=r_\pm$. Finally, it is easy to see that
\be
\frac{\mathcal{V}}{2\Omega_{D-2}}=\int_{r_{min}}^{r_{max}} \frac{g(r)}{f(r)}\left[\sqrt{h(r)^{2}r^{2(D-3)}f(r)^2+E^2}+E\right]\, dr-E\left(t_R+r_*(\infty)-r_*(r_{min})\right).
\ee
Choosing the symmetric case $t_R=t_L\equiv\tau/2$, it is straightforward to show using \eqref{rmin} that
\be
\frac{1}{\Omega_{D-2}}\frac{d\mathcal{V}}{d\tau}=-\ E=r^{D-2}\sqrt{-\mathcal{G}(r_{min})}\equiv W(r_{min})
\ee
where $\mathcal{G}(r)\equiv g(r)^{-2}$. The complexity rate of change is then
\be
\frac{d\mathcal{C}_\mathcal{V}}{d\tau}=\frac{\Omega_{D-2}}{G_NR}W(r_{min}).
\ee
To find its total dependence on time, one first notes that equation \eqref{tR} can be written as
 \be
 \frac{\tau}{2}+r_*(\infty)-r_*(r_{min})=\int_{r_{min}}^{r_{max}} \frac{g(r)}{f(r)}\left[\frac{W(r_{min})}{\sqrt{-W(r)^2+W(r_{min})^2}}+1\right]\, dr
 \ee
 from which one can solve for $r_{min}(\tau)$. However, if we focus on the late-time limit $\lim\limits_{\tau\rightarrow\infty}r_{min}(\tau)=\tilde{r}_{min}$, it can be shown that 
 \be\label{late-rmin}
 0=W'(\tilde{r}_{min})=(D-2)\tilde{r}_{min}^{D-3}\sqrt{-\mathcal{G}(\tilde{r}_{min})}-\frac{\tilde{r}_{min}^{D-2}\mathcal{G}'(\tilde{r}_{min})}{2\sqrt{-\mathcal{G}(\tilde{r}_{min})}}.
 \ee
This is because, from \eqref{rmin}, $r_{min}$ is also the largest root of
 \be
 -W(r_{min})^2+E^2=0
\ee
and as $\tau\rightarrow\infty$, $|E|$ increases until the two roots meet at the extremum of $W(r_{min})$. Therefore, 
 \be
 \frac{d\mathcal{C}_\mathcal{V}}{d\tau}\bigg|_{\tau\rightarrow\infty}=\frac{\Omega_{D-2}}{G_NR}W(\tilde{r}_{min}) \, .
 \ee
Numerical studies have shown that
 \be\label{dCvdt-infty}
 \lim\limits_{r_+/\ell\rightarrow \infty} \frac{d\mathcal{C}_\mathcal{V}}{d\tau}\bigg|_{\tau\rightarrow\infty}=\frac{8\pi}{D-1}\left[\left(M-\Omega_+ J\right)-\left(M-\Omega_- J\right)\right] \, .
 \ee 
This reduces to the result \cite{Carmi2017} found for Schwarzschild-AdS black holes, which was $8\pi M_{sch}/(D-2)$\footnote{$M_{sch}$ is the thermodynamic mass of Schwarzschild-AdS black hole, given by taking the $a\rightarrow 0$ limit of $M$ in \eqref{thermoMass}. Note that for the BTZ black hole with $D=3$, we have $M_{sch}=\frac{r_+^2}{8G_N\ell^2}$ which is different from the one naively obtained from the $D\rightarrow 3$ limit of the blackening factor of Schwarzschild-AdS black hole, giving $M_{sch}=\frac{r_+^2+\ell^2}{8G_N\ell^2}$. This is because for Schwarzschild-AdS black holes with $D>3$ we implicitly assume that the $r_+\rightarrow 0$ limit corresponds to the Neveu-Schwarz vacuum of $AdS_D$ with blackening factor $f(r)=1+\frac{r^2}{\ell^2}$ of the metric in Schwarzschild coordinates, whereas the $r_+\rightarrow 0$ limit of the BTZ black hole corresponds to the Ramond vacuum of $AdS_3$ with blackening factor $f(r)=\frac{r^2}{\ell^2}$ of the metric in Schwarzschild coordinates. For more details on this, see \cite{Coussaert:1993jp}.}, since it is straightforward to show that
\be\label{Msch}
\lim\limits_{r_-\rightarrow 0}\left(\Omega_--\Omega_+\right)J=\frac{D-1}{D-2}M_{sch}.
\ee
For illustration, we will prove \eqref{dCvdt-infty} in spacetime dimensions $D=3$ and $D=5$ below, where generalization to other spacetime dimensions follows the same methods.

\begin{figure*}[t]
\centering
\includegraphics[width=0.8\linewidth]{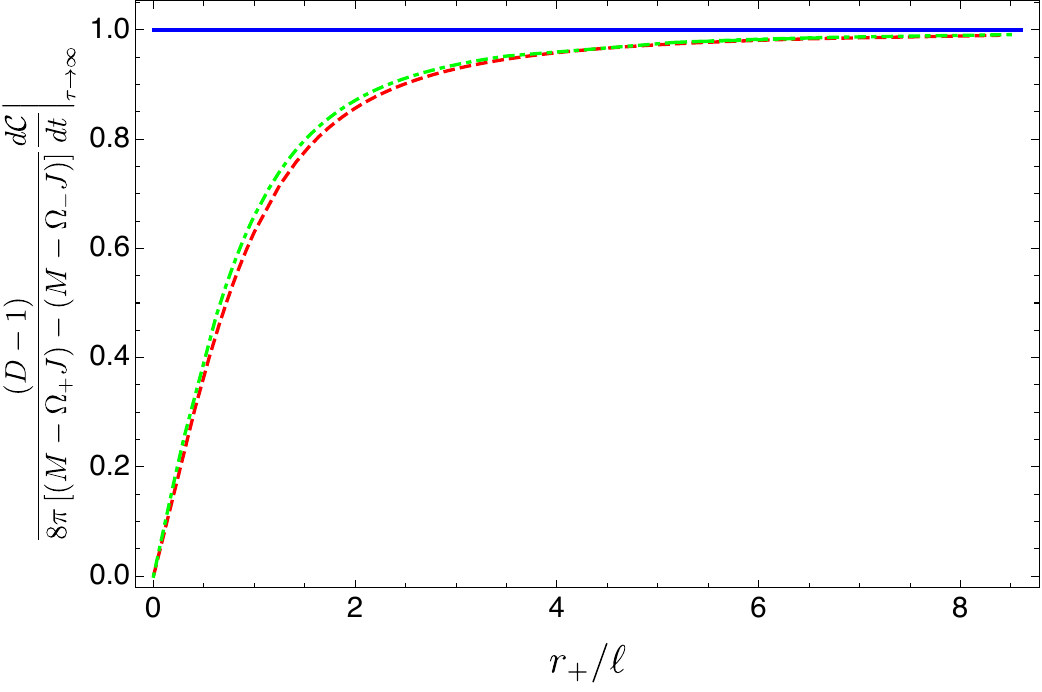}
\caption{The late-time rate of complexity growth $d\mathcal{C}_\mathcal{V}/d\tau$ is shown as a function of $r_+/\ell$ for spacetime dimensions $D=3,5,7$ (solid blue, dashed red, and dot-dashed green, respectively). It is shown that the limit \eqref{dCvdt-infty} is always approached from below. The fact that the late-time $d\mathcal{C}_\mathcal{V}/d\tau$ can be expressed in this way in terms of the thermodynamic quantities of the black hole only in the large $r_+/\ell$ limit shows one of its shortcomings compared to $d\mathcal{C}_A/d\tau$, which can expressed at late-times in terms of thermodynamic quantities of the black hole for all $r_+/\ell$.}
\label{dCvdt-fig} 
\end{figure*}

\subsubsection*{Late-time complexity growth in $D=3$}
In this case, we can explicitly solve \eqref{late-rmin} and find that
\be
\tilde{r}_{min}=\frac{\sqrt{2m\ell^2-2ma^2-\ell^2}}{\sqrt{2}}
\ee
and
\be
r_+=\sqrt{\tilde{r}_{min}^2+\sqrt{\tilde{r}_{min}^4-2ma^2\ell^2}},\quad r_-=\sqrt{\tilde{r}_{min}^2-\sqrt{\tilde{r}_{min}^4-2ma^2\ell^2}}.
\ee
Using this, it is straightforward to show that
\be
W(\tilde{r}_{min})=\frac{\sqrt{\tilde{r}_{min}^4-2ma^2\ell^2}}{\ell},\quad \left(\Omega_--\Omega_+\right)J=\frac{W(\tilde{r}_{min})}{2G_N\ell}
\ee
from which we get \eqref{dCvdt-infty} by setting $R=\ell$. In fact, as shown in figure \ref{dCvdt-fig}, the late-time rate of complexity growth $d\mathcal{C}_\mathcal{V}/d\tau$ is independent of $r_+/\ell$.

\subsubsection*{Late-time complexity growth in $D=5$}
In this case, the expression for $\tilde{r}_{min}$ is considerably more complicated. However, we can use it to expand the two sides of \eqref{dCvdt-infty} as a series in large $r_+$, from which we get
\begin{align}
\left(\Omega_--\Omega_+\right)J&=\frac{\pi}{2G_N\ell^2}r_+^4+\frac{\pi}{2G_N}r_+^2-\frac{\pi r_-^2(r_-^2+\ell^2)}{2G_N\ell^2}+\mathcal{O}\left(\frac{1}{r_+^2}\right)\nonumber,\\
\frac{\Omega_3}{G_N\ell}W(\tilde{r}_{min})&=\frac{\pi^2}{G_N\ell^2}r_+^4+\frac{\pi^2\left[(\sqrt{2}-1)\ell^2-(2-\sqrt{2})r_-^2\right]}{\sqrt{2}G_N\ell^2}r_+^2\nonumber\\
&+\frac{\pi^2\left(\ell^2+2r_-^2\right)\left((4\sqrt{2}-5)\ell^2+2(2\sqrt{2}-1)r_-^2\right)}{16G_N\ell^2}+\mathcal{O}\left(\frac{1}{r_+^2}\right).
\end{align}
Expanding the ratio of these two expressions in the large $r_+$ limit gives
\be
\frac{1}{\left(\Omega_--\Omega_+\right)J}\times\frac{\Omega_3}{G_N\ell}W(\tilde{r}_{min})=2\pi-\left[\pi\sqrt{2}\ell^2+(2-\sqrt{2})r_-^2\right]\frac{1}{r_+^2}+\mathcal{O}\left(\frac{1}{r_+^4}\right)
\ee
which yields \eqref{dCvdt-infty} as $r_+/\ell\rightarrow\infty$. Interestingly, it also shows that the limit is always approached from below, which agrees with the behaviour of $d\mathcal{C}_\mathcal{V}/d\tau$ found for Schwarzschild-AdS black holes \cite{Carmi2017}.

\section{Discussion}\label{sec6}

We have considered several aspects of the CA and CV proposals for holographic complexity in the context of rotating black holes. While the behaviour of these proposals for numerous static and/or spherically symmetric spacetimes has been thoroughly studied, their extension to rotating black holes is a somewhat nontrivial task. In large part, the difficultly arises due to the comparative lack of symmetry in rotating solutions and therefore more complicated causal structure. Here we have partly side-stepped this issue by considering equal-spinning odd-dimensional rotating black holes, which enjoy enough additional symmetry to make the computations tractable, while still revealing a number of non-trivial features. Here our focus has been devoted to understanding the complexity of formation and also the time-dependent growth rate of complexity.

%

First, we introduced the Myers-Perry-AdS spacetimes with equal angular momenta in odd dimensions and discussed the enhancement of symmetry and the associated causal structure and thermodynamic properties. In studying holographic complexity, especially within the action proposal, it is necessary to have a thorough understanding of the causal structure of the spacetime of interest. We have done this here by analysing the structure of light cones in this geometry. The enhanced symmetry of the equal-spinning case allows for us to chose $SU(N+1) \times U(1)$ invariant hypersurfaces, effectively making the causal structure two-dimensional as is the case for static, spherically symmetric black holes. This represents a significant technical simplification over the most general case.\footnote{For the most general rotating black holes the light cones can be defined using PDEs as discussed in \cite{Pretorius1998, AlBalushi2019,Imseis:2020vsw}, though they must be solved numerically.} Despite this simplification, the solutions maintain the classical features associated with  rotating black holes (such as ergoregions, for example), which allow us to rigorously study holographic complexity for rotating black holes for the first time.


Second, we studied the complexity of formation for rotating black holes in both the CA and CV conjectures. As shown in detail in appendix \ref{appStaticLimit}, there is an order of limits problem when taking the static limit of $\Delta \mathcal{C}_\mathcal{A}$. We note that there have been previous investigations where such order of limits problems have been observed for the growth rate in the CA conjecture~\cite{Goto:2018iay, Brown:2018bms, Cai2016, Cano2018, Fan:2019aoj}, however we believe this is the first observation of this for the complexity of formation. This issue can be resolved by an alternative regularization scheme where the future and past tips of the WDW patch are ignored near the singularity and at the static limit (see appendix \ref{appOtherReg}). It would be interesting to explore more deeply the implications of this alternative regularization, in particular the mechanism and/or interpretation of the regulator itself.

Perhaps the most intriguing result of our analysis concerns the scaling of the complexity of formation for large black holes. In both the CV and CA proposals we found that this behaviour is given by
\be 
\Delta \mathcal{C} \sim S \log \frac{\Omega_H}{T} + f \left( \frac{\Omega_H}{T} \right) V^{\frac{D-2}{D-1}}
\ee
where the function $f$ appearing above is dimensionless and independent of the size of the black hole. This result stands in contrast to what was previously understood about complexity of formation for static black holes. Previous work~\cite{Chapman2017Form, Carmi2017} that analysed the complexity of formation for static black holes found that in both the charged and uncharged cases the complexity of formation depends on the black hole size exclusively through entropy. Here, due to the more complicated nature of the metrics involved, we have been able to deduce that there are in fact two scaling regimes. When viewed as a function of temperature and fixed black hole size, there exists a logarithmic singularity in the complexity of formation that is governed by the entropy. This term will dominate at sufficiently low temperatures for a given fixed black hole size. An alternative case is the behaviour of  the complexity of formation at fixed temperature, viewed as a function of the black hole size. In this case, the complexity of formation will be controlled by the thermodynamic volume when the size becomes sufficiently large. In this regime, the above relationship implies that, at fixed temperature, the complexity of formation of sufficiently large black holes is controlled by the thermodynamic volume:
\be \label{DCbehave} 
\Delta \mathcal{C} = {\Sigma}_{\rm g} C_T \left(\frac{V}{V_{\rm AdS}}\right)^{\frac{D-2}{D-1}} +\dots
\ee
where $V_{\rm AdS} = \ell^{D-1}$, $\Sigma_{\rm g}$ is a factor that depends on the specific metric, dimension, etc. (but not on the size of the black hole), and 
$C_T$ is the central charge of the CFT as computed from Newton's constant $G_N$. 
 
The interpretation of thermodynamic volume in the holographic context remains to be  completely understood, but some concrete statements can be made.  From the perspective of the dual theory, variations of $\ell$ corresponds to variations in the central charge $C_T \propto \ell^{D-2}/G_N$ along with variations in the volume of the space where the field theory lives $V_{\rm CFT} \propto \ell^{D-2}$; the thermodynamic volume is the chemical potential associated to variations in these quantities~\cite{Johnson:2014yja, Karch:2015rpa, Sinamuli:2017rhp, Visser:2021eqk, Caceres:2016xjz}. Despite this identification, it is not obvious (at least to us) why such a quantity would naturally be connected to the idea of complexity of formation in the field theory. However, heuristic motivation for this connection is more transparent from the gravitational picture. It should be recalled that the original motivation for holographic complexity was to provide a holographic interpretation for the time-dependent growth of the Einstein-Rosen bridge after thermalization had occurred. In this sense, thermodynamic volume is a contender because, at least in simple scenarios, it can be related to the spacetime volume contained within the black hole horizon~\cite{Kastor:2009wy, Couch:2016exn}.  
 
Another motivation for our proposal is simplicity. Of course, it is possible to use the Smarr relation to replace the thermodynamic volume with a combination of other thermodynamic potentials. However, none of the resulting expressions appear to have a more direct holographic interpretation. In the present case of rotating black holes, use of the Smarr formula would allow the volume to be replaced by the combination
\be 
V = \frac{(D-2)\left[TS + \Omega_H J \right] - (D-3)M}{P} \, .
\ee
While the holographic interpretation of each of the terms in the numerator on the right is clear and well-established for a long time, the factor of $P$ appearing in the denominator, which is required on dimensional grounds, spoils any simpler interpretation that could be obtained. Moreover, the expression in terms of $V$ is far more economical from the gravitational perspective, involving only a single term to capture the correct scaling and dimensionful factors. 

The expression in terms of thermodynamic volume also allows a more direct comparison with what is understood about complexity of formation in the static case, where the result can be written in terms of the entropy. In those cases, our result reduces to the previously known expressions. This is because for static (charged) black holes $V^{(D-2)/(D-1)} \propto S$. The thermodynamic volume has been conjectured~\cite{Cvetic:2010jb} to obey a `reverse' isoperimetric inequality:
\be 
\mathcal{R} \equiv \left(\frac{(D-1) V}{\Omega_{D-2}} \right)^{1/(D-1)} \left(\frac{ \Omega_{D-2}}{ 4 G_N S} \right)^{1/(D-2)} \ge 1 \, .
\ee
The inequality is saturated by (charged) Schwarzschild-AdS spacetimes. Assuming the relationship~\eqref{DCbehave} is general, the reverse isoperimetric inequality becomes the statement
\be 
\Delta \mathcal{C} \ge \beta_D S
\ee
where $\beta_D$ is a positive constant that can be easily worked out from the above. This means that the complexity of formation for large black holes is \textit{bounded from below} by the entropy (equivalently, the number of degrees of freedom).

The above appears to be suggestive of a rather robust connection between complexity of formation and extended thermodynamics. The expression as we have presented it covers static black holes, rotating black holes, as well as gravitational solitons~\cite{Andrews:2019hvq}. While evidence from the field theory side remains lacking, the fact that the behaviour is observed in both CV and CA dualities is nontrivial. It is our view that the relationship \eqref{DCbehave} merits further exploration, both from the field theory perspective and from the gravitational perspective, where it could be further tested through analysis of other black hole geometries that have $S$ and $V$ independent.

Finally, we examined the time-dependent rate of complexity growth using both the CA and CV conjectures. Previous studies have shown that the late time limit of complexity growth in black holes with two horizons is bounded by the difference in internal energy between the outer and inner horizons\footnote{Though see~\cite{Jiang:2020spf} for a recent example where the situation is more subtle.}
\be\label{latetimeC}
\pi\frac{d\mathcal{C}}{d\tau}\bigg|_{\tau\rightarrow\infty}\leq\left(F_++T_+S_+\right)-\left(F_-+T_-S_-\right)=U_+-U_-.
\ee
This surprising result seems to be of near universal scope and it suggests a deep connection between complexity and black hole thermodynamics \cite{Brown:2017jil,Bernamonti:2019zyy,Bernamonti:2020bcf}. In the CV conjecture, we have shown that the complexity is a positive function of time whose late time rate of growth saturates the bound \eqref{latetimeC} in the $r_+/\ell\rightarrow\infty$ limit, up to a constant that depends on the spacetime dimension. We have also explicitly shown that the bound is always approached from below as $r_+/\ell$ is varied. In the CA conjecture, the bound \eqref{latetimeC} is always saturated, and we have shown that it is always approached from above as time is varied. Both of these results agree with the behaviour found for the charged black hole \cite{Carmi2017}.  Furthermore, we found that the arbitrary length scale $\ell_{\text{ct}}$ does not affect the late-time rate of complexity growth but does affect its early behaviour, as shown in figure \ref{Lct-effect}. 

Going forward, there are a number of directions worth exploring. Perhaps the most interesting one concerns the result~\eqref{DCbehave}. While we have not offered a definitive proof of this relationship, it reduces to known results for static black holes, holds also for large gravitational solitons~\cite{Andrews:2019hvq}, and we have provided robust evidence that it is obeyed in general for large rotating black holes. It would be interesting to test the full range of validity of this relationship, which could be done most effectively by studying other black hole solutions for which the entropy and thermodynamic volume are independent and scale differently. Such explorations could provide useful insight from which a general proof of the relationship could be deduced, or a counter-example from which its limitations could be assessed. It would also be interesting to explore this feature in light of the recently proposed first law of complexity~\cite{Bernamonti:2019zyy}.   While the holographic interpretation of thermodynamic volume has been understood for sometime, its utility in this realm has remained comparatively undeveloped (though see~\cite{Johnson:2014yja, Kastor:2014dra, Karch:2015rpa, Caceres:2016xjz, Sinamuli:2017rhp, Johnson:2018amj, Johnson:2019wcq, Rosso:2020zkk} for progress in this direction). Our results provide one concrete setting where thermodynamic volume appears to play a natural role in holography, and it is our view that this result provides further impetus to investigate in greater detail the role of thermodynamic volume in the holographic context and its relation to complexity.



It would be worthwhile to extend the analysis here to the most general class of rotating black holes, though this may be a formidable task. Exploring the implications of the known instabilities (e.g.~superradiance) of rotating black holes for complexity would also be of interest. Although our complexity calculations were done in the case of odd-dimensional equal angular momenta in each independent plane of rotation, we expect that the general family of Myers-Perry-AdS black holes should possess similar qualitative behaviour.


\acknowledgments
This work was supported in part by the Natural Sciences Engineering Research Council of Canada. The work of RAH is supported by the Natural Sciences and Engineering Research Council of Canada through the Banting postdoctoral fellowship program. HK acknowledges the support of  NSERC grant RGPIN-2018-04887. 

\appendix
\section{Fefferman-Graham form of the metric}
\label{fgForm}

In computing the complexity of formation, it is important to justify equating the cutoffs at large  distance $r_{\rm max}$ in both AdS and the black hole spacetimes. To see that this is the case, here we case the metric into the Fefferman-Graham form which will then allow us to directly compare the differences in the fall off of the metric components.

We define a new coordinate $\rho$ according to the relation
\be 
g^2 dr^2 = \frac{\ell^2}{\rho^2} d\rho^2 \, .
\ee
Directly solving this relation to obtain $r$ as a function of $\rho$ yields
\be 
r = \rho - \frac{\ell^2}{4 \rho} + \frac{\ell^2 M \Xi}{(2N+2) \rho^{2N + 1} } + \mathcal{O}\left(\rho^{-(2N+3)} \right) \, .
\ee

In terms of the coordinate $\rho$ the metric now reads
\be 
ds^2 = \frac{\ell^2 d\rho^2}{\rho^2} + \gamma_{\mu\nu} dx^\mu dx^\nu \, ,
\ee
with the metric $\gamma_{\mu\nu}$ approaching the metric on the boundary as $\rho \to \infty$, along with the relevant corrections to this from the bulk. The specific form of this metric can be easily worked out, but its exact form is not necessary here.

With this expansion at hand it is now possible to directly compare the behaviour of $r$ for the global AdS metric with that for the black hole metric. The result is, placing a UV cutoff at $\rho = \ell^2/\delta$, 
\be\label{cutoffFG}
r_{\rm max} - r_{\rm max}^{\rm AdS} = \frac{ M \Xi}{(2N+2) \ell^{4N}} \delta^{2N + 1}.
\ee
Thus, for all positive $N$ the difference in the cutoffs tends to zero in the limit where $\delta \to 0$. This justifies working directly with a cutoff $r_{\rm max}$ in both the AdS and black hole geometries. 

\section{Vanishing contribution of the GHY term}\label{GHYappend}
We will show that the GHY term \eqref{GHYterm} in the action does not contribute to the complexity of formation $\Delta\mathcal{C}_A$ and is canceled by the contribution from vacuum AdS$_D$. First, note that the GHY term for vacuum AdS$_D$ is given by replacing $g(r)^{-2}\rightarrow f_0(r)$ in \eqref{GHYterm}, where $f_0(r)$ is the blackening factor of vaccum AdS$_D$. At $r\rightarrow\infty$, the difference $I_{\text{GHY}}-I_{\text{GHY}}^{\rm AdS}$ depends only on the tortoise coordinates. Using \eqref{cutoffFG}, it is straightforward to show that
\begin{align}
(r^*_\infty - r^*(r_{\rm max}))-({r_0^*}_\infty - r_0^*(r_{\rm max}^{\rm AdS}))&=\int_{r_{\rm max}}^\infty \frac{f(r)}{g(r)}\, dr-\int_{r_{\rm max}^{\rm AdS}}^\infty  \frac{1}{f_0(r)}\, dr \nonumber\\
&=\int_{r_{\rm max}^{\rm AdS}+\mathcal{O}(\delta^{2N + 1})}^\infty \frac{f(r)}{g(r)}\, dr -\int_{r_{\rm max}^{\rm AdS}}^\infty  \frac{1}{f_0(r)}dr\nonumber\\
&=\mathcal{O}(\delta^{2N+3}).
\end{align}
Furthermore, the factor multiplying this term is of order $\mathcal{O}(1/\delta^{2N+2})$. Therefore,
\be
I_{\text{GHY}}-I_{\text{GHY}}^{\rm AdS}=\mathcal{O}(\delta)
\ee
which vanish in the limit $\delta\rightarrow 0$.

\section{Complexity of formation in the static limit}\label{appStaticLimit}

Here we consider, in arbitrary dimensions, the behaviour of the complexity of formation in the limit where $r_-/r_+ \to 0$. We compare the result with the analogous limit for charged black holes, and compare both with the results for the Schwarzschild-AdS black hole.

\subsection{Complexity of formation for Schwarzschild-AdS}

The Schwarzschild-AdS metric in $D$ spacetime dimensions reads
\be 
ds^2 = -f_{\rm Schw}(r) dt^2 + \frac{dr^2}{f_{\rm Schw}(r)} + r^2 d\Omega^2_{D-2}
\ee
where
\be 
f_{\rm Schw}(r) = k + \frac{r^2}{\ell^2} - \frac{2M}{r^{D-3}} \, .
\ee
(In the remainder of this section we will drop the ``Schw'' subscript, but will re-introduce it in later sections when confusion could arise.)
Here we will consider the complexity of formation for this geometry focusing on the $k=0, +1$ cases, essentially reviewing the discussion of~\cite{Chapman2017Form} but with a slightly different emphasis to allow straightforward comparison with our results for the rotating black holes.\footnote{We avoid the case of hyperbolic black holes (i.e. $k=-1$) here as the causal structure in that case is different and does not offer any useful insight for our interests here.} 

The calculation of the action on the WDW patch consists of a bulk term and a GHY term at the past/future singularities. Additional contributions vanish when the result is regularized by subtracting the contribution of two copies of global AdS. The calculation is carried out by focusing on a single quadrant of the WDW patch, then multiplying by a factor of four to obtain the full answer. Let us consider each of these contributions in turn.

Consider first the GHY term on the future singularity. It is straightforward to show that in this case the extrinsic curvature takes the form
\be 
K = - \frac{1}{2 \sqrt{- f_{\rm Schw}(\epsilon)}} \left[f_{\rm Schw}'(\epsilon) + \frac{2(D-2) f_{\rm Schw}(\epsilon)}{r} \right] \, .
\ee
The space-time has a four-fold reflection symmetry along the lines $t=0$, and so the computation can be performed by focussing on one quadrant of the diagram and then multiplying by four. Focusing on the top-right quadrant of the Penrose diagram, the integration for $t$ is carried out between $t = 0$ and $t = r^*_{\text{Schw,}\infty} - r_{\rm Schw}^*(\epsilon)$, where the latter corresponds to the future right boundary of the WDW patch. The idea is to send $\epsilon$ to zero at the end of the computation, yielding for the GHY term
\be\label{SchwGHY} \begin{aligned}
I_{\rm GHY}^{\rm quadrant} &= \frac{ (D-1)M \Omega_{D-2}}{8 \pi G_N} \left[r^*_{\text{Schw,}\infty}  - r_{\rm Schw}^*(0) \right] \\ &= \frac{(D-1) \Omega_{D-2} r_+^{D-3}(k + r_+^2/\ell^2)}{16 \pi G_N} \left[r^*_{\text{Schw,}\infty}  - r_{\rm Schw}^*(0) \right] \, ,
\end{aligned}
\ee
where in the second equality we replaced the mass in terms of $r_+$. This term must be multiplied by a factor of $4$ to account for the GHY contributions in each quadrant. Generally we will set $r^*_{\text{Schw,}\infty}  = 0$ by suitable choice of integration constant.

Next consider the bulk in the upper right quadrant, which takes the form
\be 
I_{\rm Bulk}^{\rm Schw, quadrant} = \frac{ \Omega_{D-2}}{8 \pi G_N} \int_0^{r_{\rm max}}  \left[\frac{d\mathcal{I}_{\rm Schw}}{dr}\left(r^*_{\text{Schw,}\infty}  - r_{\rm Schw}^*(0) \right) \right] \, dr \, .
\ee
where
\be 
\mathcal{I}_{\rm Schw} = -\frac{r^{D-1}}{\ell^2} \, .
\ee
Note that, just as in the main text, we have cut the integration off at $r= r_{\rm max}$ since the integral diverges otherwise. We will send $r_{\rm max} \to \infty$ after subtracting the contributions of the AdS vacuum, which will render the integral convergent. It is generally hard to evaluate the tortoise coordinate, and so a simpler form for the bulk integral is obtained using integration by parts:
\be 
I_{\rm Bulk}^{\rm Schw, quadrant} = \frac{  \Omega_{D-2}}{8 \pi G_N} \int_0^{r_{\rm max}} \frac{\mathcal{I}_{\rm Schw}}{f_{\rm Schw}(r)} dr \, .
\ee
This can be further simplified by isolating and separately dealing with the pole contribution at the black hole horizon. Doing this, writing
\be 
f_{\rm Schw}(r) = F_{\rm Schw}(r)(r^2-r_+^2) \, ,
\ee 
we obtain 
\begin{align} 
I_{\rm Bulk}^{\rm Schw, quadrant} &= \frac{  \Omega_{D-2}}{8  \pi G_N} \left[ \frac{\mathcal{I}_{\rm Schw}(r_+)}{2 r_+ F_{\rm Schw}(r_+)} \log \frac{|r-r_+|}{r+r_+} \bigg|_0^\infty  + \int_0^{r_{\rm max}} \bigg( \frac{\mathcal{I}_{\rm Schw}(r)}{F_{\rm Schw}(r)(r^2-r_+^2)} \right.
\nonumber\\
&\left.- \frac{\mathcal{I}_{\rm Schw}(r_+)}{F_{\rm Schw}(r_+)(r^2-r_+^2)} \bigg) dr \right]
\nonumber\\
&= \frac{ \Omega_{D-2}}{8 \pi G_N} \left[ \int_0^{r_{\rm max}} \left( \frac{\mathcal{I}_{\rm Schw}(r)}{F_{\rm Schw}(r)(r^2-r_+^2)} - \frac{\mathcal{I}_{\rm Schw}(r_+)}{F_{\rm Schw}(r_+)(r^2-r_+^2)} \right) dr \right].
\end{align}
In the first term involving the logarithm, we have extended the integration to infinity since that term is convergent. The remaining integral is completely well-behaved at the horizon and can easily be evaluated numerically. (It can be evaluated analytically in certain dimensions, or in the case of planar $k=0$ black holes~\cite{Chapman2017Form}.)

The complexity of formation is then written as four times the sum of the GHY and bulk terms studied above, along with a subtraction of two copies of global AdS. The final result is
\begin{align}
\pi \Delta \mathcal{C}^{\rm Schw}_{\rm form} =& \, - \frac{(D-1) \Omega_{D-2} r_+^{D-3}(k + r_+^2/\ell^2)}{4 \pi G_N} r_{\rm Schw}^*(0) 
\nonumber\\
&+ \frac{ \Omega_{D-2}}{2 \pi  G_N} \left[ \int_0^{\infty} \left( \frac{\mathcal{I}_{\rm Schw}(r)}{F_{\rm Schw}(r)(r^2-r_+^2)} - \frac{\mathcal{I}_{\rm Schw}(r_+)}{F_{\rm Schw}(r_+)(r^2-r_+^2)}  - \frac{\mathcal{I}_0(r)}{f_0(r)} \right) dr \right] \, .
\end{align}
Here we have explicitly set $r^*_{\text{Schw,}\infty} = 0$, which we will do also throughout the remainder of this appendix.

%
%
%

\subsection{Charged black holes \& the neutral limit}

Let us consider here the complexity of formation for charged black holes, as it will be insightful to compare the results for charged solutions with the results for the rotating solutions studied in this work. The charged solutions are given by the following metrics
\be 
ds^2 = - f_Q(r) dt^2 + \frac{dr^2}{f_Q(r)} + r^2 d \Omega^2_{D-2}
\ee
where
\be 
f_Q(r) = k + \frac{r^2}{\ell^2} - \frac{2M}{r^{D-3}} + \frac{q^2}{r^{2(D-3)}}\, .
\ee
We will be concerned here with the planar and spherical solutions, i.e.  the $k=0, 1$ ones. 

Our objective is to understand how the complexity of formation for these solutions behaves in the limit $q \to 0$. The causal structure of the charged black holes is qualitatively identical 
to the equal-spinning rotating holes considered in this work --- see~\cite{Carmi2017} for a full discussion. Since here we are only interested in the neutral limit, we will not consider the counterterm for null boundaries, as its contribution is subleading and vanishing in that limit. Moreover, just as for the rotating solutions, a GHY term at large distances is unimportant as it cancels when the subtraction relative to global AdS is performed. Therefore the complexity of formation consists of two ingredients: the bulk action and two corner terms where the past/future sheets of the WDW meet. 

Let us consider first the corner terms. The analysis is qualitatively similar to that performed already in the rotating case (and we refer the reader to~\cite{Carmi2017} for a full discussion of these terms in the charged case), leading to the final result:
\be 
I^Q_{\rm jnt} = -\frac{\Omega_{D-2}}{8\pi G_N} r_{m_0}^{D-2} \log \frac{|f_Q(r_{m_0})|}{\alpha^2} \, ,
\ee
where we have included a constant $\alpha$ that keeps track of the parameterization of the null geodesics normal to the sheets of the WDW patch. This accounts for the contribution of the future joint, the joint term at the past meeting point is identical and so the above should be multiplied by two when including it in the complexity of formation.

The parameter $r_{m_0}$ appearing in the above is the value of the radial coordinate where the sheets of the WDW patch meet. It is obtained by solving the condition 
\be 
r^*_\infty - r^*(r_{m_0}) = 0
\ee
where $r^*$ is the tortoise coordinate for the charged black hole. Here, introducing 
\be 
f_Q(r) = F_Q(r)(r^2-r_+^2)(r^2-r_-^2),
\ee
to allow the problematic pieces at the horizons to be isolated and treated separately, we find it has the form
\begin{align} 
r^*(r) =& \, \frac{1}{2 r_+ F_Q(r_+) (r_+^2-r_-^2)} \log \frac{|r-r_+|}{r+r_+} - \frac{1}{2 r_- F_Q(r_-) (r_+^2-r_-^2)} \log \frac{|r-r_-|}{r+r_-}  + \mathcal{R}_Q(r)
\end{align}
where in the above we have chosen an integration constant such that $r^*_\infty = 0$ and have introduced
\be 
\mathcal{R}_Q(r) =  \int_\infty^r \left[\frac{1}{F_Q(r')(r'^2-r_+^2)(r'^2-r_-^2)} - \frac{1}{F_Q(r_+)(r'^2-r_+^2)(r_+^2-r_-^2)} + \frac{1}{F_Q(r_-)(r'^2-r_-^2)(r_+^2-r_-^2)} \right] dr' \, .
\ee

Consider next the bulk contribution. After some massaging, the bulk action for charged black holes can be written in the form\footnote{Unlike the other solutions in this manuscript, the charged solutions are, of course, not vacuum. We follow here exactly the conventions of~\cite{Carmi2017} for the electromagnetic terms in the action.}
\be 
\Delta I_{\rm Bulk}^Q = \frac{\Omega}{2 \pi G_N} \int_{r_{m_0}}^\infty \left[ \frac{\mathcal{I}_Q(r)}{f_Q(r)} - \frac{\mathcal{I}_0(r)}{f_0(r)} \right] dr - \frac{\Omega}{2 \pi G_N} \int_0^{r_{m_0}} \frac{\mathcal{I}_0(r)}{f_0(r)} dr
\ee
where
\be 
\mathcal{I}_Q(r) = - \frac{r^{D-1}}{\ell^2} - \frac{q^2}{r^{D-3}}
\ee
and the subscript ``$0$'' denotes this quantity and the metric for the AdS vacuum.  Note that since the AdS contribution has been subtracted here, making the integral convergent, we have taken the limit of integration to infinity. The bulk term can be massaged in a manner similar to the tortoise coordinate we considered early in the manuscript. We first write the metric function as
\be 
f_Q(r) = F_Q(r)(r^2-r_+^2)(r^2-r_-^2) \, ,
\ee
as before. Then, the integrand of the bulk term can be split up according to
\begin{align} 
\frac{\mathcal{I}_Q(r)}{f(r)} &= \left[\frac{\mathcal{I}_Q(r)}{F_Q(r)(r^2-r_+^2)(r^2-r_-^2)} - \frac{\mathcal{I}_Q(r_+)}{F_Q(r_+)(r^2-r_+^2)(r_+^2-r_-^2)} + \frac{\mathcal{I}_Q(r_-)}{F_Q(r_-)(r^2-r_-^2)(r_+^2-r_-^2)}  \right] 
\nonumber\\
&+   \frac{\mathcal{I}_Q(r_+)}{F_Q(r_+)(r^2-r_+^2)(r_+^2-r_-^2)} - \frac{\mathcal{I}_Q(r_-)}{F_Q(r_-)(r^2-r_-^2)(r_+^2-r_-^2)}  \, .
\end{align}
This decomposition of the integral allows us to isolate the contributions at the horizons which require special care. We can integrate these terms explicitly, and then arrive at the following expression for the bulk:
\begin{align}
\Delta I_{\rm Bulk}^Q &=  \frac{\Omega}{2 \pi G_N} \left[-\frac{\mathcal{I}_Q(r_+)}{2 r_+ F_Q(r_+)(r_+^2-r_-^2)} \log \frac{|r_{m_0} - r_+|}{r_{m_0} + r_+} \right.
\nonumber\\
&\left. + \frac{\mathcal{I}_Q(r_-)}{2 r_- F_Q(r_-)(r_+^2-r_-^2)}  \log \frac{|r_{m_0} - r_-|}{r_{m_0} + r_-} \right] + \mathfrak{I}_Q(r_{m_0})
\end{align}
where we have defined
\begin{align}\label{chargedOtherCont} 
\mathfrak{I}_Q(r_{m_0}) &= \frac{\Omega}{2 \pi G_N} \int_{r_{m_0}}^\infty \bigg[\frac{\mathcal{I}_Q(r)}{F_Q(r)(r^2-r_+^2)(r^2-r_-^2)} - \frac{\mathcal{I}_Q(r_+)}{F_Q(r_+)(r^2-r_+^2)(r_+^2-r_-^2)} 
\nonumber\\
&+ \frac{\mathcal{I}_Q(r_-)}{F_Q(r_-)(r^2-r_-^2)(r_+^2-r_-^2)}  - \frac{\mathcal{I}_0}{f_0(r)} \bigg] dr - \frac{\Omega}{2 \pi G_N} \int_0^{r_{m_0}} \frac{\mathcal{I}_0(r)}{f_0(r)} dr . 
\end{align}
This term is convergent and completely regular, requiring no special treatment at the horizons. It can be straightforwardly integrated numerically (or analytically in certain special cases).

The complexity of formation then takes the final form
\be 
\pi \Delta \mathcal{C}^Q_{\rm form} = \Delta I^Q_{\rm Bulk} + 2 I^Q_{\rm jnt}  \, .
\ee
We want to understand how this quantity behaves in the limit $r_-/r_+ \to 0$. For this we must first understand the asymptotic behaviour of $r_{m_0}$ in this limit. In general dimensions, writing $r_{m_0} = y r_+(1+\epsilon)$ we find that
\be 
r^*(r) \sim - \frac{y^{D-2} r_+ \ell^2 }{(D-3) (k\ell^2+r_+^2)} \log \frac{\epsilon}{2} + r^*_{\rm Schw}(0)
\ee
where $y = r_-/r_+$ and $r^*_{\rm Schw}(0)$ is the value of the tortoise coordinate for the static solution at the origin (recall that we have set the integration constant so that $r^*_\infty = 0$). Explicitly, this term takes the form
\be\label{schwTort0} 
r^*_{\rm Schw}(0) = \int_0^\infty \frac{f_{\rm Schw}(r) - F_{\rm Schw}(r_+)(r^2 - r_+^2)}{f_{\rm Schw}(r) F_{\rm Schw}(r_+)(r^2-r_+^2)} \,  dr \, , \quad F_{\rm Schw}(r) \equiv \frac{f_{\rm Schw}(r)}{r^2-r_+^2} \, .
\ee
We then deduce the asymptotic form of the meeting location\footnote{The factor of $2$ in front of the exponential differs from~\cite{Carmi2017}, where this factor is unity. The difference comes from the fact that we defined $f(r) = F(r)(r^2 -r_+^2)(r^2-r_-^2)$ whereas those authors defined $f(r) = F(r)(r-r_+)(r - r_-)$. The prefactor of the exponential is completely unimportant for the $y \to 0$ limit, and the same results are obtained for $r_{m_0} = y r_+ (1 + A \epsilon)$ for any choice of parameter $A$. It is the argument of the exponential that is important.} 
\be 
r_{m_0} = y r_+ \left[1 + 2 \exp \left( \frac{(D-3)(k \ell^2 + r_+^2) r^*_{\rm Schw}(0)}{r_+ \ell^2 y^{D-2} } \right) \right] \, .
\ee
Using this asymptotic result along with the fact that near the inner horizon we have $|f_Q(r)| \approx |f_Q'(r_-)|(r-r_-)$ it is rather straightforward to show that 
\be 
\lim_{y \to 0} 2 I^Q_{\rm jnt} = - \frac{(D-3) \Omega_{D-2}}{4 \pi \ell^2 G_N} r_+^{D-3} (k \ell^2 + r_+^2) r^*_{\rm Schw}(0) \, .
\ee
Comparing with the results for the neutral~\eqref{SchwGHY} case we see that
\be 
\frac{\lim_{y \to 0} 2 I^Q_{\rm jnt}}{I_{\rm GHY}^{\rm Schw}} = \frac{D-3}{D-1} \, .
\ee
Note that this limit is independent of the parametrization of the null normals to the WDW patch, as indicated by the absence of $\alpha$ in the final expression.\footnote{As we mentioned earlier, inclusion of the counterterm for null boundaries changes the structure of the joint term, but this addition has no effect on the $y \to 0$ limit. For this reason, to keep the complexity of the expressions at a minimum, we did not include that term in the analysis presented here.}

The limit of the bulk term is more difficult. It is easy to deal with the logarithm terms in this limit --- one of them simply vanishes, while the other yields a finite result. We have:
\begin{align}
\lim_{y \to 0} \Delta I^Q_{\rm Bulk} = - \frac{ \Omega_{D-2} r_+^{D-3}(r_+^2 + k \ell^2) r^*_{\rm Schw}(0)}{2 \pi \ell^2G_N} + \mathfrak{I}_Q(0) \, .
\end{align}
Determining the value of $\mathfrak{I}_Q(0)$ is the tricky part. However, after careful examination of~\eqref{chargedOtherCont} it can be shown that this term can be expressed as
\be 
\mathfrak{I}_Q(0) = \frac{ \Omega_{D-2}}{2 \pi  G_N} \left[ \int_0^{\infty} \left( \frac{\mathcal{I}_{\rm Schw}(r)}{F_{\rm Schw}(r)(r^2-r_+^2)} - \frac{\mathcal{I}_{\rm Schw}(r_+)}{F_{\rm Schw}(r_+)(r^2-r_+^2)}  - \frac{\mathcal{I}_0(r)}{f_0(r)} \right) dr \right] = \Delta I_{\rm Bulk}^{\rm Schw} \, .
\ee
Thus, we conclude that the limit of the bulk action is
\be 
\lim_{y\to 0} \Delta I^Q_{\rm Bulk} = \frac{2 I_{\rm GHY}^{\rm Schw}}{D-1} + \Delta I_{\rm Bulk}^{\rm Schw} \, .
\ee
It can be further shown that
\be 
\Delta I_{\rm Bulk}^{\rm Schw}  = \frac{\Omega_{D-2} r_+^{D-1}r^*_{\rm Schw}(0)}{2 \pi \ell^2G_N} = - \frac{2I_{\rm GHY}^{\rm Schw}}{D-1} \quad \text{when} \quad k = 0 \, .
\ee
The conclusion is that, when $k=0$, the limit of the bulk part of the action $\Delta I^Q_{\rm Bulk}$ vanishes in all dimensions. This is consistent with the analysis of~\cite{Carmi2017} where the $D = 5$ case was studied. However, the bulk term $\Delta I^Q_{\rm Bulk}$ \textit{does not} vanish when $k = 1$, as the equation just above does not hold in that case. However, the way in which the particular terms combine yields in general
\be 
\lim_{y\to 0} \pi \Delta \mathcal{C}_{\rm form}^Q =  \frac{D-3}{D-1} I_{\rm GHY}^{\rm Schw} + \frac{2}{D-1} I_{\rm GHY}^{\rm Schw}+ \Delta I_{\rm Bulk}^{\rm Schw} = I_{\rm GHY}^{\rm Schw} + \Delta I_{\rm Bulk}^{\rm Schw} = \pi  \Delta C_{\rm form}^{\rm Schw}\, .
\ee
Thus, in the charged case the $y \to 0$ limit of the complexity of formation matches the complexity of formation for the Schwarzschild AdS solution, irrespective of the horizon topology. However, note the non-trivial way in which this limit is achieved, with the corner term producing one fraction of the GHY term and the bulk action for the charged solution producing the other fraction of the GHY term while at the same time giving the full Schwarzschild-AdS bulk contribution.


\subsection{Rotating black holes \& the static limit}

Let us finally consider in detail the static limit of the rotating black holes that have been our focus here in this work. We are interested once again in determining the limit of the bulk and joint terms in the action in the limit $y \equiv r_-/r_+ \to 0$. We work in general (odd) dimensions.

Consider first the joint term. The relevant part of this term is
\be \label{funnyCorner}
I_{\rm jnt} = -\frac{\Omega_{2N + 1}}{4 \pi G_N} (r_{m_0})^{2N} h(r_{m_0}) \log |f(r_{m_0})^2|  \, .
\ee
Here we have neglected the term $\ell_{\rm ct}^2 \Theta^2/\alpha^2$ inside the logarithm for simplicity of presentation as it will have no effect on our discussion as it is subleading. Note also that here we have included the overall factor of $2$ to account for both the past and future joints. Our objective is to understand the behaviour of this term as $r_-/r_+ \to 0$. 

In order to understand the behaviour of this corner term as $y \to 0$ we need to understand the behaviour of $r_{m_0}$. Working in the limit of small $y$, and writing $r = y r_+ (1 + \epsilon)$, it is easy to show that the tortoise coordinate~\eqref{goodTort} behaves as
\be 
r^* = -\frac{r_+ \ell y^{N+1}}{2 \sqrt{\ell^2+ r_+^2}} \log \frac{ \epsilon}{2} + r^*_{\rm Schw}(0) \, .
\ee
where $r^*_{\rm Schw}(0)$ is the value of the Schwarzschild-AdS tortoise coordinate at the origin --- see eq.~\eqref{schwTort0}. In deriving this expression it is useful to note that
\be\label{limGH} 
G(r_-) \sim \frac{\ell^2 r_+^4 y^{2N+2}}{\ell^2 + r_+^2} \quad \text{and} \quad h(r_-) \sim \frac{r_+ \sqrt{r_+^2 + \ell^2}}{\ell y^{N-1}} \, ,
\ee
as $y \to 0$. We can then deduce that the meeting point behaves as
\be\label{rMeetAsympRot} 
r_{m_0} = r_+ y \left[1 + 2 \exp \left( \frac{2 r_{\rm Schw}^*(0) \sqrt{\ell^2 + r_+^2} }{r_+ \ell y^{N+1}} \right) \right] \, 
\ee
in the limit $y \to 0$.

Near the inner horizon we can expand 
\be 
f^2(r_{m_0}) \approx (f^2)'(r_-)(r_{m_0} - r_-) \, . 
\ee
Subsituting this into~\eqref{funnyCorner} and taking the limit $y \to 0$, we obtain the following result:
\be 
\lim_{y \to 0} I_{\rm jnt} = - \frac{\Omega_{2N+1}}{2 \pi \ell^2 G_N} r_+^{2N} (\ell^2 + r_+^2) r^*_{\rm Schw}(0) \, .
\ee
Noting that $D = 2N + 3$ we see that
\be 
\lim_{y \to 0} I_{\rm jnt} = \frac{2 I_{\rm GHY}^{\rm Schw}}{D-1} = \frac{I_{\rm GHY}^{\rm Schw}}{N+1}  \, .
\ee
This limit is different in structure than the limit in the charged case.\footnote{Though note that for the special case of $D = 5$ (i.e. $N  =1$) the limit of the joint term matches in the two cases.} The reason partly has to do with the behaviour of $h(r_{m_0})$ in the limit $y \to 0$ which approaches a constant --- or blows up --- rather than behaving $\sim y$ in this limit (as it would for the charged solution).

Next let us consider the behaviour of the bulk. Again, it is useful to split the bulk into pieces, isolating the parts that are divergent at the horizon. Doing this we can write the bulk term as
\begin{align}
\Delta I_{\rm Bulk} =&\, \frac{\Lambda \Omega_{2N+1}}{2(N+1)(2N+1) \pi G_N} \bigg\{- \frac{r_+^{2N}G(r_+) h(r_+)}{2 (r_+^2 - r_-^2)} \log \frac{|r_{m_0} - r_+|}{r_{m_0} + r_+} + \frac{r_-^{2N}G(r_-) h(r_-)}{2 (r_+^2 - r_-^2)} \log \frac{|r_{m_0} - r_-|}{r_{m_0} + r_-}  
\nonumber\\
&+ \int_{r_{m_0}}^\infty \left[\frac{r^{2N+1} G(r) h(r)}{(r^2-r_+^2)(r^2-r_-^2)} -  \frac{r_+^{2N+1} G(r_+) h(r_+)}{(r^2-r_+^2)(r_+^2-r_-^2)} + \frac{r_-^{2N+1} G(r_-) h(r_-)}{(r^2-r_-^2)(r_+^2-r_-^2)} - \frac{r^{2N+2}}{r^2+\ell^2} \right] dr 
\nonumber\\
&- \int_0^{r_{m_0}} \frac{r^{2N+2}}{r^2+\ell^2} dr 
\bigg\} .
\end{align}
Again, the last integral is convergent and its argument completely regular. As in the charged case, we can now easily study the limit of the logarithmic terms and then carefully consider the remaining integral. As before, the logarithmic term involving $r_+$ vanishes in this limit, and we must only consider the contribution from the logarithmic term involving $r_-$. However, here a crucial difference from the charged case arises. In the rotating case, we have
\be 
\frac{r_-^{2N}G(r_-) h(r_-)}{2 (r_+^2 - r_-^2)} \sim \mathcal{O}(y^{3N+3})
\ee
from the limiting behaviour of $G(r_-)$ and $h(r_-)$ presented in eq.~\eqref{limGH} above. Meanwhile, the logarithm goes like
\be 
\log (r_{m_0} - r_-) \sim \mathcal{O}(y^{-(N+1)}) \, ,
\ee
based on the behaviour of $r_{m_0}$ presented in eq.~\ref{rMeetAsympRot}. We therefore see that the logarithmic contributions to the bulk \textit{vanishes} in the limit $y \to 0$! We then must only consider the remaining integral in the bulk. However, this term behaves just as it did in the charged case, producing the following final limit for the bulk term:
\be 
\Delta I_{\rm Bulk} = \frac{ \Omega_{2N+1}}{2 \pi  G_N} \left[ \int_0^{\infty} \left( \frac{\mathcal{I}_{\rm Schw}(r)}{F_{\rm Schw}(r)(r^2-r_+^2)} - \frac{\mathcal{I}_{\rm Schw}(r_+)}{F_{\rm Schw}(r_+)(r^2-r_+^2)}  - \frac{\mathcal{I}_0(r)}{f_0(r)} \right) dr \right] = \Delta I_{\rm Bulk}^{\rm Schw} \, .
\ee
The combined joint and bulk terms give
\be 
\lim_{y\to 0} \pi \Delta \mathcal{C}_{\rm form} = \frac{I_{\rm GHY}^{\rm Schw}}{N+1} + \Delta I_{\rm Bulk}^{\rm Schw} \neq \pi  \Delta \mathcal{C}_{\rm form}^{\rm Schw}  
\ee
which is the order of limits problem in the rotating case.

\section{Alternate regularization of the WDW patch}\label{appOtherReg}

\begin{figure}[t]
\centering
\includegraphics[width=0.45\textwidth]{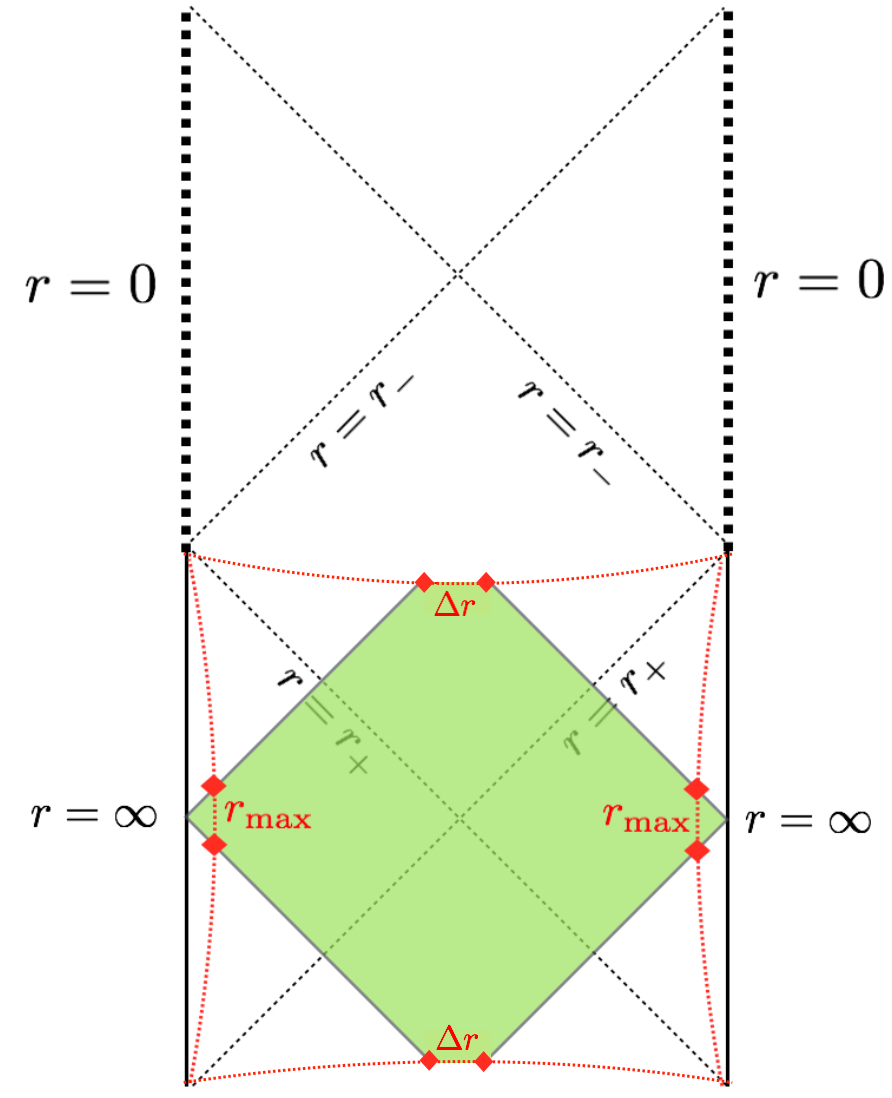}
\caption{The WDW patch with the alternative regularization scheme, where the future and past tips are cut off and replaced with a regularization surface at $r=\Delta r$. This introduces four new joints that are shown in the figure as red diamonds.}
\label{otherRegPlot}
\end{figure}

Here we consider an alternate regularization of the WDW patch to examine the limiting behaviour of the complexity of formation as $y = r_-/r_+ \to 0$. We do so by cutting off the future and past tips of the WDW patch at $r = r_{m_0} + \rdiff$ and introducing the appropriate GHY and joint terms to accommodate this (see figure \ref{otherRegPlot}). This amounts to introducing two corner terms and one GHY term at the future tip of the WDW patch, and likewise at the past tip. 

Consider first the GHY term on the right side of the future cutoff surface. This can be worked out to be
\be 
I_{\rm GHY}^{F, R} = \frac{\Omega_{2N+1}}{16 \pi G_N} \rD^{2N+1} \left[ (g^{-2})'(\rD) + \frac{2(2N+1) }{\rD g^{2}(\rD)} \right] r^*(\rD) \, .
\ee
where we have denoted $\rD = r_{m_0} + \Delta r$. There are four contributions, all identical to this one, and so the final result for the GHY contribution is
\be 
I_{\rm GHY} = \frac{\Omega_{2N+1}}{4 \pi G_N} \rD^{2N+1} \left[ (g^{-2})'(\rD) + \frac{2(2N+1) }{\rD g^{2}(\rD)}  \right] r^*(\rD) \, .
\ee

Consider next the corner terms that occur where the boundaries of the WDW patch intersect the cutoff surface at $\rD$. Focussing on the contribution on the right side of the future boundary of the WDW patch, the relevant null normal is
\be 
k_F = \alpha (dt + dr^*) \, .
\ee
To determine the relevant dot products appearing in the joint term we need the form of the auxillary future/outward pointing unit vector $\hat{s}$. In the present case $\hat{s} = \sqrt{|f^2(\rD)|} dt$ is the appropriate choice. We can then work out the sign $\varepsilon$ appearing in the definition of the joint term --- see eq.~\eqref{ajntterm2}. We find here that $\varepsilon = +1$. We then find the following result for the joint term
\be 
I_{\rm jnt}^{F, R} = - \frac{\Omega_{2N+1}}{16 \pi G_N} \rD^{2N} h(\rD) \log  \frac{|f^2(\rD)|}{\alpha^2} \, .
\ee
There are four joints of this kind, giving the total
\be 
I_{\rm jnt} = - \frac{\Omega_{2N+1}}{4 \pi G_N} \rD^{2N} h(\rD) \log  \frac{|f^2(\rD)|}{\alpha^2} \, .
\ee
The idea, then, is to replace the corner term appearing in section~\ref{sec4} with the combination of joint and GHY terms shown above. Note for our purposes here we will not consider the contribution of the null boundary counterterm. This is because we are interested in the limit $y = r_-/r_+ \to 0$ and the null boundary counterterm vanishes in this limit. We now examine this limit keeping $\rD$ small but finite until after the limit $y \to 0$ has been performed. 

The GHY term limits to precisely the GHY term in the static case,
\be 
\lim_{\Delta r \to 0} \lim_{y \to 0} I_{\rm GHY} = I_{\rm GHY}^{\rm Schw} \, ,
\ee
while the joint term vanishes in the limit
\be
\lim_{\Delta r \to 0} \lim_{y \to 0} I_{\rm jnt} = 0 \, .
\ee
It must be emphasized that the order of limits here is important. The $y \to 0$ limit must be taken prior to taking the $\Delta r \to 0$ limit. The entire issue associated with the order of limits problem is that this limit does not commute. Said another way, effectively what this conclusion means is that the future and past `tips' of the WDW patch contain the following amount of action:
\[
I_{\rm tip} = -\frac{N}{N+1}I_{\rm GHY}^{\rm Schw} \, , 
\]
in a vanishing amount of volume. Interestingly, this is exactly the limit of the corner term in the charged case. Thus, in this alternate regularization of the WDW patch the limit agrees with the Schwarzschild-AdS result. Note that for any finite $y$ the two approaches will agree, as in that case the limits considered above will commute. 

\section{Behaviour of complexity of formation for large black holes}\label{appExtremal}

Here we present additional details for the behaviour of the complexity of formation in the limit of large black holes. For the cases of charged black holes and also the rotating black holes considered here there are two independent limits that are of interest. The first involves holding fixed the size of the black hole, $r_+/\ell$, while exploring the extremal limit $r_-/r_+ \to 1$. The second is to hold fixed $r_-/r_+$ while examining the behaviour of the complexity of formation for $r_+/\ell \to \infty$. 

In previous work that focused on five-dimensional charged black holes~\cite{Carmi2017}, it was demonstrated that the entropy controls the behaviour of the complexity of formation in either limit when the black holes are large enough. In particular, those authors found that the complexity of formation diverges logarithmically as extremality is approached with a prefactor proportional to the entropy when the black holes are large. Moreover, the subleading terms in a near extremal expansion were also found to be related to the entropy. Here we wish to examine those conclusions in more detail and extend them to higher dimensions. We will then contrast them with the rotating case where it is found that different thermodynamic potentials control the different limits.

\subsection{Charged black holes: complexity equals volume}

To understand our results in the rotating case, it will be important to have an understanding of how the relevant computations play out for charged black holes. In this case, the complexity of formation is given by the following integral:
\be 
\Delta \mathcal{C}_\mathcal{V} = \frac{\Omega_{D-2}}{2 G_N R} \lim_{r_{\rm max} \to \infty} \left[\int_{r_+}^{r_{\rm max}} \frac{r^{D-2} }{\sqrt{f_Q(r)}}dr - \int_{0}^{r_{\rm max}} \frac{r^{D-2}}{\sqrt{f_{0}(r)}}dr \right] \, .
\ee
To illustrate a particular example, we consider the five dimensional case. In five dimensions, the above integrals can be worked out to be
\begin{align}\label{CformCharged}
\frac{2 G_N R}{\Omega_{D-2}} \Delta \mathcal{C}_\mathcal{V} = \ell^4 \alpha^4 \int_{1}^\infty &x^3 \bigg[\frac{x^2}{\sqrt{(x^2-1)(1+x-\epsilon)(-1+x + \epsilon)} \sqrt{k + \alpha^2 (2 + x^2 + \epsilon(\epsilon-2))}} 
\nonumber\\
&- \frac{1}{\sqrt{k+\alpha^2 x^2}} \bigg] dx - \frac{\ell^4}{3} \left[2 k^{3/2}  + (\alpha^2-2k) \sqrt{k + \alpha^2} \right] 
\end{align}
where we have defined
\be 
x \equiv \frac{r}{r_+} \,, \quad \alpha = \frac{r_+}{\ell} \,, \quad \epsilon \equiv 1- \frac{r_-}{r_+} \, .
\ee
Our main objective here will be to try to understand how the resulting integral scales with $\alpha$. While this is not so hard for these charged black holes, it will be considerably more involved for the rotating ones. So we will use the simpler setting of charged black holes to illustrate our ideas.

Although it is not our main focus, let us mention here the case of planar charged black holes. For these solutions, the dependence of complexity of formation on the quantity $\alpha = r_+/\ell$ completely factors out of the integral, leaving a result dependent only on $\epsilon =  1- r_-/r_+$. In five dimensions the remaining integral can be evaluated explicitly, giving the final result:
\be 
\Delta \mathcal{C}_\mathcal{V}^{k=0, D=5} = \frac{S \ell}{R} \frac{(1-\epsilon + \epsilon^2)(3 - 3 \epsilon + \epsilon^2)}{6 \sqrt{\epsilon(2-\epsilon)}} E \left[\frac{3 - 4 \epsilon + 2 \epsilon^2}{\epsilon(\epsilon-2)} \right] \, .
\ee
Here $S$ is the black hole entropy, while $E(X)$ refers to the elliptic integral of the first kind. We see clearly here that, for planar black holes, the only dependence on the black hole size is through the entropy. This property extends directly to all higher dimensions, though the resulting integrals no longer yield such a simple final result.

\begin{figure}
\centering
\includegraphics[width=0.6\textwidth]{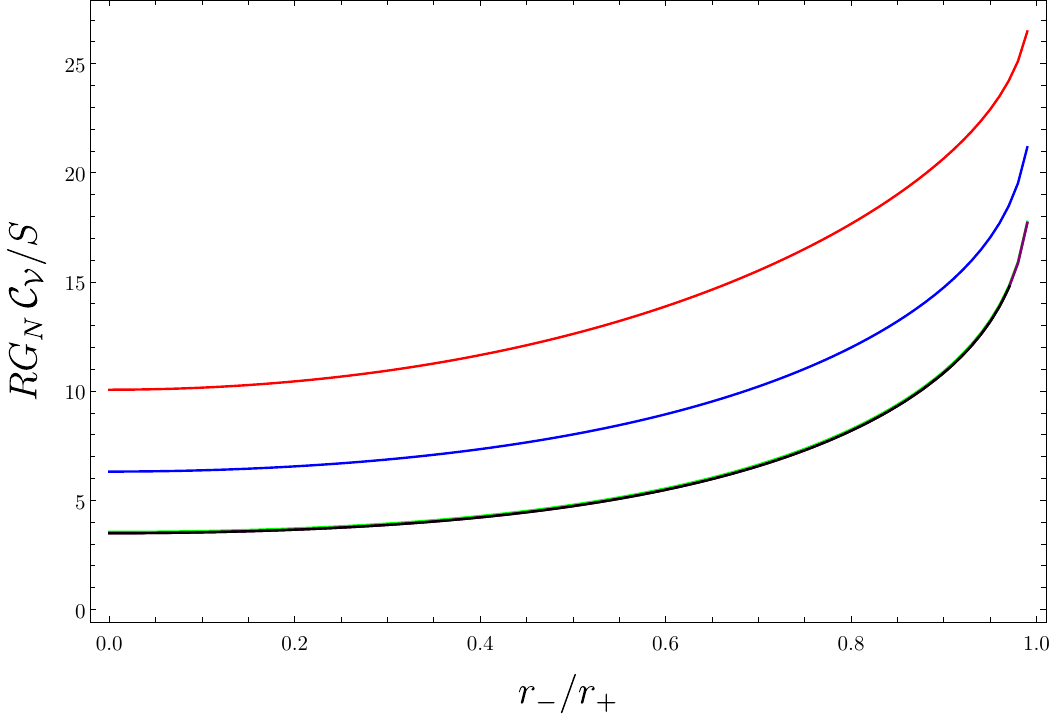}
\caption{A plot of the complexity of formation within the CV conjecture for five dimensional, spherical $(k=+1)$ black holes. We have normalized the complexity of formation by the entropy and the curves shown correspond to $r_+/\ell = 1/2, 1, 10, 50, 100$ in order from top to bottom. The last three curves are visually indistinguishable. Imposed on the plot in a black curve is the complexity of formation for the planar $k=0$ charged black hole. This curve coincides with the last three plots for the spherical black holes.}
\label{CVform-charged}
\end{figure}

From a heuristic examination of the integrals above, it is not too hard to become convinced that as $\alpha \to \infty$ the behaviour of the spherical $(k=+1)$ black holes will match that of the planar black holes. We illustrate this with a numerical evaluation of the complexity of formation in figure~\ref{CVform-charged}. In this figure we have normalized the complexity of formation by dividing by the entropy
\be 
S = \frac{\Omega_{3} r_+^3}{4G_N}
\ee
and have shown the result as a function of $r_-/r_+$ for several values of $r_+/\ell$. The plot illustrates that when $r_+/\ell$ is small the curves can be distinguished. However, as $r_+/\ell$ becomes large the results all converge to the planar case (shown here as the black curve). This illustrates that, for large black holes at fixed $\epsilon = r_-/r_+ - 1$, the entropy completely controls the complexity of formation. 

For charged black holes it is also not too difficult to confirm this conclusion analytically. Expanding \eqref{CformCharged} in the large $\alpha$ limit for five-dimensional spherical ($k=+1$) black holes gives 
\be
\Delta \mathcal{C}_\mathcal{V}= \Delta \mathcal{C}_\mathcal{V}^{k=0, D = 5}  + \mathcal{O}(S^{2/3}) \, .
\ee
While an analytic study is possible in the charged case, it will turn out to be much more difficult in the rotating case. For this reason we will discuss a numerical approach to determine the dependence of the complexity of formation on the horizon radius for large black holes. Suppose that 
\be 
\Delta \mathcal{C}_\mathcal{V} \sim (r_+/\ell)^\gamma
\ee
for some power $\gamma$. A convenient way to determine the value of $\gamma$ is the following. We consider the ratio
\be 
R(\beta) = \frac{ R G_N \Delta \mathcal{C}_\mathcal{V}}{(r_+/\ell)^\beta} \sim (r_+/\ell)^{\gamma - \beta} \, .
\ee
We then take the logarithm of this ratio treated as a function of both $(r_+/\ell)$ and $\beta$. For each value of $\beta$, we compute $R(\beta)$ for several (large) values of $r_+/\ell$ and fit the resulting data to a linear model, and extract the slope of the numerical model. We explore the $\beta$ parameter range until the slope determined in this way is zero. The value of $\beta$ for which the slope vanishes corresponds to the case $\beta = \gamma$, allowing us to extract how the complexity of formation depends on the size of the black holes. 

\begin{figure}
\centering
\includegraphics[width=0.6\textwidth]{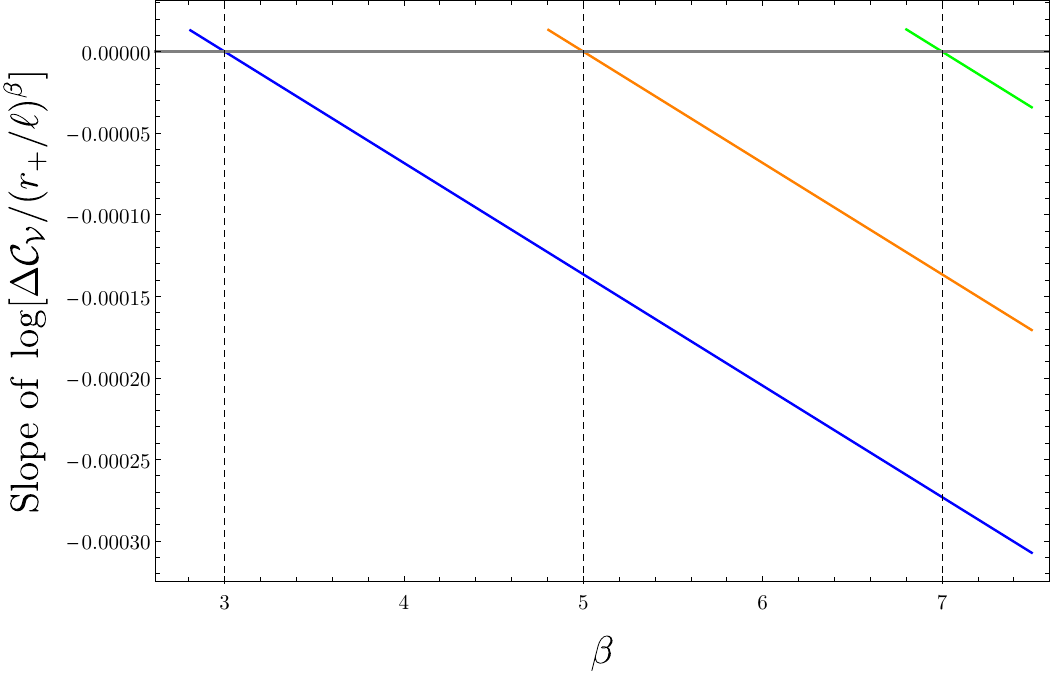}
\caption{The slope of the logarithm of the ratio $R(\beta)$ for several dimensions. The curves correspond to 5 dimensions (blue, left), 7 dimensions (orange, center), and 9 dimensions (green, right). For each value of $\beta$ the integrals have been evaluated for 500 points laying between $r_+/\ell = 10,000$ and $r_+/\ell = 20,000$. The slope is extracted by performing a linear fit to this data.}
\label{chargedSlope}
\end{figure}

This scheme is illustrated in figure~\ref{chargedSlope} for five, seven, and nine dimensions. In each case it is clear from the plot that the slopes vanish for $\beta = 3, 5, 7$, respectively (but this can be confirmed to much higher precision numerically). This numerical finding is consistent with the results discussed above: In general dimensions, the complexity of formation for large charged black holes is controlled by the entropy and nothing more.

\subsection{Rotating black holes: complexity equals volume}

Let us now consider the rotating black holes, which are the main topic of our interest here. Once again for ease of presentation we will present detailed equations only in the five dimensional case and will comment how the situation plays out in higher (odd) dimensions.

The complexity of formation for rotating black holes according to the CV conjecture is
\begin{align}
\Delta\mathcal{C}_\mathcal{V}= \lim_{r_{\rm max} \to \infty}\frac{2\Omega_{D-2}}{G_NR}\left[\int_{r_{+}}^{r_{\rm max}}h(r)r^{D-3}g(r)\, dr-\int_{0}^{r_{\rm max}^{\rm AdS}}\frac{r^{D-2}}{\sqrt{f_0(r)}} \, dr\right] \, .
\end{align}
As in the charged case, there are two limits that are interesting to consider here. We can consider holding the size of the black hole $r_+/\ell$ fixed and examine the extremal limit $\epsilon = (1 - r_-/r_+) \to 0$ or vice versa. Let us first consider the former. 

To understand the leading behaviour in the extremal limit we split the integrand for the black hole into two parts:
\begin{align} 
\int_{r_{+}}^{r_{\rm max}} h(r)r^{D-3}g(r)\, dr =& \int_{r_{+}}^{r_{\rm max}}\frac{h(r_+)r_+^{D-3}\sqrt{G(r_+)}}{\sqrt{(r^2-r_+^2)(r^2-r_-^2)}} \, dr
\nonumber\\
 &+ \int_{r_{+}}^{r_{\rm max}} \frac{\left[h(r)r^{D-3}\sqrt{G(r)} - h(r_+) r_+^{D-3} \sqrt{G(r_+)} \right]}{\sqrt{(r^2-r_+^2)(r^2-r_-^2)}}\,dr \, .
\end{align}
In the first term we have isolated a part of the integral that will behave like $\sim 1/(r-r_+)$ in the extremal limit, and so we expect a logarithmic singularity for this term. The second term does not exhibit such behaviour in the extremal limit: the behaviour of the numerator near $r = r_+$ will cancel the blow up due to the denominator. Therefore, near $\epsilon = 0$, it is the asymptotics of the first integral that we must understand.

The first integral converges when integrated between $r_+$ and $\infty$, and so we extend the integration domain $r_{\rm max} \to \infty$. The result can then be expressed in terms of elliptic integrals:
\be 
\int_{r_{+}}^{\infty} \frac{ h(r_+)r_+^{D-3}\sqrt{G(r_+)}}{\sqrt{(r^2-r_+^2)(r^2-r_-^2)}} \, dr= h(r_+)r_+^{D-4} \sqrt{G(r_+)} E(1-\epsilon)
\ee
where $E$ is the elliptic integral of the first kind. The remaining integrals cannot be evaluated in a simple closed form, but luckily this will not trouble us here (yet). Expanding this expression near $\epsilon = 0$ and noting that this will be the dominant contribution to the complexity of formation in this limit, we find that in all dimensions
\be 
\Delta \mathcal{C}_\mathcal{V} \underset{\epsilon \to 0}{\approx} \frac{\Omega_{D-2} h(r_+) r_+^{D-4} \sqrt{G(r_+)}}{G_N R} \log \frac{8}{\epsilon} + \mathcal{O}(\epsilon, \epsilon \log \epsilon).
\ee
It is tempting to expand the prefactor appearing here to understand how it behaves for large black holes. The behaviour is given by
\be 
\frac{\Omega_{D-2} h(r_+) r_+^{D-4} \sqrt{G(r_+)}}{G_N R} \underset{r_+/\ell \to \infty}{\approx} \frac{4 \sqrt{2} \, S }{\sqrt{(N+1)(N+2)}}
\ee
where $S$ is the black hole entropy.  So it is tempting to conclude that the complexity of formation (at least near extremality) is controlled by the entropy. However, the situation is more subtle. First, while the expansion just presented above holds provided $\epsilon \to 0$, it does not follow that the subleading terms in the $\epsilon$ expansion will always be subleading for sufficiently large $r_+/\ell$. What is true is that, for fixed $r_+/\ell$, one can find an $\epsilon$ that is small enough such that the entropy will control the behaviour near extremality. However, in the general situation the entropy does not control the complexity of formation, as we will now explain.

\begin{figure}
\centering
\includegraphics[width=0.6\textwidth]{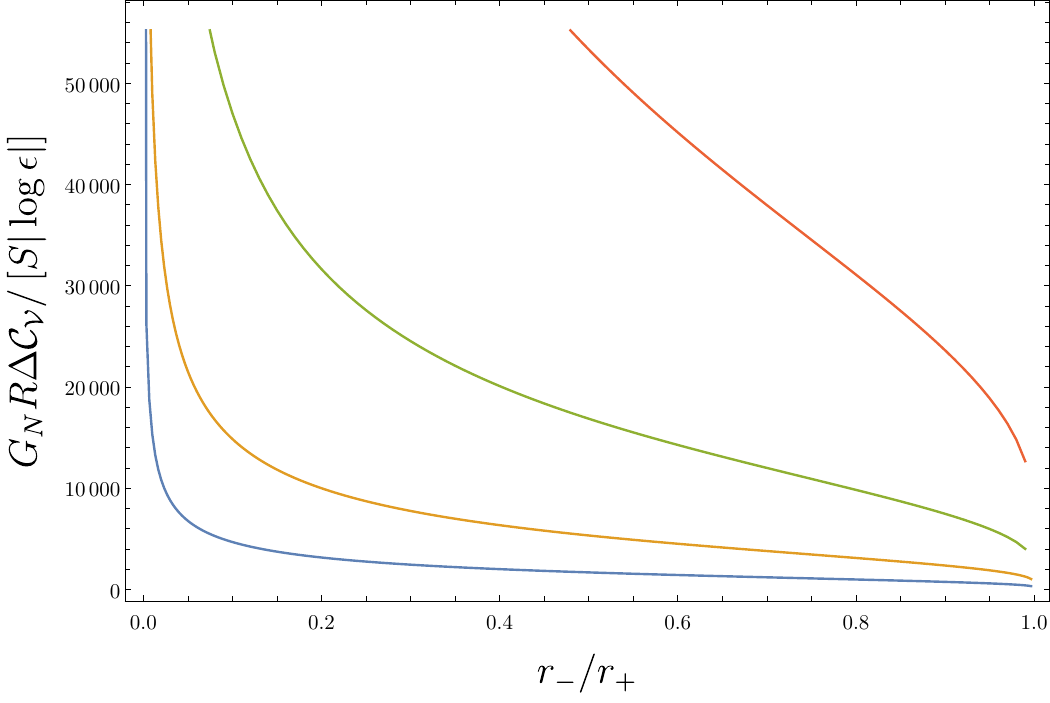}
\caption{A plot of $\Delta \mathcal{C}_\mathcal{V}$ normalized by the entropy in five dimensions. The graph displays four curves corresponding to $r_+/\ell = 10^5, 10^6, 10^7, 10^8$. The curves are plotted as a function of $\epsilon = 1-r_-/r_+$. The value of $r_+/\ell$ increases from the blue curve to the red curve.}
\label{CV-rot-ent}
\end{figure}

The process of understanding the behaviour of the complexity of formation for large black holes involves extracting the leading $r_+/\ell$ dependence of the integrals presented above. Despite a number of attempts, we have been unable to understand this problem from an analytical perspective, and therefore we resort to numerics. In figure~\ref{CV-rot-ent} we show the ratio of the complexity of formation normalized by the entropy for several large values of $r_+/\ell$. It becomes clear that the entropy \textit{does not} control the complexity of rotating formation for large black holes. This figure should be compared with figure~\ref{CVform-charged} to see the stark difference relative to the charged case.

Note that the entropy can be written as
\be 
S = \frac{\Omega_{2N+1}}{4} r_+^{2N+1} \sqrt{(1+ \frac{r_+^2}{\ell^2}(\epsilon-1)^2) P(\epsilon)} 
\ee
where $P(\epsilon)$ is a polynomial in $\epsilon$ that becomes  rather complicated in higher dimensions and the general form is not important. This means that the entropy interpolates between two different scaling regimes. In the limit of slow rotation ($\epsilon \to 1$) the entropy scales as
\be 
S \underset{\epsilon \to 1}{\sim} \left(\frac{r_+}{\ell} \right)^{2N+1} = \left(\frac{r_+}{\ell} \right)^{D-2} 
\ee
for large black holes,  
while in the near extremal limit the entropy scales like
\be 
S \underset{\epsilon \to 0}{\sim} \left(\frac{r_+}{\ell} \right)^{2N+2}  = \left(\frac{r_+}{\ell} \right)^{D-1} 
\ee
for large rotating black holes. Although it is not immediately clear from figure~\ref{CV-rot-ent}, the entropy does match the scaling decently near $r_-/r_+ \approx 0$ --- which is expected since this scaling holds for the Schwarzschild-AdS black hole~\cite{Chapman2017Form} --- but fails miserably closer to extremality.

\begin{figure}
\centering
\includegraphics[width=0.6\textwidth]{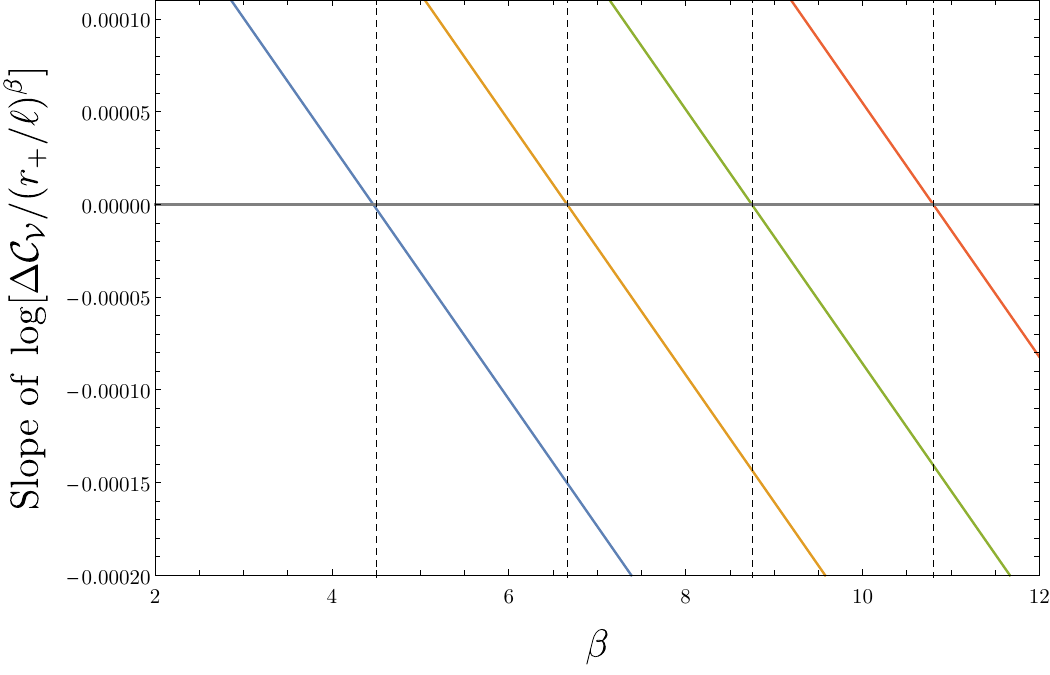}
\caption{The slope of the logarithm of the ratio $R(\beta)$ for rotating black holes in several dimensions. The curves correspond to 5,7, 9, 11 dimensions from left to right, respectively. For each value of $\beta$ the integrals have been evaluated for 500 points laying between $r_+/\ell = 10,000$ and $r_+/\ell = 20,000$. The slope is extracted by performing a linear fit to this data. In all cases we have set $\epsilon = 10^{-10}$ to probe close to extremality. Vertical dashed lines have been added to aid in seeing where the slopes cross the horizontal axis.}
\label{rotatingSlope}
\end{figure}

\begin{table}
\centering
\begin{tabular}{SSS} \toprule
{Dimension} & {Numerical value of $\beta$} & {Thermodynamic volume scaling $V^{(D-2)/(D-1)}$} \\ \midrule
5 & {$4.50000$} & {$9/2 = 4.5$}  \\
7 & {$6.66667$} &  {$20/3 \approx 6.66667$} \\
9 & {$8.75000$} & {$35/4 = 8.75$} \\
11 & {$10.80000$}& {$54/5 = 10.8$}  \\ \midrule
13 & {$12.83333$} & {$77/6 \approx 12.83333$}  \\
15 & {$14.85714$} & {$104/7 \approx 14.85714 $} \\
17 & {$16.87500$} & {$135/8 \approx 16.87500 $} \\
19 & {$18.88889$} &{$170/9 \approx 18.88889$} \\ \midrule
21 & {$20.90000$}& {$209/10 = 20.9$} \\
23 & {$22.90909$} & {$252/11 \approx 22.90909 $}   \\
25 & {$24.91667$} &  {$299/12 \approx 24.91667$}\\
27 & {$26.92308$} & {$350/13 \approx 26.92308$} \\
 \bottomrule
\end{tabular}
\caption{Table of numerically calculated values of $\beta$ compared with the scaling of the thermodynamic volume $V^{(D-2)/(D-1)}$ for large $r_+/\ell$. Here we have computed  numerically the values of $\beta$ according to the method outlined in the text. The data is obtained by evaluating the complexity of formation between $r_+/\ell = 10^{10}$ and $r_+/\ell = 10^{20}$ and we have fixed $\epsilon = 10^{-10}$, so we are considering the situation very close to extremality. The numerical values agree with the scaling of the thermodynamic volume to at least five decimal places in all cases. By pushing the domain of $r_+/\ell$ to large values, the agreement becomes even better. Note that in all cases the scaling differs from the scaling of the entropy which behaves like $(r_+/\ell)^{D-1}$ for large $r_+/\ell$ at fixed $\epsilon$ near extremality.}
\label{powerTab}
\end{table}

Using the same numerical scheme described in the previous section for charged black holes we can understand how the complexity of formation behaves as a function of $r_+/\ell$ for large black holes. The objective is to understand this scaling close to extremality where the departure from entropic scaling is most severe. To briefly recap, the process involves studying the ratio
\be 
R(\beta) = \frac{\Delta \mathcal{C}_\mathcal{V}}{(r_+/\ell)^\beta}
\ee
and numerically determining the value of $\beta$ so that $R(\beta)$ exhibits no dependence on $r_+/\ell$ (when $r_+/\ell$ is large). We show a sample of this numerical scheme in figure~\ref{rotatingSlope}, and tabulate the results up to 27 dimensions in table~\ref{powerTab}. The conclusion is that in spacetime dimension $D$ the complexity of formation scales like
\be 
\Delta \mathcal{C}_\mathcal{V} \underset{\epsilon \to 0}{\sim} \left(\frac{r_+}{\ell} \right)^{(D+1)(D-2)/(D-1)} \, 
\ee
for large black holes near extremality.

\begin{figure}
\centering
\includegraphics[width=0.6\textwidth]{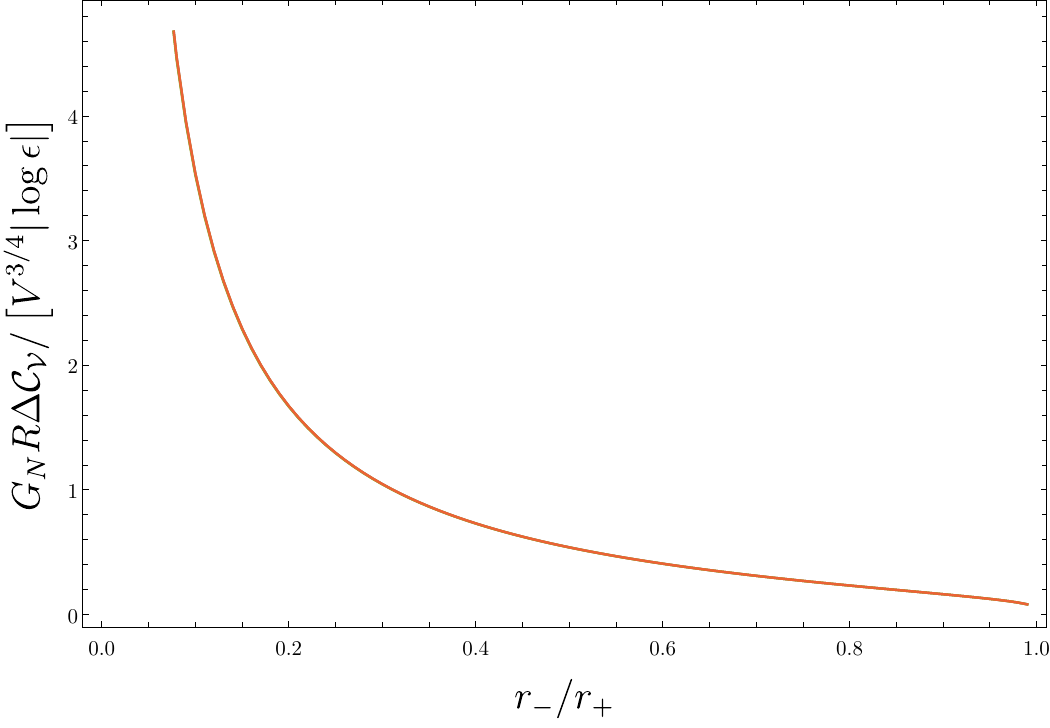}
\caption{ A plot of $\Delta \mathcal{C}_\mathcal{V}$ normalized by the thermodynamic volume to the appropriate power in five dimensions. The graph displays four curves corresponding to $r_+/\ell = 10^5, 10^6, 10^7, 10^8$, the same as those shown in figure~\ref{CV-rot-ent}, but these curves cannot be distinguished from one another here. The curves are plotted as a function of $\epsilon = 1-r_-/r_+$. The value of $r_+/\ell$ increases from the blue curve to the red curve.}
\label{CV-logPlot}
\end{figure}

It is obvious from table~\ref{powerTab} that the scaling of $\Delta \mathcal{C}_\mathcal{V}$ is different from the scaling of the entropy. The question then becomes whether or not there is a thermodynamic parameter that \textit{does} have this scaling. As already hinted in table~\ref{powerTab}, the answer is that the thermodynamic volume possesses this scaling for large black holes. Isolating the dependence on $r_+$, the thermodynamic volume can be written schematically as
\be 
V = \frac{\Omega_{2N+1}}{3(N+1)}\left[r_+^{2N+2} H(\epsilon) + \frac{r_+^{2N+4}}{\ell^2} (\epsilon-1)^2 K(\epsilon)  \right]
\ee
where again $H(\epsilon)$ and $K(\epsilon)$ are messy polynomials in $\epsilon$ whose form does not matter for the information we need here. These polynomials vanish nowhere on the range $\epsilon \in [0, 1]$. We therefore see that the thermodynamic volume also has two scaling regimes, behaving as $\epsilon \to 1$ like
\be 
V \underset{\epsilon \to 1}{\sim} \left(\frac{r_+}{\ell} \right)^{2N+2} = \left(\frac{r_+}{\ell} \right)^{D-1} 
\ee 
for large black holes, while near extremality it scales like
\be 
V \underset{\epsilon \to 0}{\sim} \left(\frac{r_+}{\ell} \right)^{2N+4} = \left(\frac{r_+}{\ell} \right)^{D+1} \, .
\ee 
If we then notice that a power of the thermodynamic volume has the appropriate scaling:
\be 
V^{(D-2)/(D-1)} \underset{\epsilon \to 1}{\sim} \left(\frac{r_+}{\ell} \right)^{D-2} \quad \text{and} \quad V^{(D-2)/(D-1)} \underset{\epsilon \to 0}{\sim} \left(\frac{r_+}{\ell} \right)^{(D-2)(D+1)/(D-1)} \, .
\ee
The scaling of the thermodynamic volume to this power interpolates precisely between the two scaling regimes of the complexity of formation. We show this graphically for five dimensions in figure~\ref{CV-logPlot}.

There are a few important things to note here:
\begin{itemize}
\item The power of thermodynamic volume is natural. Recall that the thermodynamic volume has dimensionality $[{\rm length}]^{D-1}$, therefore to obtain a quantity that has the correct dimensions of $[{\rm length}]^{D-2}$ requires precisely this power.

\item The scaling with thermodynamic volume is consistent with the entropic scaling observed for charged black holes and the Schwarzschild black hole~\cite{Chapman2017Form, Carmi2017}. This is because those solutions satisfy
\be 
S \sim V^{(D-2)/(D-1)} \, .
\ee
In other words, for those solutions the thermodynamic volume and the entropy are not independent and so the results can be written in terms of either quantity. For the rotating black holes these quantities are truly independent and we observe that it is actually the expression written in terms of the thermodynamic volume that prevails.

\item The convergence to ``volumetric scaling'' is slower for rotating black holes than it is for charged black holes. In the charged case the subleading terms die off at least as fast as $\ell/r_+$, while in the rotating case they die off like $\sqrt{\ell/r_+}$.

\item To the best of our knowledge there is no {\it a priori} reason to expect that the thermodynamic volume should be related to an extremal volume in a black hole spacetime. However, deriving such a relationship could contribute to a proof of our relationship for the complexity of formation in general situations.

\item The conjectured reverse isoperimetric inequality \cite{Cvetic:2010jb} bounds the entropy in terms of the thermodynamic volume:
\be 
\mathcal{R}^{D-2} =  \left(\frac{(D-1) V}{\Omega_{D-2}}\right)^{(D-2)/(D-1)} \left(\frac{\Omega_{D-2}}{4G_N S} \right) \ge  1\,  .
\ee
If our result is general, {\it i.e.} the complexity of formation generally scales with the volume for large black holes, then the reverse isoperimetric inequality can be interpreted as the statement that the entropy provides a lower bound for the complexity of formation. This bound is saturated for static black holes, but more complicated black holes have a larger complexity of formation than naively suggested by their degrees of freedom (entropy).

\end{itemize}

\subsection{Rotating black holes: complexity equals action}
It is now natural to ask whether this scaling with the thermodynamic volume is universal to both complexity proposals, or if it is a peculiar behaviour associated with the CV proposal. Recall that, as shown in section \ref{sec4.1}, the complexity of formation in the CA conjecture is given by
\begin{align}
\pi \Delta \mathcal{C}_\mathcal{A} =& \frac{\Lambda \Omega_{2N+1}}{2 (N+1) (2N+1) \pi G_N} \bigg[
\int_{r_{m_0}}^{\infty} r^{2N+1} \left( g(r)^2 h(r)  - \frac{r}{1 + r^2/\ell^2} \right)  dr - \int_0^{r_{m_0}} \frac{r^{2(N+1)}}{1 + r^2/\ell^2} dr \bigg] \nonumber\\&-\frac{\Omega_{2N+1} (r_{m_0})^{2N+1}}{2 \pi G_N (2N+1)}
 -\frac{\Omega_{2N + 1}}{4 \pi G_N} (r_{m_0})^{2N} h(r_{m_0}) \log \ell_{\rm ct}^2 \Theta(r_{m_0})^2 |f(r_{m_0})^2|  
\nonumber\\ &- \frac{\Omega_{2N+1}}{2 \pi G_N} \int_{r_{m_0}}^{\infty}  \, r^{2N} \left[  h(r) \frac{\Theta'}{\Theta} + 1 \right]dr .
\end{align}
\begin{figure}
\centering
\includegraphics[width=0.6\textwidth]{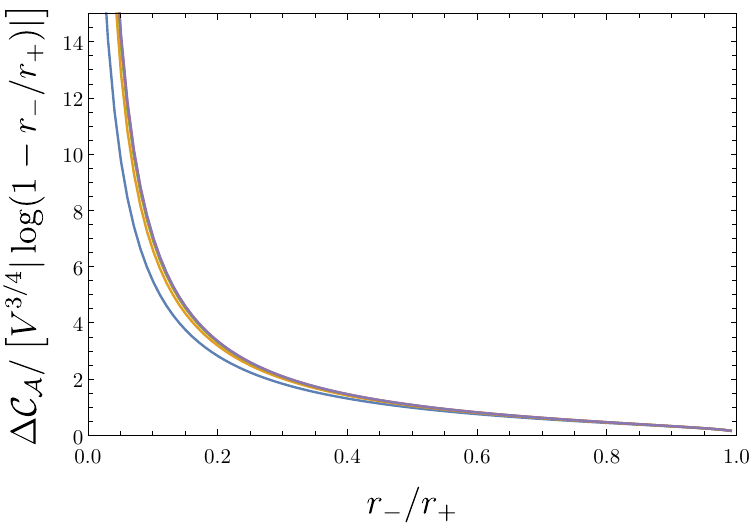}
\caption{A plot showing the CA complexity of formation normalized by the thermodynamic volume as a function of the ratio $r_-/r_+$ in five dimensions. The plot shows curves for fixed $r_+/\ell = 10, 10^2, 10^3, 10^4, 10^5, 10^6$ and $10^7$, however after $r_+/\ell = 1000$ the curves are visually indistinguishable. Here we have set $\ell_{\rm ct} = \ell$.}
\label{CAplot}
\end{figure}
The most difficult part of the CA computation is the determination of $r_{m_0}$. In some instances, particularly in the limit $r_-/r_+ \to 0$, accurate determination of this parameter requires hundreds of digits of precision in the numerics. This technicality has limited our ability to probe the behaviour of the complexity of formation within the CA conjecture as broadly as the CV conjecture. However, we show in Fig.~\ref{CAplot} the result of the action computation in five dimensions. The plot makes clear that the thermodynamic volume controls the scaling of $\Delta \mathcal{C}_\mathcal{A}$ for large black holes, just as in the CV conjecture. While it was possible to compute the behaviour in various higher dimensions for the CV case, this is more difficult in the CA scenario. Nonetheless, we have confirmed the scaling with thermodynamic volume in seven dimensions, which suggests the same trend holds in general for CA.

\bibliographystyle{JHEP}
\bibliography{myrefs2}
\end{document}